\documentclass[12pt,a4paper]{article}
\pdfoutput=1
\usepackage{jheppub}
\usepackage[utf8]{inputenc}
\usepackage{mathtools}
\usepackage{caption}
\usepackage{subcaption}
\usepackage{graphicx}
\usepackage{booktabs} %nice tables

\usepackage{xkeyval}

\usepackage{tikz,fp}
\usetikzlibrary{external,fixedpointarithmetic,decorations.pathmorphing,positioning,calc,fadings,decorations.markings,decorations.pathreplacing,patterns,arrows}
\tikzset{
  label distance=-3pt,
  >=stealth,
  inner sep=1.5pt,
  boson/.style={
    decoration={snake, segment length=2mm, amplitude=0.5mm},
    decorate,
  },
  quark/.style={
    postaction={decorate},
    decoration={markings,mark=at position .5 with {\draw[->] (-0.01,0) -- (0.07,0);}}
  },
  aquark/.style={
    postaction={decorate},
    decoration={markings,mark=at position .5 with {\draw[->] (0.01,0) -- (-0.07,0);}}
  },
  gluonOld/.style={
    line width=0.7pt,
    decorate,
    decoration={coil,aspect=-0.5,amplitude=-2pt,segment length=2.2pt}
  },
  gluon/.style={
    line width=0.5pt,
    decorate,
    decoration={coil,aspect=-0.75,amplitude=-1.5pt,segment length=2pt}
  },
  agluon/.style={
    line width=0.5pt,
    decorate,
    decoration={coil,aspect=0.75,amplitude=1.5pt,segment length=2pt}
  },
  scalar/.style={
    dashed
  },
  dscalar/.style={
    dashed,
    postaction={decorate},
    decoration={markings,mark=at position 0.65 with {\draw[->] (-0.01,0) -- (0.07,0);}}
  },
  adscalar/.style={
    dashed,
    postaction={decorate},
    decoration={markings,mark=at position 0.65 with {\draw[->] (-0.01,0) -- (0.07,0);}}
  },
  line/.style={
  },
  doubleline/.style={
    double
  },
  middleArrow/.style={
    draw=white,
    decoration={markings,mark=at position 0.65 with{\arrow[scale=1,black]{>}}},
    decorate
  },
  dot/.style={
  	postaction={decorate},
    decoration={markings,mark=between positions 0.5 and 0.5 step 1 with {\draw[fill] (0,0) circle [radius=1.5pt];}}  
  },
  dots/.style={
  	postaction={decorate},
    decoration={markings,mark=between positions 0.3 and 0.6 step 0.3 with {\draw[fill] (0,0) circle [radius=1.5pt];}}  
  },
  dotss/.style={
  	postaction={decorate},
    decoration={markings,mark=between positions 0.25 and 0.75 step 0.25 with {\draw[fill] (0,0) circle [radius=1.5pt];}}  
  },
  blob/.style={
  	pattern=north west lines,
    draw=black
  }
}

\makeatletter
\def\gSetStdExtTypes#1{\presetkeys{gTypes}{eA=#1,eB=#1,eC=#1,eD=#1,eE=#1}{}}
\def\gSetStdIntTypes#1{\presetkeys{gTypes}{iA=#1,iB=#1,iC=#1,iD=#1,iE=#1,iF=#1,iG=#1,iH=#1,iI=#1,iJ=#1}{}}
\def\gSetStdTypes#1{\gSetStdExtTypes{#1}\gSetStdIntTypes{#1}}

\define@cmdkeys{general}{scale}

\define@cmdkeys{general}{rotate}
\define@cmdkeys{general}{font}
\define@cmdkeys{general}{yshift}
\presetkeys{general}{scale=1, rotate=0, font=\scriptsize, yshift=0pt}{}

\define@cmdkeys{gTypes}{eA,eB,eC,eD,eE,iA,iB,iC,iD,iE,iF,iG,iH,iI,iJ}
\gSetStdTypes{gluon}
\define@key{pre}{all}{\gSetStdTypes{#1}}
\define@key{pre}{allE}{\gSetStdExtTypes{#1}}
\define@key{pre}{allI}{\gSetStdIntTypes{#1}}

\define@cmdkeys{gLabels}{eLA,eLB,eLC,eLD,eLE,iLA,iLB,iLC,iLD,iLE,iLF,iLG,iLH,iLI,iLJ}
\presetkeys{gLabels}{eLA={},eLB={},eLC={},eLD={},eLE={},iLA={},iLB={},iLC={},iLD={},iLE={},iLF={},iLG={},iLH={},iLI={},iLJ={}}{}

\def\gTreeTri{\@ifnextchar[\@gTreeTri{\@gTreeTri[]}}
\def\@gTreeTri[#1]#2{%
  \setkeys*{pre}{#1}%
  \setrmkeys{general,gTypes,gLabels}%
  \begin{tikzpicture}
    [line width=1pt,
    baseline={([yshift=-0.5ex+\cmdKV@general@yshift]current bounding box.center)},
    scale=\cmdKV@general@scale,
    rotate=\cmdKV@general@rotate,
    font=\cmdKV@general@font]
    \draw[\cmdKV@gTypes@eA] (0.7,0.3) -- (0.9,0.6) node[label=(\cmdKV@general@rotate):\cmdKV@gLabels@eLA] {};
    \draw[\cmdKV@gTypes@eB] (0.7,0.3) -- (0.9,0) node[label=(\cmdKV@general@rotate):\cmdKV@gLabels@eLB] {};
    \draw[\cmdKV@gTypes@eC] (0.7,0.3) -- (0.4,0.3) node[label=(-180+\cmdKV@general@rotate):\cmdKV@gLabels@eLC] {};
    #2
  \end{tikzpicture}
  \gSetStdTypes{gluon}
}

\def\gTreeS{\@ifnextchar[\@gTreeS{\@gTreeS[]}}
\def\@gTreeS[#1]#2{%
  \setkeys*{pre}{#1}%
  \setrmkeys{general,gTypes,gLabels}%
  \begin{tikzpicture}
    [line width=1pt,
    baseline={([yshift=-0.5ex+\cmdKV@general@yshift]current bounding box.center)},
    scale=\cmdKV@general@scale,
    rotate=\cmdKV@general@rotate,
    font=\cmdKV@general@font]
    \draw[\cmdKV@gTypes@eA] (0.7,0.3) -- (0.9,0.6) node[label=(\cmdKV@general@rotate):\cmdKV@gLabels@eLA] {};
    \draw[\cmdKV@gTypes@eB] (0.7,0.3) -- (0.9,0) node[label=(\cmdKV@general@rotate):\cmdKV@gLabels@eLB] {};
    \draw[\cmdKV@gTypes@eC] (0.2,0.3) -- (0,0) node[label=(-180+\cmdKV@general@rotate):\cmdKV@gLabels@eLC] {};
    \draw[\cmdKV@gTypes@eD] (0.2,0.3) -- (0,0.6) node[label=(180+\cmdKV@general@rotate):\cmdKV@gLabels@eLD] {};
    \draw[label distance=0,\cmdKV@gTypes@iA] (0.7,0.3) -- node[label=(90+\cmdKV@general@rotate):\cmdKV@gLabels@iLA] {} (0.2,0.3);
    #2
  \end{tikzpicture}
  \gSetStdTypes{gluon}
}

\def\gTreeT{\@ifnextchar[\@gTreeT{\@gTreeT[]}}
\def\@gTreeT[#1]#2{%
  \setkeys*{pre}{#1}%
  \setrmkeys{general,gTypes,gLabels}%
  \begin{tikzpicture}
    [line width=1pt,
    baseline={([yshift=-0.5ex+\cmdKV@general@yshift]current bounding box.center)},
    scale=\cmdKV@general@scale,
    rotate=\cmdKV@general@rotate,
    font=\cmdKV@general@font]
    \draw[\cmdKV@gTypes@eA] (0.45,0.5) -- (0.9,0.6) node[label=(\cmdKV@general@rotate):\cmdKV@gLabels@eLA] {};
    \draw[\cmdKV@gTypes@eB] (0.45,0.1) -- (0.9,0) node[label=(\cmdKV@general@rotate):\cmdKV@gLabels@eLB] {};
    \draw[\cmdKV@gTypes@eC] (0.45,0.1) -- (0,0) node[label=(-180+\cmdKV@general@rotate):\cmdKV@gLabels@eLC] {};
    \draw[\cmdKV@gTypes@eD] (0.45,0.5) -- (0,0.6) node[label=(180+\cmdKV@general@rotate):\cmdKV@gLabels@eLD] {};
    \draw[label distance=0,\cmdKV@gTypes@iA] (0.45,0.5) -- node[label=(\cmdKV@general@rotate):\cmdKV@gLabels@iLA] {} (0.45,0.1);
    #2
  \end{tikzpicture}
  \gSetStdTypes{gluon}
}

\def\gTreeU{\@ifnextchar[\@gTreeU{\@gTreeU[]}}
\def\@gTreeU[#1]#2{%
  \setkeys*{pre}{#1}%
  \setrmkeys{general,gTypes,gLabels}%
  \begin{tikzpicture}
    [line width=1pt,
    baseline={([yshift=-0.5ex+\cmdKV@general@yshift]current bounding box.center)},
    scale=\cmdKV@general@scale,
    rotate=\cmdKV@general@rotate,
    font=\cmdKV@general@font]
    \draw[\cmdKV@gTypes@eA] (0.45,0.5) -- (0.9,0.6) node[label=(\cmdKV@general@rotate):\cmdKV@gLabels@eLA] {};
    \draw[\cmdKV@gTypes@eB] (0.45,0.1) -- (0.9,0) node[label=(\cmdKV@general@rotate):\cmdKV@gLabels@eLB] {};
    \draw[\cmdKV@gTypes@eC] (0.45,0.5) -- (0,0) node[label=(-180+\cmdKV@general@rotate):\cmdKV@gLabels@eLC] {};
    \draw[\cmdKV@gTypes@eD] (0.45,0.1) -- (0,0.6) node[label=(180+\cmdKV@general@rotate):\cmdKV@gLabels@eLD] {};
    \draw[label distance=0,\cmdKV@gTypes@iA] (0.45,0.5) -- node[label=(\cmdKV@general@rotate):\cmdKV@gLabels@iLA] {} (0.45,0.1);
    #2
  \end{tikzpicture}
  \gSetStdTypes{gluon}
}

\def\gTadA{\@ifnextchar[\@gTadA{\@gTadA[]}}
\def\@gTadA[#1]#2{%
  \setkeys*{pre}{#1}%
  \setrmkeys{general,gTypes,gLabels}%
  \begin{tikzpicture}
    [line width=1pt,
    baseline={([yshift=-0.5ex+\cmdKV@general@yshift]current bounding box.center)},
    scale=\cmdKV@general@scale,
    rotate=\cmdKV@general@rotate,
    font=\cmdKV@general@font]
    \draw[\cmdKV@gTypes@eA] (0.7,0.3) -- (0.9,0.6) node[label=(\cmdKV@general@rotate):\cmdKV@gLabels@eLA] {};
    \draw[\cmdKV@gTypes@eB] (0.7,0.3) -- (0.9,0) node[label=(\cmdKV@general@rotate):\cmdKV@gLabels@eLB] {};
    \draw[\cmdKV@gTypes@eC] (0.3,0.3) -- (0.2,0) node[label=(-180+\cmdKV@general@rotate):\cmdKV@gLabels@eLC] {};
    \draw[\cmdKV@gTypes@eD] (-0.15,0.45) -- (-0.5,0.6) node[label=(180+\cmdKV@general@rotate):\cmdKV@gLabels@eLD] {};
    \draw[label distance=0,\cmdKV@gTypes@iA] (0.7,0.3) -- node[label=(90+\cmdKV@general@rotate):\cmdKV@gLabels@iLA] {} (0.3,0.3);
    \draw[label distance=0,\cmdKV@gTypes@iB] (0.3,0.3) -- node[label=(-100+\cmdKV@general@rotate):\cmdKV@gLabels@iLB] {} (-0.15,0.45);
    \draw[label distance=0,\cmdKV@gTypes@iC] (-0.15,0.45) -- node[label=(135+\cmdKV@general@rotate):\cmdKV@gLabels@iLC] {} (0.18,0.69);
    \draw[label distance=-0.5,\cmdKV@gTypes@iD] (0.18,0.69) .. controls ($(0.18,0.69) + (-0.12,0.12)$) and ($(0.35,0.87) + (-0.12,0.12)$) .. (0.35,0.87) node[label=(\cmdKV@general@rotate):\cmdKV@gLabels@iLD] {} .. controls ($(0.35,0.87) + (0.12,-0.12)$) and ($(0.18,0.69) + (0.12,-0.12)$) .. (0.18,0.69);
    #2
  \end{tikzpicture}
  \gSetStdTypes{gluon}
}

\def\gTadB{\@ifnextchar[\@gTadB{\@gTadB[]}}
\def\@gTadB[#1]#2{%
  \setkeys*{pre}{#1}%
  \setrmkeys{general,gTypes,gLabels}%
  \begin{tikzpicture}
    [line width=1pt,
    baseline={([yshift=-0.5ex+\cmdKV@general@yshift]current bounding box.center)},
    scale=\cmdKV@general@scale,
    rotate=\cmdKV@general@rotate,
    font=\cmdKV@general@font]
    \draw[\cmdKV@gTypes@eA] (0.7,0.3) -- (0.9,0.6) node[label=(\cmdKV@general@rotate):\cmdKV@gLabels@eLA] {};
    \draw[\cmdKV@gTypes@eB] (0.7,0.3) -- (0.9,0) node[label=(\cmdKV@general@rotate):\cmdKV@gLabels@eLB] {};
    \draw[\cmdKV@gTypes@eC] (0,0.3) -- (-0.2,0) node[label=(-180+\cmdKV@general@rotate):\cmdKV@gLabels@eLC] {};
    \draw[\cmdKV@gTypes@eD] (0,0.3) -- (-0.2,0.6) node[label=(180+\cmdKV@general@rotate):\cmdKV@gLabels@eLD] {};
    \draw[label distance=0,\cmdKV@gTypes@iA] (0.7,0.3) -- node[label=(-90+\cmdKV@general@rotate):\cmdKV@gLabels@iLA] {} (0.35,0.3);
    \draw[label distance=0,\cmdKV@gTypes@iB] (0.35,0.3) -- node[label=(\cmdKV@general@rotate):\cmdKV@gLabels@iLB] {} (0.35,0.6);
    \draw[label distance=-0.5,\cmdKV@gTypes@iC] (0.35,0.6) .. controls (0.18,0.6) and (0.18,0.85) .. (0.35,0.85) .. controls (0.52,0.85) and (0.52,0.6) .. node[label=(\cmdKV@general@rotate):\cmdKV@gLabels@iLC] {} (0.35,0.6);
    \draw[label distance=0,\cmdKV@gTypes@iD] (0.35,0.3) -- node[label=(-90+\cmdKV@general@rotate):\cmdKV@gLabels@iLD] {} (0,0.3);
    #2
  \end{tikzpicture}
  \gSetStdTypes{gluon}
}
\def\gBubA{\@ifnextchar[\@gBubA{\@gBubA[]}}
\def\@gBubA[#1]#2{%
  \setkeys*{pre}{#1}%
  \setrmkeys{general,gTypes,gLabels}%
  \begin{tikzpicture}
    [line width=1pt,
    baseline={([yshift=-0.5ex+\cmdKV@general@yshift]current bounding box.center)},
    scale=\cmdKV@general@scale,
    rotate=\cmdKV@general@rotate,
    font=\cmdKV@general@font]
    \draw[\cmdKV@gTypes@eA] (0.9,0.7) -- (1.2,0.7) node[label=(\cmdKV@general@rotate):\cmdKV@gLabels@eLA] {};
    \draw[\cmdKV@gTypes@eB] (0.5,0.7) -- (0.2,0.7) node[label=(180+\cmdKV@general@rotate):\cmdKV@gLabels@eLB] {};
    \draw[label distance=0,\cmdKV@gTypes@iA] (0.9,0.7) .. controls (0.9,0.45) and (0.5,0.45) .. node[label=(-90+\cmdKV@general@rotate):\cmdKV@gLabels@iLA] {} (0.5,0.7);
    \draw[label distance=0,\cmdKV@gTypes@iB] (0.5,0.7) .. controls (0.5,0.95) and (0.9,0.95) .. node[label=(90+\cmdKV@general@rotate):\cmdKV@gLabels@iLB] {} (0.9,0.7);
    #2
  \end{tikzpicture}
  \gSetStdTypes{gluon}
}
\def\gBubB{\@ifnextchar[\@gBubB{\@gBubB[]}}
\def\@gBubB[#1]#2{%
  \setkeys*{pre}{#1}%
  \setrmkeys{general,gTypes,gLabels}%
  \begin{tikzpicture}
    [line width=1pt,
    baseline={([yshift=-0.5ex+\cmdKV@general@yshift]current bounding box.center)},
    scale=\cmdKV@general@scale,
    rotate=\cmdKV@general@rotate,
    font=\cmdKV@general@font]
    \draw[\cmdKV@gTypes@eA] (1.2,0.7) -- (1.4,1) node[label=(\cmdKV@general@rotate):\cmdKV@gLabels@eLA] {};
    \draw[\cmdKV@gTypes@eB] (1.2,0.7) -- (1.4,0.4) node[label=(\cmdKV@general@rotate):\cmdKV@gLabels@eLB] {};
    \draw[\cmdKV@gTypes@eC] (0.2,0.7) -- (0,0.4) node[label=(-180+\cmdKV@general@rotate):\cmdKV@gLabels@eLC] {};
    \draw[\cmdKV@gTypes@eD] (0.2,0.7) -- (0,1) node[label=(180+\cmdKV@general@rotate):\cmdKV@gLabels@eLD] {};
    \draw[label distance=0,\cmdKV@gTypes@iA] (1.2,0.7) -- node[label=(90+\cmdKV@general@rotate):\cmdKV@gLabels@iLA] {} (0.9,0.7);
    \draw[label distance=0,\cmdKV@gTypes@iB] (0.9,0.7) .. controls (0.9,0.45) and (0.5,0.45) .. node[label=(-90+\cmdKV@general@rotate):\cmdKV@gLabels@iLB] {} (0.5,0.7);
    \draw[label distance=0,\cmdKV@gTypes@iC] (0.5,0.7) .. controls (0.5,0.95) and (0.9,0.95) .. node[label=(90+\cmdKV@general@rotate):\cmdKV@gLabels@iLC] {} (0.9,0.7);
    \draw[label distance=0,\cmdKV@gTypes@iD] (0.5,0.7) -- node[label=(90+\cmdKV@general@rotate):\cmdKV@gLabels@iLD] {} (0.2,0.7);
    #2
  \end{tikzpicture}
  \gSetStdTypes{gluon}
}

\def\gBubC{\@ifnextchar[\@gBubC{\@gBubC[]}}
\def\@gBubC[#1]#2{%
  \setkeys*{pre}{#1}%
  \setrmkeys{general,gTypes,gLabels}%
  \begin{tikzpicture}
    [line width=1pt,
    baseline={([yshift=-0.5ex+\cmdKV@general@yshift]current bounding box.center)},
    scale=\cmdKV@general@scale,
    rotate=\cmdKV@general@rotate,
    font=\cmdKV@general@font]
    \draw[\cmdKV@gTypes@eA] (0.7,0.3) -- (0.9,0.6) node[label=(\cmdKV@general@rotate):\cmdKV@gLabels@eLA] {};
    \draw[\cmdKV@gTypes@eB] (0.7,0.3) -- (0.9,0) node[label=(\cmdKV@general@rotate):\cmdKV@gLabels@eLB] {};
    \draw[\cmdKV@gTypes@eC] (0.3,0.3) -- (0.2,0) node[label=(-180+\cmdKV@general@rotate):\cmdKV@gLabels@eLC] {};
    \draw[\cmdKV@gTypes@eD] (-0.3,0.5) -- (-0.5,0.6) node[label=(180+\cmdKV@general@rotate):\cmdKV@gLabels@eLD] {};
    \draw[label distance=0,\cmdKV@gTypes@iA] (0.7,0.3) -- node[label=(90+\cmdKV@general@rotate):\cmdKV@gLabels@iLA] {} (0.3,0.3);
    \draw[label distance=0,\cmdKV@gTypes@iB] (0.3,0.3) -- node[label=(80+\cmdKV@general@rotate):\cmdKV@gLabels@iLB] {} (0,0.4);
    \draw[label distance=0,\cmdKV@gTypes@iC] (0,0.4) .. controls (-0.08,0.21) and (-0.38,0.3) .. node[label=(-90+\cmdKV@general@rotate):\cmdKV@gLabels@iLC] {} (-0.3,0.5);
    \draw[label distance=0,\cmdKV@gTypes@iD] (-0.3,0.5) .. controls (-0.22,0.65) and (0.03,0.6) .. node[label=(90+\cmdKV@general@rotate):\cmdKV@gLabels@iLD] {} (0,0.4);
    #2
  \end{tikzpicture}
  \gSetStdTypes{gluon}
}
\def\gTriA{\@ifnextchar[\@gTriA{\@gTriA[]}}
\def\@gTriA[#1]#2{%
  \setkeys*{pre}{#1}%
  \setrmkeys{general,gTypes,gLabels}%
  \begin{tikzpicture}
    [line width=1pt,
    baseline={([yshift=-0.5ex+\cmdKV@general@yshift]current bounding box.center)},
    scale=\cmdKV@general@scale,
    rotate=\cmdKV@general@rotate,
    font=\cmdKV@general@font]
    \draw[\cmdKV@gTypes@eA] (0:0.3) -- (0:0.55) node[label=(\cmdKV@general@rotate):\cmdKV@gLabels@eLA] {};
    \draw[\cmdKV@gTypes@eB] (-120:0.3) -- (-120:0.55) node[label=(-180+\cmdKV@general@rotate):\cmdKV@gLabels@eLB] {};
    \draw[\cmdKV@gTypes@eC] (120:0.3) -- (120:0.55) node[label=(180+\cmdKV@general@rotate):\cmdKV@gLabels@eLC] {};
    \draw[label distance=0,\cmdKV@gTypes@iA] (0:0.3) -- node[label=(-60+\cmdKV@general@rotate):\cmdKV@gLabels@iLA] {} (-120:0.3);
    \draw[label distance=0,\cmdKV@gTypes@iB] (-120:0.3) -- node[label=(180+\cmdKV@general@rotate):\cmdKV@gLabels@iLB] {} (120:0.3);
    \draw[label distance=0,\cmdKV@gTypes@iC] (120:0.3) -- node[label=(60+\cmdKV@general@rotate):\cmdKV@gLabels@iLC] {} (0:0.3);
    #2
  \end{tikzpicture}
  \gSetStdTypes{gluon}
}
\def\gTriC{\@ifnextchar[\@gTriC{\@gTriC[]}}
\def\@gTriC[#1]#2{%
  \setkeys*{pre}{#1}%
  \setrmkeys{general,gTypes,gLabels}%
  \begin{tikzpicture}
    [line width=1pt,
    baseline={([yshift=-0.5ex+\cmdKV@general@yshift]current bounding box.center)},
    scale=\cmdKV@general@scale,
    rotate=\cmdKV@general@rotate,
    font=\cmdKV@general@font]
    \draw[\cmdKV@gTypes@eA] (1,0.7) -- (1.3,1.1) node[label=(\cmdKV@general@rotate):\cmdKV@gLabels@eLA] {};
    \draw[\cmdKV@gTypes@eB] (1,0.7) -- (1.3,0.3) node[label=(\cmdKV@general@rotate):\cmdKV@gLabels@eLB] {};
    \draw[\cmdKV@gTypes@eC] (0.3,0.45) -- (0,0.3) node[label=(-180+\cmdKV@general@rotate):\cmdKV@gLabels@eLC] {};
    \draw[\cmdKV@gTypes@eD] (0.3,0.95) -- (0,1.1) node[label=(180+\cmdKV@general@rotate):\cmdKV@gLabels@eLD] {};
    \draw[label distance=0,\cmdKV@gTypes@iA] (1,0.7) -- node[label=(90+\cmdKV@general@rotate):\cmdKV@gLabels@iLA] {} (0.7,0.7);
    \draw[label distance=0,\cmdKV@gTypes@iB] (0.7,0.7) -- node[label=(-70+\cmdKV@general@rotate):\cmdKV@gLabels@iLB] {} (0.3,0.45);
    \draw[label distance=0,\cmdKV@gTypes@iC] (0.3,0.45) -- node[label=(180+\cmdKV@general@rotate):\cmdKV@gLabels@iLC] {} (0.3,0.95);
    \draw[label distance=0,\cmdKV@gTypes@iD] (0.3,0.95) -- node[label=(70+\cmdKV@general@rotate):\cmdKV@gLabels@iLD] {} (0.7,0.7);
    #2
  \end{tikzpicture}
  \gSetStdTypes{gluon}
}
\def\gBox{\@ifnextchar[\@gBox{\@gBox[]}}
\def\@gBox[#1]#2{%
  \setkeys*{pre}{#1}%
  \setrmkeys{general,gTypes,gLabels}%
  \begin{tikzpicture}
    [line width=1pt,
    baseline={([yshift=-0.5ex+\cmdKV@general@yshift]current bounding box.center)},
    scale=\cmdKV@general@scale,
    rotate=\cmdKV@general@rotate,
    font=\cmdKV@general@font]
    \draw[\cmdKV@gTypes@eA] (0.75,0.75) -- (1,1) node[label=(\cmdKV@general@rotate):\cmdKV@gLabels@eLA] {};
    \draw[\cmdKV@gTypes@eB] (0.75,0.25) -- (1,0) node[label=(\cmdKV@general@rotate):\cmdKV@gLabels@eLB] {};
    \draw[\cmdKV@gTypes@eC] (0.25,0.25) -- (0,0) node[label=(-180+\cmdKV@general@rotate):\cmdKV@gLabels@eLC] {};
    \draw[\cmdKV@gTypes@eD] (0.25,0.75) -- (0,1) node[label=(180+\cmdKV@general@rotate):\cmdKV@gLabels@eLD] {};
    \draw[label distance=0,\cmdKV@gTypes@iA] (0.75,0.75) -- node[label=(\cmdKV@general@rotate):\cmdKV@gLabels@iLA] {} (0.75,0.25);
    \draw[label distance=0,\cmdKV@gTypes@iB] (0.75,0.25) -- node[label=(-90+\cmdKV@general@rotate):\cmdKV@gLabels@iLB] {} (0.25,0.25);
    \draw[label distance=0,\cmdKV@gTypes@iC] (0.25,0.25) -- node[label=(180+\cmdKV@general@rotate):\cmdKV@gLabels@iLC] {} (0.25,0.75);
    \draw[label distance=0,\cmdKV@gTypes@iD] (0.25,0.75) -- node[label=(90+\cmdKV@general@rotate):\cmdKV@gLabels@iLD] {} (0.75,0.75);
    #2
  \end{tikzpicture}
  \gSetStdTypes{gluon}
}
\def\gPenta{\@ifnextchar[\@gPenta{\@gPenta[]}}
\def\@gPenta[#1]#2{%
  \setkeys*{pre}{#1}%
  \setrmkeys{general,gTypes,gLabels}%
  \begin{tikzpicture}
    [line width=1pt,
    baseline={([yshift=-0.5ex+\cmdKV@general@yshift]current bounding box.center)},
    scale=\cmdKV@general@scale,
    rotate=\cmdKV@general@rotate,
    font=\cmdKV@general@font]
    \draw[\cmdKV@gTypes@eA] (72:0.35) -- (72:0.75) node[label=(72+\cmdKV@general@rotate):\cmdKV@gLabels@eLA] {};
    \draw[\cmdKV@gTypes@eB] (0:0.35) -- (0:0.75) node[label=(\cmdKV@general@rotate):\cmdKV@gLabels@eLB] {};
    \draw[\cmdKV@gTypes@eC] (-72:0.35) -- (-72:0.75) node[label=(-72+\cmdKV@general@rotate):\cmdKV@gLabels@eLC] {};
    \draw[\cmdKV@gTypes@eD] (-144:0.35) -- (-144:0.75) node[label=(-144+\cmdKV@general@rotate):\cmdKV@gLabels@eLD] {};
    \draw[\cmdKV@gTypes@eE] (144:0.35) -- (144:0.75) node[label=(144+\cmdKV@general@rotate):\cmdKV@gLabels@eLE] {};
    \draw[label distance=0,\cmdKV@gTypes@iA] (72:0.35) -- node[label=(36+\cmdKV@general@rotate):\cmdKV@gLabels@iLA] {} (0:0.35);
    \draw[label distance=0,\cmdKV@gTypes@iB] (0:0.35) -- node[label=(-36+\cmdKV@general@rotate):\cmdKV@gLabels@iLB] {} (-72:0.35);
    \draw[label distance=0,\cmdKV@gTypes@iC] (-72:0.35) -- node[label=(-108+\cmdKV@general@rotate):\cmdKV@gLabels@iLC] {} (-144:0.35);
    \draw[label distance=0,\cmdKV@gTypes@iD] (-144:0.35) -- node[label=(180+\cmdKV@general@rotate):\cmdKV@gLabels@iLD] {} (144:0.35);
    \draw[label distance=0,\cmdKV@gTypes@iE] (144:0.35) -- node[label=(108+\cmdKV@general@rotate):\cmdKV@gLabels@iLE] {} (72:0.35);
    #2
  \end{tikzpicture}
  \gSetStdTypes{gluon}
}

\def\gBoxBox{\@ifnextchar[\@gBoxBox{\@gBoxBox[]}}
\def\@gBoxBox[#1]#2{%
  \setkeys*{pre}{#1}%
  \setrmkeys{general,gTypes,gLabels}%
  \begin{tikzpicture}
    [line width=1pt,
    baseline={([yshift=-0.5ex+\cmdKV@general@yshift]current bounding box.center)},
    scale=\cmdKV@general@scale,
    rotate=\cmdKV@general@rotate,
    font=\cmdKV@general@font]
    \draw[\cmdKV@gTypes@eA] (1.5,0.9) -- (1.8,1.2) node[label=(\cmdKV@general@rotate):\cmdKV@gLabels@eLA] {};
    \draw[\cmdKV@gTypes@eB] (1.5,0.3) -- (1.8,0) node[label=(\cmdKV@general@rotate):\cmdKV@gLabels@eLB] {};
    \draw[\cmdKV@gTypes@eC] (0.3,0.3) -- (0,0) node[label=(-180+\cmdKV@general@rotate):\cmdKV@gLabels@eLC] {};
    \draw[\cmdKV@gTypes@eD] (0.3,0.9) -- (0,1.2) node[label=(180+\cmdKV@general@rotate):\cmdKV@gLabels@eLD] {};
    \draw[label distance=0,\cmdKV@gTypes@iA] (1.5,0.9) -- node[label=(\cmdKV@general@rotate):\cmdKV@gLabels@iLA] {} (1.5,0.3);
    \draw[label distance=0,\cmdKV@gTypes@iB] (1.5,0.3) -- node[label=(-90+\cmdKV@general@rotate):\cmdKV@gLabels@iLB] {} (0.9,0.3);
    \draw[label distance=0,\cmdKV@gTypes@iC] (0.9,0.3) -- node[label=(-90+\cmdKV@general@rotate):\cmdKV@gLabels@iLC] {} (0.3,0.3);
    \draw[label distance=0,\cmdKV@gTypes@iD] (0.3,0.3) -- node[label=(180+\cmdKV@general@rotate):\cmdKV@gLabels@iLD] {} (0.3,0.9);
    \draw[label distance=0,\cmdKV@gTypes@iE] (0.3,0.9) -- node[label=(90+\cmdKV@general@rotate):\cmdKV@gLabels@iLE] {} (0.9,0.9);
    \draw[label distance=0,\cmdKV@gTypes@iF] (0.9,0.9) -- node[label=(90+\cmdKV@general@rotate):\cmdKV@gLabels@iLF] {} (1.5,0.9);
    \draw[label distance=0,\cmdKV@gTypes@iG] (0.9,0.9) -- node[label=(\cmdKV@general@rotate):\cmdKV@gLabels@iLG] {} (0.9,0.3);
    #2
  \end{tikzpicture}
  \gSetStdTypes{gluon}
}
\def\gBoxBoxNP{\@ifnextchar[\@gBoxBoxNP{\@gBoxBoxNP[]}}
\def\@gBoxBoxNP[#1]#2{%
  \setkeys*{pre}{#1}%
  \setrmkeys{general,gTypes,gLabels}%
  \begin{tikzpicture}
    [line width=1pt,
    baseline={([yshift=-0.5ex+\cmdKV@general@yshift]current bounding box.center)},
    scale=\cmdKV@general@scale,
    rotate=\cmdKV@general@rotate,
    font=\cmdKV@general@font]
    \draw[\cmdKV@gTypes@eA] (1.9,1.1) -- (2.2,1.4) node[label=(\cmdKV@general@rotate):\cmdKV@gLabels@eLA] {};
    \draw[\cmdKV@gTypes@eB] (1.9,0.3) -- (2.2,0) node[label=(\cmdKV@general@rotate):\cmdKV@gLabels@eLB] {};
    \draw[\cmdKV@gTypes@eC] (0.3,0.7) -- (0,0.7) node[label=(180+\cmdKV@general@rotate):\cmdKV@gLabels@eLC] {};
    \draw[\cmdKV@gTypes@eD] (1.1,0.7) -- (1.4,0.7) node[label=(\cmdKV@general@rotate):\cmdKV@gLabels@eLD] {};
    \draw[label distance=0,\cmdKV@gTypes@iA] (1.9,1.1) -- node[label=(\cmdKV@general@rotate):\cmdKV@gLabels@iLA] {} (1.9,0.3);
    \draw[label distance=0,\cmdKV@gTypes@iB] (1.9,0.3) -- node[label=(-90+\cmdKV@general@rotate):\cmdKV@gLabels@iLB] {} (0.7,0.3);
    \draw[label distance=0,\cmdKV@gTypes@iC] (0.7,0.3) -- node[label=(-135+\cmdKV@general@rotate):\cmdKV@gLabels@iLC] {} (0.3,0.7);
    \draw[label distance=0,\cmdKV@gTypes@iD] (0.3,0.7) -- node[label=(135+\cmdKV@general@rotate):\cmdKV@gLabels@iLD] {} (0.7,1.1);
    \draw[label distance=0,\cmdKV@gTypes@iE] (0.7,1.1) -- node[label=(90+\cmdKV@general@rotate):\cmdKV@gLabels@iLE] {} (1.9,1.1);
    \draw[label distance=0,\cmdKV@gTypes@iF] (0.7,1.1) -- node[label=(45+\cmdKV@general@rotate):\cmdKV@gLabels@iLF] {} (1.1,0.7);
    \draw[label distance=0,\cmdKV@gTypes@iG] (1.1,0.7) -- node[label=(-45+\cmdKV@general@rotate):\cmdKV@gLabels@iLG] {} (0.7,0.3);
    #2
  \end{tikzpicture}
  \gSetStdTypes{gluon}
}
\def\gTriPenta{\@ifnextchar[\@gTriPenta{\@gTriPenta[]}}
\def\@gTriPenta[#1]#2{%
  \setkeys*{pre}{#1}%
  \setrmkeys{general,gTypes,gLabels}%
  \begin{tikzpicture}
    [line width=1pt,
    baseline={([yshift=-0.5ex+\cmdKV@general@yshift]current bounding box.center)},
    scale=\cmdKV@general@scale,
    rotate=\cmdKV@general@rotate,
    font=\cmdKV@general@font]
    \draw[\cmdKV@gTypes@eA] (72:0.5) -- (45:1) node[label=(35+\cmdKV@general@rotate):\cmdKV@gLabels@eLA] {};
    \draw[\cmdKV@gTypes@eB] (0:0.5) -- (0:1) node[label=(\cmdKV@general@rotate):\cmdKV@gLabels@eLB] {};
    \draw[\cmdKV@gTypes@eC] (-72:0.5) -- (-45:1) node[label=(-35+\cmdKV@general@rotate):\cmdKV@gLabels@eLC] {};
    \draw[\cmdKV@gTypes@eD] (-180:1.2) -- (-180:1.5) node[label=(180+\cmdKV@general@rotate):\cmdKV@gLabels@eLD] {};
    \draw[label distance=0,\cmdKV@gTypes@iA] (72:0.5) -- node[label=(45+\cmdKV@general@rotate):\cmdKV@gLabels@iLA] {} (0:0.5);
    \draw[label distance=0,\cmdKV@gTypes@iB] (0:0.5) -- node[label=(-45+\cmdKV@general@rotate):\cmdKV@gLabels@iLB] {} (-72:0.5);
    \draw[label distance=0,\cmdKV@gTypes@iC] (-72:0.5) -- node[label=(-108+\cmdKV@general@rotate):\cmdKV@gLabels@iLC] {} (-144:0.5);
    \draw[label distance=0,\cmdKV@gTypes@iD] (-144:0.5) -- node[label=(-108+\cmdKV@general@rotate):\cmdKV@gLabels@iLD] {} (-180:1.2);
    \draw[label distance=2,\cmdKV@gTypes@iE] (-180:1.2) -- node[label=(90+\cmdKV@general@rotate):\cmdKV@gLabels@iLE] {} (144:0.5);
    \draw[label distance=2,\cmdKV@gTypes@iF] (144:0.5) -- node[label=(90+\cmdKV@general@rotate):\cmdKV@gLabels@iLF] {} (72:0.5);
    \draw[label distance=0,\cmdKV@gTypes@iG] (144:0.5) -- node[label=(\cmdKV@general@rotate):\cmdKV@gLabels@iLG] {} (-144:0.5);
    #2
  \end{tikzpicture}
  \gSetStdTypes{gluon}
}
\def\gTriTri{\@ifnextchar[\@gTriTri{\@gTriTri[]}}
\def\@gTriTri[#1]#2{%
  \setkeys*{pre}{#1}%
  \setrmkeys{general,gTypes,gLabels}%
  \begin{tikzpicture}
    [line width=1pt,
    baseline={([yshift=-0.5ex+\cmdKV@general@yshift]current bounding box.center)},
    scale=\cmdKV@general@scale,
    rotate=\cmdKV@general@rotate,
    font=\cmdKV@general@font]
    \draw[\cmdKV@gTypes@eA] (1.9,1.1) -- (2.2,1.4) node[label=(\cmdKV@general@rotate):\cmdKV@gLabels@eLA] {};
    \draw[\cmdKV@gTypes@eB] (1.9,0.3) -- (2.2,0) node[label=(\cmdKV@general@rotate):\cmdKV@gLabels@eLB] {};
    \draw[\cmdKV@gTypes@eC] (0.3,0.3) -- (0,0) node[label=(-180+\cmdKV@general@rotate):\cmdKV@gLabels@eLC] {};
    \draw[\cmdKV@gTypes@eD] (0.3,1.1) -- (0,1.4) node[label=(180+\cmdKV@general@rotate):\cmdKV@gLabels@eLD] {};
    \draw[label distance=0,\cmdKV@gTypes@iA] (1.9,1.1) -- node[label=(\cmdKV@general@rotate):\cmdKV@gLabels@iLA] {} (1.9,0.3);
    \draw[label distance=0,\cmdKV@gTypes@iB] (1.9,0.3) -- node[label=(-135+\cmdKV@general@rotate):\cmdKV@gLabels@iLB] {} (1.3,0.7);
    \draw[label distance=0,\cmdKV@gTypes@iC] (1.3,0.7) -- node[label=(135+\cmdKV@general@rotate):\cmdKV@gLabels@iLC] {} (1.9,1.1);
    \draw[label distance=0,\cmdKV@gTypes@iD] (1.3,0.7) -- node[label=(90+\cmdKV@general@rotate):\cmdKV@gLabels@iLD] {} (0.9,0.7);
    \draw[label distance=0,\cmdKV@gTypes@iE] (0.9,0.7) -- node[label=(-45+\cmdKV@general@rotate):\cmdKV@gLabels@iLE] {} (0.3,0.3);
    \draw[label distance=0,\cmdKV@gTypes@iF] (0.3,0.3) -- node[label=(180+\cmdKV@general@rotate):\cmdKV@gLabels@iLF] {} (0.3,1.1);
    \draw[label distance=0,\cmdKV@gTypes@iG] (0.3,1.1) -- node[label=(45+\cmdKV@general@rotate):\cmdKV@gLabels@iLG] {} (0.9,0.7);
    #2
  \end{tikzpicture}
  \gSetStdTypes{gluon}
}

\def\gBoxBubA{\@ifnextchar[\@gBoxBubA{\@gBoxBubA[]}}
\def\@gBoxBubA[#1]#2{%
  \setkeys*{pre}{#1}%
  \setrmkeys{general,gTypes,gLabels}%
  \begin{tikzpicture}
    [line width=1pt,
    baseline={([yshift=-0.5ex+\cmdKV@general@yshift]current bounding box.center)},
    scale=\cmdKV@general@scale,
    rotate=\cmdKV@general@rotate,
    font=\cmdKV@general@font]
    \draw[\cmdKV@gTypes@eA] (0.9,0.9) -- (1.1,1.1) node[label=(\cmdKV@general@rotate):\cmdKV@gLabels@eLA] {};
    \draw[\cmdKV@gTypes@eB] (0.9,0.1) -- (1.1,-0.1) node[label=(\cmdKV@general@rotate):\cmdKV@gLabels@eLB] {};
    \draw[\cmdKV@gTypes@eC] (0.1,0.1) -- (-0.1,-0.1) node[label=(-180+\cmdKV@general@rotate):\cmdKV@gLabels@eLC] {};
    \draw[\cmdKV@gTypes@eD] (0.1,0.9) -- (-0.1,1.1) node[label=(180+\cmdKV@general@rotate):\cmdKV@gLabels@eLD] {};
    \draw[label distance=0,\cmdKV@gTypes@iA] (0.9,0.9) -- node[label=(\cmdKV@general@rotate):\cmdKV@gLabels@iLA] {} (0.9,0.1);
    \draw[label distance=0,\cmdKV@gTypes@iB] (0.9,0.1) -- node[label=(-90+\cmdKV@general@rotate):\cmdKV@gLabels@iLB] {} (0.1,0.1);
    \draw[label distance=0,\cmdKV@gTypes@iC] (0.1,0.1) -- node[label=(180+\cmdKV@general@rotate):\cmdKV@gLabels@iLC] {} (0.1,0.9);
    \draw[label distance=0,\cmdKV@gTypes@iD] (0.1,0.9) -- node[label=(90+\cmdKV@general@rotate):\cmdKV@gLabels@iLD] {} (0.32,0.9);
    \draw[label distance=0,\cmdKV@gTypes@iE] (0.32,0.9) .. controls (0.32,1.15) and (0.68,1.15) .. node[label=(90+\cmdKV@general@rotate):\cmdKV@gLabels@iLE] {} (0.68,0.9);
    \draw[label distance=0,\cmdKV@gTypes@iF] (0.68,0.9) .. controls (0.68,0.65) and (0.32,0.65) .. node[label=(-90+\cmdKV@general@rotate):\cmdKV@gLabels@iLF] {} (0.32,0.9);
    \draw[label distance=0,\cmdKV@gTypes@iG] (0.68,0.9) -- node[label=(90+\cmdKV@general@rotate):\cmdKV@gLabels@iLG] {} (0.9,0.9);
    #2
  \end{tikzpicture}
  \gSetStdTypes{gluon}
}

\def\gBoxBubB{\@ifnextchar[\@gBoxBubB{\@gBoxBubB[]}}
\def\@gBoxBubB[#1]#2{%
  \setkeys*{pre}{#1}%
  \setrmkeys{general,gTypes,gLabels}%
  \begin{tikzpicture}
    [line width=1pt,
    baseline={([yshift=-0.5ex+\cmdKV@general@yshift]current bounding box.center)},
    scale=\cmdKV@general@scale,
    rotate=\cmdKV@general@rotate,
    font=\cmdKV@general@font]
    \draw[label distance=-1,\cmdKV@gTypes@eA] (0,0.4) -- (0,0.7) node[label=(\cmdKV@general@rotate):\cmdKV@gLabels@eLA] {};
    \draw[\cmdKV@gTypes@eB] (0.4,0) -- (0.7,0) node[label=(\cmdKV@general@rotate):\cmdKV@gLabels@eLB] {};
    \draw[label distance=-1,\cmdKV@gTypes@eC] (0,-0.4) -- (0,-0.7) node[label=(\cmdKV@general@rotate):\cmdKV@gLabels@eLC] {};
    \draw[\cmdKV@gTypes@eD] (-1.1,0) -- (-1.4,0) node[label=(180+\cmdKV@general@rotate):\cmdKV@gLabels@eLD] {};
    \draw[label distance=0,\cmdKV@gTypes@iA] (0,0.4) -- node[label=(45+\cmdKV@general@rotate):\cmdKV@gLabels@iLA] {} (0.4,0);
    \draw[label distance=0,\cmdKV@gTypes@iB] (0.4,0) -- node[label=(-45+\cmdKV@general@rotate):\cmdKV@gLabels@iLB] {} (0,-0.4);
    \draw[label distance=0,\cmdKV@gTypes@iC] (0,-0.4) -- node[label=(-135+\cmdKV@general@rotate):\cmdKV@gLabels@iLC] {} (-0.4,0);
    \draw[label distance=0,\cmdKV@gTypes@iD] (-0.4,0) -- node[label=(135+\cmdKV@general@rotate):\cmdKV@gLabels@iLD] {} (0,0.4);
    \draw[label distance=0,\cmdKV@gTypes@iE] (-0.4,0) --  node[label=(90+\cmdKV@general@rotate):\cmdKV@gLabels@iLE] {} (-0.7,0);
    \draw[label distance=0,\cmdKV@gTypes@iF] (-0.7,0) .. controls (-0.7,-0.26) and (-1.1,-0.26) .. node[label=(-90+\cmdKV@general@rotate):\cmdKV@gLabels@iLF] {} (-1.1,0);
    \draw[label distance=0,\cmdKV@gTypes@iG] (-1.1,0) .. controls (-1.1,0.26) and (-0.7,0.26) ..  node[label=(90+\cmdKV@general@rotate):\cmdKV@gLabels@iLG] {} (-0.7,0);
    #2
  \end{tikzpicture}
  \gSetStdTypes{gluon}
}

\def\gPentaTad{\@ifnextchar[\@gPentaTad{\@gPentaTad[]}}
\def\@gPentaTad[#1]#2{%
  \setkeys*{pre}{#1}%
  \setrmkeys{general,gTypes,gLabels}%
  \begin{tikzpicture}
    [line width=1pt,
    baseline={([yshift=-0.5ex+\cmdKV@general@yshift]current bounding box.center)},
    scale=\cmdKV@general@scale,
    rotate=\cmdKV@general@rotate,
    font=\cmdKV@general@font]
    \draw[label distance=-0.5,\cmdKV@gTypes@eA] (108:0.35) -- (108:0.75) node[label=(180+\cmdKV@general@rotate):\cmdKV@gLabels@eLA] {};
    \draw[\cmdKV@gTypes@eB] (36:0.35) -- (36:0.75) node[label=(36+\cmdKV@general@rotate):\cmdKV@gLabels@eLB] {};
    \draw[\cmdKV@gTypes@eC] (-36:0.35) -- (-36:0.75) node[label=(-36+\cmdKV@general@rotate):\cmdKV@gLabels@eLC] {};
    \draw[label distance=-0.5,\cmdKV@gTypes@eD] (-108:0.35) -- (-108:0.75) node[label=(-180+\cmdKV@general@rotate):\cmdKV@gLabels@eLD] {};
    \draw[label distance=0,\cmdKV@gTypes@iA] (108:0.35) -- node[label=(72+\cmdKV@general@rotate):\cmdKV@gLabels@iLA] {} (36:0.35);
    \draw[label distance=0,\cmdKV@gTypes@iB] (36:0.35) -- node[label=(\cmdKV@general@rotate):\cmdKV@gLabels@iLB] {} (-36:0.35);
    \draw[label distance=0,\cmdKV@gTypes@iC] (-36:0.35) -- node[label=(-72+\cmdKV@general@rotate):\cmdKV@gLabels@iLC] {} (-108:0.35);
    \draw[label distance=0,\cmdKV@gTypes@iD] (-108:0.35) -- node[label=(-144+\cmdKV@general@rotate):\cmdKV@gLabels@iLD] {} (-180:0.35);
    \draw[label distance=0,\cmdKV@gTypes@iE] (180:0.35) -- node[label=(144+\cmdKV@general@rotate):\cmdKV@gLabels@iLE] {} (108:0.35);
    \draw[label distance=0,\cmdKV@gTypes@iF] (180:0.35) -- node[label=(90+\cmdKV@general@rotate):\cmdKV@gLabels@iLF] {} (180:0.75);
    \draw[label distance=-0.5,\cmdKV@gTypes@iG] (180:0.75) .. controls ($(180:0.75)+(0,-0.18)$) and ($(180:1)+(0,-0.18)$) .. (180:1) node[label=(180+\cmdKV@general@rotate):\cmdKV@gLabels@iLG] {} .. controls ($(180:1)+(0,0.18)$) and ($(180:0.75)+(0,0.18)$) .. (180:0.75);
    #2
  \end{tikzpicture}
  \gSetStdTypes{gluon}
}

\def\gBoxTadA{\@ifnextchar[\@gBoxTadA{\@gBoxTadA[]}}
\def\@gBoxTadA[#1]#2{%
  \setkeys*{pre}{#1}%
  \setrmkeys{general,gTypes,gLabels}%
  \begin{tikzpicture}
    [line width=1pt,
    baseline={([yshift=-0.5ex+\cmdKV@general@yshift]current bounding box.center)},
    scale=\cmdKV@general@scale,
    rotate=\cmdKV@general@rotate,
    font=\cmdKV@general@font]
    \draw[label distance=-0.5,\cmdKV@gTypes@eA] (0,0.4) -- (0,0.7) node[label=(\cmdKV@general@rotate):\cmdKV@gLabels@eLA] {};
    \draw[\cmdKV@gTypes@eB] (0.4,0) -- (0.7,0) node[label=(\cmdKV@general@rotate):\cmdKV@gLabels@eLB] {};
    \draw[label distance=-0.5,\cmdKV@gTypes@eC] (0,-0.4) -- (0,-0.7) node[label=(\cmdKV@general@rotate):\cmdKV@gLabels@eLC] {};
    \draw[\cmdKV@gTypes@eD] (-0.7,0) -- (-0.7,-0.3) node[label=(-90+\cmdKV@general@rotate):\cmdKV@gLabels@eLD] {};
    \draw[label distance=0,\cmdKV@gTypes@iA] (0,0.4) -- node[label=(45+\cmdKV@general@rotate):\cmdKV@gLabels@iLA] {} (0.4,0);
    \draw[label distance=0,\cmdKV@gTypes@iB] (0.4,0) -- node[label=(-45+\cmdKV@general@rotate):\cmdKV@gLabels@iLB] {} (0,-0.4);
    \draw[label distance=0,\cmdKV@gTypes@iC] (0,-0.4) -- node[label=(-135+\cmdKV@general@rotate):\cmdKV@gLabels@iLC] {} (-0.4,0);
    \draw[label distance=0,\cmdKV@gTypes@iD] (-0.4,0) -- node[label=(135+\cmdKV@general@rotate):\cmdKV@gLabels@iLD] {} (0,0.4);
    \draw[label distance=0,\cmdKV@gTypes@iE] (-0.4,0) --  node[label=(90+\cmdKV@general@rotate):\cmdKV@gLabels@iLE] {} (-0.7,0);
    \draw[label distance=0,\cmdKV@gTypes@iF] (-0.7,0) -- node[label=(90+\cmdKV@general@rotate):\cmdKV@gLabels@iLF] {} (-1.1,0);
    \draw[label distance=-0.5,\cmdKV@gTypes@iG] (-1.1,0) .. controls (-1.1,-0.18) and (-1.35,-0.18) .. (-1.35,0) node[label=(180+\cmdKV@general@rotate):\cmdKV@gLabels@iLG] {} .. controls (-1.35,0.18) and (-1.1,0.18) .. (-1.1,0);
    #2
  \end{tikzpicture}
  \gSetStdTypes{gluon}
}

\def\gBoxBoxBoxA{\@ifnextchar[\@gBoxBoxBoxA{\@gBoxBoxBoxA[]}}
\def\@gBoxBoxBoxA[#1]#2{%
  \setkeys*{pre}{#1}%
  \setrmkeys{general,gTypes,gLabels}%
  \begin{tikzpicture}
    [line width=1pt,
    baseline={([yshift=-0.5ex+\cmdKV@general@yshift]current bounding box.center)},
    scale=\cmdKV@general@scale,
    rotate=\cmdKV@general@rotate,
    font=\cmdKV@general@font]
    \draw[\cmdKV@gTypes@eA] (1.5,0.9) -- (1.8,1.2) node[label=(\cmdKV@general@rotate):\cmdKV@gLabels@eLA] {};
    \draw[\cmdKV@gTypes@eB] (1.5,0.3) -- (1.8,0) node[label=(\cmdKV@general@rotate):\cmdKV@gLabels@eLB] {};
    \draw[\cmdKV@gTypes@eC] (-0.3,0.3) -- (-0.6,0) node[label=(-180+\cmdKV@general@rotate):\cmdKV@gLabels@eLC] {};
    \draw[\cmdKV@gTypes@eD] (-0.3,0.9) -- (-0.6,1.2) node[label=(180+\cmdKV@general@rotate):\cmdKV@gLabels@eLD] {};
    \draw[label distance=0,\cmdKV@gTypes@iA] (1.5,0.9) -- node[label=(\cmdKV@general@rotate):\cmdKV@gLabels@iLA] {} (1.5,0.3);
    \draw[label distance=0,\cmdKV@gTypes@iB] (1.5,0.3) -- node[label=(-90+\cmdKV@general@rotate):\cmdKV@gLabels@iLB] {} (0.9,0.3);
    \draw[label distance=0,\cmdKV@gTypes@iC] (0.9,0.3) -- node[label=(-90+\cmdKV@general@rotate):\cmdKV@gLabels@iLC] {} (0.3,0.3);
    \draw[label distance=0,\cmdKV@gTypes@iD] (0.3,0.3) -- node[label=(-90+\cmdKV@general@rotate):\cmdKV@gLabels@iLD] {} (-0.3,0.3);
    \draw[label distance=0,\cmdKV@gTypes@iE] (-0.3,0.3) -- node[label=(180+\cmdKV@general@rotate):\cmdKV@gLabels@iLE] {} (-0.3,0.9);
    \draw[label distance=0,\cmdKV@gTypes@iF] (-0.3,0.9) -- node[label=(90+\cmdKV@general@rotate):\cmdKV@gLabels@iLF] {} (0.3,0.9);
    \draw[label distance=0,\cmdKV@gTypes@iG] (0.3,0.9) -- node[label=(90+\cmdKV@general@rotate):\cmdKV@gLabels@iLG] {} (0.9,0.9);
    \draw[label distance=0,\cmdKV@gTypes@iH] (0.9,0.9) -- node[label=(90+\cmdKV@general@rotate):\cmdKV@gLabels@iLH] {} (1.5,0.9);
    \draw[label distance=0,\cmdKV@gTypes@iI] (0.9,0.9) -- node[label=(\cmdKV@general@rotate):\cmdKV@gLabels@iLI] {} (0.9,0.3);
    \draw[label distance=0,\cmdKV@gTypes@iJ] (0.3,0.9) -- node[label=(\cmdKV@general@rotate):\cmdKV@gLabels@iLJ] {} (0.3,0.3);
    #2
  \end{tikzpicture}
  \gSetStdTypes{gluon}
}

\def\gBoxBoxBoxB{\@ifnextchar[\@gBoxBoxBoxB{\@gBoxBoxBoxB[]}}
\def\@gBoxBoxBoxB[#1]#2{%
  \setkeys*{pre}{#1}%
  \setrmkeys{general,gTypes,gLabels}%
  \begin{tikzpicture}
    [line width=1pt,
    baseline={([yshift=-0.5ex+\cmdKV@general@yshift]current bounding box.center)},
    scale=\cmdKV@general@scale,
    rotate=\cmdKV@general@rotate,
    font=\cmdKV@general@font]
    \draw[\cmdKV@gTypes@eA] (1.5,1.5) -- (1.8,1.8) node[label=(\cmdKV@general@rotate):\cmdKV@gLabels@eLA] {};
    \draw[\cmdKV@gTypes@eB] (1.5,0.3) -- (1.8,0) node[label=(\cmdKV@general@rotate):\cmdKV@gLabels@eLB] {};
    \draw[\cmdKV@gTypes@eC] (0.3,0.3) -- (0,0) node[label=(-180+\cmdKV@general@rotate):\cmdKV@gLabels@eLC] {};
    \draw[\cmdKV@gTypes@eD] (0.3,1.5) -- (0,1.8) node[label=(180+\cmdKV@general@rotate):\cmdKV@gLabels@eLD] {};
    \draw[label distance=0,\cmdKV@gTypes@iA] (1.5,1.5) -- node[label=(\cmdKV@general@rotate):\cmdKV@gLabels@iLA] {} (1.5,0.3);
    \draw[label distance=0,\cmdKV@gTypes@iB] (1.5,0.3) -- node[label=(-90+\cmdKV@general@rotate):\cmdKV@gLabels@iLB] {} (0.9,0.3);
    \draw[label distance=0,\cmdKV@gTypes@iC] (0.9,0.3) -- node[label=(-90+\cmdKV@general@rotate):\cmdKV@gLabels@iLC] {} (0.3,0.3);
    \draw[label distance=0,\cmdKV@gTypes@iD] (0.3,0.3) -- node[label=(180+\cmdKV@general@rotate):\cmdKV@gLabels@iLD] {} (0.3,0.9);
    \draw[label distance=0,\cmdKV@gTypes@iE] (0.3,0.9) -- node[label=(180+\cmdKV@general@rotate):\cmdKV@gLabels@iLE] {} (0.3,1.5);
    \draw[label distance=0,\cmdKV@gTypes@iF] (0.3,1.5) -- node[label=(90+\cmdKV@general@rotate):\cmdKV@gLabels@iLF] {} (0.9,1.5);
    \draw[label distance=0,\cmdKV@gTypes@iG] (0.9,1.5) -- node[label=(90+\cmdKV@general@rotate):\cmdKV@gLabels@iLG] {} (1.5,1.5);
    \draw[label distance=0,\cmdKV@gTypes@iH] (0.9,1.5) -- node[label=(\cmdKV@general@rotate):\cmdKV@gLabels@iLH] {} (0.9,0.9);
    \draw[label distance=0,\cmdKV@gTypes@iI] (0.9,0.9) -- node[label=(\cmdKV@general@rotate):\cmdKV@gLabels@iLI] {} (0.9,0.3);
    \draw[label distance=0,\cmdKV@gTypes@iJ] (0.9,0.9) -- node[label=(90+\cmdKV@general@rotate):\cmdKV@gLabels@iLJ] {} (0.3,0.9);
    #2
  \end{tikzpicture}
  \gSetStdTypes{gluon}
}

\def\cBoxA{\@ifnextchar[\@cBoxA{\@cBoxA[]}}
\def\@cBoxA[#1]#2{%
  \setkeys*{pre}{#1}%
  \setrmkeys{general,gTypes,gLabels}%
  \begin{tikzpicture}
    [line width=1pt,
    baseline={([yshift=-0.5ex+\cmdKV@general@yshift]current bounding box.center)},
    scale=\cmdKV@general@scale,
    rotate=\cmdKV@general@rotate,
    font=\cmdKV@general@font]
    \fill[blob] (0.4,0) ellipse (0.2 and 0.4);
    \fill[blob] (-0.4,0) ellipse (0.2 and 0.4);
    \draw[\cmdKV@gTypes@eA] ($ (0.4,0) + (45:0.26) $) -- ($ (0.4,0) + (45:0.7) $) node[label=(\cmdKV@general@rotate):\cmdKV@gLabels@eLA] {};
    \draw[\cmdKV@gTypes@eB] ($ (0.4,0) + (-45:0.26) $) -- ($ (0.4,0) + (-45:0.7) $) node[label=(\cmdKV@general@rotate):\cmdKV@gLabels@eLB] {};
    \draw[\cmdKV@gTypes@eC] ($ (-0.4,0) + (-135:0.26) $) -- ($ (-0.4,0) + (-135:0.7) $) node[label=(-180+\cmdKV@general@rotate):\cmdKV@gLabels@eLC] {};
    \draw[\cmdKV@gTypes@eD] ($ (-0.4,0) + (135:0.26) $) -- ($ (-0.4,0) + (135:0.7) $) node[label=(180+\cmdKV@general@rotate):\cmdKV@gLabels@eLD] {};
    \draw[label distance=1,\cmdKV@gTypes@iA] ($ (0.4,0) + (-135:0.26) $) -- node[label=(-90+\cmdKV@general@rotate):\cmdKV@gLabels@iLA] {}($ (-0.4,0) + (-45:0.26) $);
    \draw[label distance=1,\cmdKV@gTypes@iB] ($ (-0.4,0) + (45:0.26) $) -- node[label=(90+\cmdKV@general@rotate):\cmdKV@gLabels@iLB] {}($ (0.4,0) + (135:0.26) $);
    \node[label=(\cmdKV@general@rotate):\cmdKV@gLabels@iLC] at (0.6,0) {};
    \node[label=(180+\cmdKV@general@rotate):\cmdKV@gLabels@iLD] at (-0.6,0) {};
    #2
  \end{tikzpicture}
  \gSetStdTypes{gluon}
}

\def\cBoxB{\@ifnextchar[\@cBoxB{\@cBoxB[]}}
\def\@cBoxB[#1]#2{%
  \setkeys*{pre}{#1}%
  \setrmkeys{general,gTypes,gLabels}%
  \begin{tikzpicture}
    [line width=1pt,
    baseline={([yshift=-0.5ex+\cmdKV@general@yshift]current bounding box.center)},
    scale=\cmdKV@general@scale,
    rotate=\cmdKV@general@rotate,
    font=\cmdKV@general@font]
    \fill[blob] (0,0.3) ellipse (0.4 and 0.2);
    \fill[blob] (0,-0.3) ellipse (0.4 and 0.2);
    \draw[\cmdKV@gTypes@eA] ($ (0,0.3) + (0:0.4) $) -- ($ (0,0.3) + (10:0.7) $) node[label=(\cmdKV@general@rotate):\cmdKV@gLabels@eLA] {};
    \draw[\cmdKV@gTypes@eB] ($ (0,-0.3) + (0:0.4) $) -- ($ (0,-0.3) + (-10:0.7) $) node[label=(\cmdKV@general@rotate):\cmdKV@gLabels@eLB] {};
    \draw[\cmdKV@gTypes@eC] ($ (0,-0.3) + (180:0.4) $) -- ($ (0,-0.3) + (-170:0.7) $) node[label=(180+\cmdKV@general@rotate):\cmdKV@gLabels@eLC] {};
    \draw[\cmdKV@gTypes@eD] ($ (0,0.3) + (180:0.4) $) -- ($ (0,0.3) + (170:0.7) $) node[label=(180+\cmdKV@general@rotate):\cmdKV@gLabels@eLD] {};
    \draw[label distance=1,\cmdKV@gTypes@iA] ($ (0,0.3) + (-45:0.26) $) -- node[label=(\cmdKV@general@rotate):\cmdKV@gLabels@iLA] {}($ (0,-0.3) + (45:0.26) $);
    \draw[label distance=1,\cmdKV@gTypes@iA] ($ (0,-0.3) + (135:0.26) $) -- node[label=(-180+\cmdKV@general@rotate):\cmdKV@gLabels@iLB] {}($ (0,0.3) + (-135:0.26) $);
    \node[label=(-90+\cmdKV@general@rotate):\cmdKV@gLabels@iLC] at (0,-0.5) {};
    \node[label=(90+\cmdKV@general@rotate):\cmdKV@gLabels@iLD] at (0,0.5) {};
    #2
  \end{tikzpicture}
  \gSetStdTypes{gluon}
}

\def\cBoxBoxA{\@ifnextchar[\@cBoxBoxA{\@cBoxBoxA[]}}
\def\@cBoxBoxA[#1]#2{%
  \setkeys*{pre}{#1}%
  \setrmkeys{general,gTypes,gLabels}%
  \begin{tikzpicture}
    [line width=1pt,
    baseline={([yshift=-0.5ex+\cmdKV@general@yshift]current bounding box.center)},
    scale=\cmdKV@general@scale,
    rotate=\cmdKV@general@rotate,
    font=\cmdKV@general@font]
    \fill[blob] (0.4,0) ellipse (0.2 and 0.4);
    \fill[blob] (-0.4,0) ellipse (0.2 and 0.4);
    \fill[blob] (1.2,0) ellipse (0.2 and 0.4);
    \draw[\cmdKV@gTypes@eA] ($ (1.2,0) + (45:0.26) $) -- ($ (1.2,0) + (45:0.7) $) node[label=(\cmdKV@general@rotate):\cmdKV@gLabels@eLA] {};
    \draw[\cmdKV@gTypes@eB] ($ (1.2,0) + (-45:0.26) $) -- ($ (1.2,0) + (-45:0.7) $) node[label=(\cmdKV@general@rotate):\cmdKV@gLabels@eLB] {};
    \draw[\cmdKV@gTypes@eC] ($ (-0.4,0) + (-135:0.26) $) -- ($ (-0.4,0) + (-135:0.7) $) node[label=(-180+\cmdKV@general@rotate):\cmdKV@gLabels@eLC] {};
    \draw[\cmdKV@gTypes@eD] ($ (-0.4,0) + (135:0.26) $) -- ($ (-0.4,0) + (135:0.7) $) node[label=(180+\cmdKV@general@rotate):\cmdKV@gLabels@eLD] {};
    \draw[label distance=1,\cmdKV@gTypes@iA] ($ (1.2,0) + (-135:0.26) $) -- node[label=(-90+\cmdKV@general@rotate):\cmdKV@gLabels@iLA] {}($ (0.4,0) + (-45:0.26) $);
    \draw[label distance=1,\cmdKV@gTypes@iB] ($ (0.4,0) + (-135:0.26) $) -- node[label=(-90+\cmdKV@general@rotate):\cmdKV@gLabels@iLB] {}($ (-0.4,0) + (-45:0.26) $);
    \draw[label distance=1,\cmdKV@gTypes@iC] ($ (-0.4,0) + (45:0.26) $) -- node[label=(90+\cmdKV@general@rotate):\cmdKV@gLabels@iLC] {}($ (0.4,0) + (135:0.26) $);
    \draw[label distance=1,\cmdKV@gTypes@iD] ($ (0.4,0) + (45:0.26) $) -- node[label=(90+\cmdKV@general@rotate):\cmdKV@gLabels@iLD] {}($ (1.2,0) + (135:0.26) $);
    \node[label=(\cmdKV@general@rotate):\cmdKV@gLabels@iLE] at (1.4,0) {};
    \node[label=(\cmdKV@general@rotate):\cmdKV@gLabels@iLF] at (0.6,0) {};
    \node[label=(180+\cmdKV@general@rotate):\cmdKV@gLabels@iLG] at (-0.6,0) {};
    #2
  \end{tikzpicture}
  \gSetStdTypes{gluon}
}

\def\cBoxBoxB{\@ifnextchar[\@cBoxBoxB{\@cBoxBoxB[]}}
\def\@cBoxBoxB[#1]#2{%
  \setkeys*{pre}{#1}%
  \setrmkeys{general,gTypes,gLabels}%
  \begin{tikzpicture}
    [line width=1pt,
    baseline={([yshift=-0.5ex+\cmdKV@general@yshift]current bounding box.center)},
    scale=\cmdKV@general@scale,
    rotate=\cmdKV@general@rotate,
    font=\cmdKV@general@font]
    \fill[blob] (0,0.3) ellipse (0.4 and 0.2);
    \fill[blob] (0,-0.3) ellipse (0.4 and 0.2);
    \fill[blob] (1,0) ellipse (0.2 and 0.4);
    \draw[\cmdKV@gTypes@eA] ($ (1,0) + (45:0.26) $) -- ($ (1,0) + (45:0.7) $) node[label=(\cmdKV@general@rotate):\cmdKV@gLabels@eLA] {};
    \draw[\cmdKV@gTypes@eB] ($ (1,0) + (-45:0.26) $) -- ($ (1,0) + (-45:0.7) $) node[label=(\cmdKV@general@rotate):\cmdKV@gLabels@eLB] {};
    \draw[\cmdKV@gTypes@eC] ($ (0,-0.3) + (180:0.4) $) -- ($ (0,-0.3) + (-170:0.7) $) node[label=(180+\cmdKV@general@rotate):\cmdKV@gLabels@eLC] {};
    \draw[\cmdKV@gTypes@eD] ($ (0,0.3) + (180:0.4) $) -- ($ (0,0.3) + (170:0.7) $) node[label=(180+\cmdKV@general@rotate):\cmdKV@gLabels@eLD] {};
    \draw[label distance=1,\cmdKV@gTypes@iA] ($ (0,-0.3) + (0:0.4) $) -- node[label=(-90+\cmdKV@general@rotate):\cmdKV@gLabels@iLA] {} ($ (1,0) + (-135:0.26) $);
    \draw[label distance=1,\cmdKV@gTypes@iB] ($ (0,-0.3) + (135:0.26) $) -- node[label=(-180+\cmdKV@general@rotate):\cmdKV@gLabels@iLB] {}($ (0,0.3) + (-135:0.26) $);
    \draw[label distance=1,\cmdKV@gTypes@iC] ($ (0,0.3) + (0:0.4) $) -- node[label=(90+\cmdKV@general@rotate):\cmdKV@gLabels@iLC] {} ($ (1,0) + (135:0.26) $);
    \draw[label distance=1,\cmdKV@gTypes@iD] ($ (0,0.3) + (-45:0.26) $) -- node[label=(\cmdKV@general@rotate):\cmdKV@gLabels@iLD] {}($ (0,-0.3) + (45:0.26) $);
    \node[label=(\cmdKV@general@rotate):\cmdKV@gLabels@iLE] at (1.2,0) {};
    \node[label=(-90+\cmdKV@general@rotate):\cmdKV@gLabels@iLF] at (0,-0.5) {};
    \node[label=(90+\cmdKV@general@rotate):\cmdKV@gLabels@iLG] at (0,0.5) {};
    #2
  \end{tikzpicture}
  \gSetStdTypes{gluon}
}

\makeatother

\usepackage{tikzexternal}
\tikzsetexternalprefix{figures/}
\tikzsetfigurename{bcj-figure}

\definecolor{myblue}{rgb}{0,0,0.7}

\definecolor{mygreen}{rgb}{0,0.4,0}

\definecolor{myred}{rgb}{0.4,0,0}

\def\Res{\mathop{\textrm{Res}}}

\def\ket#1{|#1 \rangle}

\def\braket#1{\langle #1 \rangle}

\def\o#1{\overline{#1}}

\newcommand{\widebar}{\overline}

\def\wPhi{\widebar{\Phi}}

\def\kT{\hat{\kappa}}

\def\Res_#1{\operatorname*{Res}_{#1}}

\def\ttr{\operatorname*{tr}}
\def\trp{\operatorname*{tr}\,\!\!_{+}}
\def\trm{\operatorname*{tr}\,\!\!_{-}}
\def\trpm{\operatorname*{tr}\,\!\!_{\pm}}

\def\tr5{\operatorname*{tr}\,\!\!_{5}}
\def\sgn{\operatorname*{sgn}}

\def\d{\mathrm{d}}

\def\ie{i.e. }
\def\eg{e.g. }

\def\eqn#1{eq.~\eqref{#1}}
\def\eqns#1#2{eqs.~\eqref{#1} and~\eqref{#2}}
\def\eqnss#1#2#3{eqs.~\eqref{#1}, \eqref{#2} and~\eqref{#3}}

\def\fig#1{figure~{\ref{#1}}}

\def\tab#1{table~{\ref{#1}}}

\def\sec#1{section~{\ref{#1}}}
\def\secs#1#2{sections~{\ref{#1}} and~{\ref{#2}}}

\def\app#1{appendix~{\ref{#1}}}

\def\rcite#1{ref.~{\cite{#1}}}
\def\rcites#1{refs.~{\cite{#1}}}

\def\be{\begin{equation}}
\def\ee{\end{equation}}
\def\bea{\begin{eqnarray}}
\def\eea{\end{eqnarray}}
\def\beal{\begin{equation}\begin{aligned}}
\def\eeal{\end{aligned}\end{equation}}
\def\nn{\nonumber}

\def\cV{\mathcal{V}}
\def\cA{\mathcal{A}}

\def\cN{\mathcal{N}}

\def\tf{\tilde{f}}

\def\eps{\epsilon}

\def\MHVb{\overline{\text{MHV}}}

\usepackage{stackengine}
\newcommand{\oset}[3][0.3ex]{\stackon[#1]{$#3$}{$\scriptstyle#2$}}
\newcommand{\uset}[3][0.3ex]{\stackunder[#1]{$#3$}{$\scriptstyle#2$}}

\begin{document}
%\tikzset{external/force remake}%
\preprint{UUITP-54/18}
\title{Two-loop \texorpdfstring{$\cN=2$}{N=2} SQCD amplitudes
       with external matter from iterated cuts}

\author[a]{Gregor K\"alin,}
\author[a]{Gustav Mogull,}
\author[b]{and Alexander Ochirov}
\affiliation[a]{Department of Physics and Astronomy, Uppsala University,
75108 Uppsala, Sweden}
\affiliation[b]{ETH Z\"urich, Institut f\"ur Theoretische Physik,
Wolfgang-Pauli-Str. 27, 8093 Z\"urich, Switzerland}
\emailAdd{gregor.kaelin@physics.uu.se, gustav.mogull@physics.uu.se, aochirov@phys.ethz.ch}
\abstract{
We develop an iterative method for constructing four-dimensional generalized unitarity cuts in $\cN=2$ supersymmetric Yang-Mills (SYM) theory coupled to fundamental matter hypermultiplets ($\cN=2$ SQCD).
For iterated two-particle cuts, specifically those involving only four-point amplitudes, this implies simple diagrammatic rules for assembling the cuts to any loop order, reminiscent of the rung rule in $\cN=4$ SYM.
By identifying physical poles, the construction simplifies the task of extracting complete integrands.
In combination with the duality between color and kinematics we construct all four-point massless MHV-sector scattering amplitudes up to two loops in $\cN=2$ SQCD, including those with matter on external legs.
Our results reveal chiral infrared-finite integrands closely related to those found using loop-level BCFW recursion.
The integrands are valid in $D\leq6$ dimensions with external states in a four-dimensional subspace; the upper bound is dictated by our use of six-dimensional chiral $\cN=(1,0)$ SYM as a means of dimensionally regulating loop integrals.
}


\setcounter{tocdepth}{2}
\maketitle
\newpage

%%%%%%%%%%%%%%%%%%%%%%%%%%%%%%%%%%%%%%%%%%%%%%%%%%
%%%%%%%%%%%%%%%%%%%%%%%%%%%%%%%%%%%%%%%%%%%%%%%%%%
\section{Introduction}

\label{sec:intro}

Supersymmetric Yang-Mills (SYM) theories are well known
to have simpler scattering amplitudes
than the most physically interesting gauge theory ---
quantum chromodynamics (QCD).
%Calculating scattering amplitudes in perturbative quantum chromodynamics (QCD) is notoriously hard --- much harder than its supersymmetrized cousins.
In planar $\cN=4$ SYM theory, for instance,
modern methods have enabled five-loop six-point amplitude
computations~\cite{Caron-Huot:2016owq}, with the four- and five- point amplitudes known
to all loop orders for more than a decade~\cite{Anastasiou:2003kj,Bern:2005iz}.
Even more is known about amplitude integrands in $\cN=4$ SYM,
with all-loop $n$-point results in the maximally helicity-violating (MHV)
sector~\cite{ArkaniHamed:2010kv,ArkaniHamed:2010gh,ArkaniHamed:2012nw},
their two-loop extensions beyond MHV~\cite{Bourjaily:2015jna},
and higher-loop but lower-point results beyond the leading-color (planar) limit~\cite{Bern:2007hh,Bern:2010tq,
Carrasco:2011mn,Bern:2012uc,Bern:2015ple,Henn:2016jdu}.

Such studies of SYM theories, together with impressive developments beyond next-to-leading order~\cite{
  Anastasiou:2000kg,Anastasiou:2000ue,Anastasiou:2001sv,Glover:2001af,%
  Garland:2001tf,Garland:2002ak,Catani:2011qz,Gehrmann:2011aa,%
  Czakon:2013goa,Grazzini:2013bna,Cascioli:2014yka,Gehrmann:2014fva,Chen:2014gva,%
  Caola:2014iua,Czakon:2014xsa,%
  Gehrmann:2015ora,Caola:2015ila,vonManteuffel:2015msa,Grazzini:2015nwa,%
  Boughezal:2015dva,Boughezal:2015dra,Boughezal:2015aha,Ridder:2015dxa,%
  Anastasiou:2015vya},
have helped mature
modern on-shell methods~\cite{Bern:1994zx,Bern:1994cg,Britto:2004nc,Britto:2004ap,Britto:2005fq,
  Forde:2007mi,Anastasiou:2006jv,Giele:2008ve}.
Together with other techniques, these methods are widely used in current state-of-the-art QCD calculations ---
nowadays involving two-loop five-parton amplitudes~\cite{Badger:2013gxa,Badger:2015lda,Gehrmann:2015bfy,Badger:2017jhb,
Abreu:2017hqn,Chawdhry:2018awn,Badger:2018gip,Abreu:2018jgq}.
Moreover, amplitudes in supersymmetric gauge theories can often be viewed
as specific contributions to QCD amplitudes,
at least at tree~\cite{Dixon:1996wi,Dixon:2010ik,Melia:2013epa}
and one-loop level~\cite{Bern:1993mq,Bern:1994zx,Bern:2002zk}.
In these ways SYM calculations have paved the way to new results in QCD.

Recent all-loop BCFW constructions
of four-dimensional amplitude integrands in $\cN=4$ SYM~\cite{ArkaniHamed:2010kv,ArkaniHamed:2010gh,ArkaniHamed:2012nw}
were preceded by a more pedestrian way of constructing integrands,
often referred to as the ``rung rule''~\cite{Bern:1997nh,Bern:1998ug}.
It is based on an analysis of two- and three-particle
unitarity cuts and their iterative structure.
The idea is to directly obtain $(L+1)$- from $L$-loop integrands
by attaching rungs to the individual diagrams:
\begin{equation}
  %\tikzset{external/force remake}%
  \begin{tikzpicture}
    [line width=1pt,
    baseline={([yshift=-0.5ex]current bounding box.center)},
    font=\scriptsize]
    \draw[dotted] (-0.3,0.4) -- (0,0.4);
    \draw (0,0.4) -- node[above] {$\uset{\rightarrow}{\ell_1}$} (0.5,0.4);
    \draw[dotted] (0.5,0.4) -- (0.9,0.4);
    \draw[dotted] (-0.3,0) -- (0,0);
    \draw (0,0) -- node[below] {$\oset{\rightarrow}{\ell_2}$} (0.5,0);
    \draw[dotted] (0.5,0) -- (0.9,0);
  \end{tikzpicture}
  \rightarrow -i(\ell_1+\ell_2)^2\times
  %\tikzset{external/force remake}%
  \begin{tikzpicture}
    [line width=1pt,
    baseline={([yshift=-0.5ex]current bounding box.center)},
    font=\scriptsize]
    \draw[dotted] (-0.3,0.4) -- (0,0.4);
    \draw (0.0,0.4) node[above] {$\uset{\rightarrow}{\ell_1}$} -- (0.5,0.4);
    \draw[dotted] (0.5,0.4) -- (0.9,0.4);
    \draw[dotted] (-0.3,0) -- (0,0);
    \draw (0,0) node[below] {$\oset{\rightarrow}{\ell_2}$} -- (0.5,0);
    \draw[dotted] (0.5,0) -- (0.9,0);
    \draw (0.25,0.4) -- (0.25,0);
  \end{tikzpicture}
\end{equation}
where each rung comes with a kinematic factor.
Despite its known shortcomings
(it does not give unique representations of the integrand,
and if one desires integrands obeying color-kinematics duality
\cite{Bern:2008qj,Bern:2010ue,Johansson:2015oia}
then the results are not always compatible)
the rung rule has been instrumental to initial progress in $\cN=4$ SYM.

In this paper we develop an iterative approach for computing generalized 
unitarity cuts in $\cN=2$ supersymmetric QCD (SQCD),\footnote{%
  Our methods are in particular inspired by supersum technologies developed in \rcite{Bern:2010tq}.
}
which is reminiscent of the rung rule and helps us construct amplitude integrands.
This theory is equivalent to $\cN=2$ SYM coupled to $N_f$ copies
of massless $\cN=2$ matter multiplets in the (anti-)fundamental representation
of the (arbitrary) gauge group $G$.
It is therefore more similar to ordinary QCD than $\cN=4$ SYM,
while retaining considerable simplifications with respect to the former.
This makes it an ideal theory from which to study the effect of reducing supersymmetry on
the analytic structure of gauge theories ---
an open question that is crucial should one wish
to extend the impressive progress in $\cN=4$ SYM to QCD.

Our concrete results,
obtained using the rung-rule-like iterative structure of the unitarity cuts,
are the complete set of massless four-point MHV amplitude integrands up to two loops,
including those with external matter states.
The one-loop amplitude with four external matter states has already
been computed using an orbifold construction~\cite{Chiodaroli:2013upa},
the one- and two-loop amplitudes with four external gluons were
determined in~\rcites{Johansson:2014zca,Johansson:2017bfl},
and the rest were previously unknown.
All of these full-color amplitudes are obtained in a form that respects
color-kinematics duality~\cite{Bern:2008qj,Bern:2010ue,Johansson:2015oia}.
They can therefore be used to produce amplitudes
in $\cN \geq 2$ pure or matter-coupled supergravities
via an array of related double-copy constructions~\cite{Bern:2008qj,Bern:2010ue,
Bern:2011rj,Carrasco:2012ca,Chiodaroli:2013upa,Johansson:2014zca,Chiodaroli:2014xia,
Chiodaroli:2015rdg,Chiodaroli:2015wal,Anastasiou:2016csv,Johansson:2017srf,
Chiodaroli:2017ehv,Johansson:2018ues}.
Other four-point one- and two-loop results in $\cN=2$ SQCD include
\rcites{Glover:2008tu,Andree:2010na,Leoni:2014fja,Leoni:2015zxa}.

An intriguing new aspect of our approach
is the appearance of Dirac traces in the kinematic numerators,
which make infrared (IR) properties manifest.
Their structure echoes the BCFW-derived expressions in $\cN=4$ SYM,
which are known for having well-behaved IR structure;
they are also similar to traces appearing in the planar two-loop all-plus amplitudes in
non-supersymmetric Yang-Mills theory~\cite{Badger:2016ozq},
which are well known thanks to their one-loop-like simplicity
\cite{Bern:2000dn,Badger:2013gxa,Badger:2015lda,
Gehrmann:2015bfy,Dunbar:2016aux,Dunbar:2016cxp,Dunbar:2016gjb}.\footnote{In fact,
local-integrand representations of all-plus amplitudes
were inspired by those of $\cN=4$ SYM amplitudes based on a dimension-shifting relationship between their one-loop integrands~\cite{Bern:1996ja} which persists at two loops.}
A careful exposition of the IR properties of the two-loop $\cN=2$
SQCD integrands will be reported elsewhere, while in this paper
we limit ourselves to explanatory comments during the derivation of our results.

The paper is organized as follows.
In \sec{sec:review} we review the relevant aspects of $\cN=2$ SQCD and
its scattering amplitudes, and introduce the necessary tools to deal with color-kinematics duality in this theory.
In \sec{sec:rungrule} we compare the structure of iterated two-particle cuts
in this theory with that in $\cN=4$ SYM,
and formulate diagrammatic generalized rung rules for the former.
We use these rules in \secs{sec:oneloop}{sec:twoloop}
to motivate --- and, in some cases, fully derive --- the kinematic numerators
of all four-point amplitudes in $\cN=2$ SQCD,
first at one loop and then at two loops.
In \sec{sec:multiparticlecuts} we show the limits of applicability
of our rung rules by studying more general unitarity cuts.
We conclude in \sec{sec:outro} by discussing the interesting features
of our results and their derivations, and outline our next steps
in the analysis of the integrand structure of (S)QCD.

%%%%%%%%%%%%%%%%%%%%%%%%%%%%%%%%%%%%%%%%%%%%%%%%%%
%%%%%%%%%%%%%%%%%%%%%%%%%%%%%%%%%%%%%%%%%%%%%%%%%%
\section{Review: \texorpdfstring{$\cN=2$}{N=2} SQCD}
\label{sec:review}

In this section we explain our approach to scattering amplitudes in
$\cN=2$ SQCD,
to a considerable degree following \rcites{Johansson:2014zca,Johansson:2017bfl}
but updating the notation as necessary to prepare for later sections.
In particular, we introduce a new notation
to compactly write four-point tree-level amplitudes
involving fundamental hypermultiplets on external legs.
We also summarize the off-shell constraints to be placed on kinematic numerators,
in addition to those required by color-kinematics duality~\cite{Bern:2008qj,Bern:2010ue}.

%%%%%%%%%%%%%%%%%%%%%%%%%%%%%%%%%%%%%%%%%%%%%%%%%%
\subsection{On-shell particle content}
\label{sec:particlecontent}

The on-shell content of four-dimensional $\cN=2$ SQCD is most easily described by comparison with that of $\cN=4$ SYM. The latter contains $2^4=16$ states, and forms a vector supermultiplet~\cite{Nair:1988bq}:
\be\label{eq:neq4multiplet}
  \cV_{\cN=4}(\eta^I)=A^++\eta^I\psi^+_I+\frac12\eta^I\eta^J\varphi_{IJ}+
  \frac1{3!}\eps_{IJKL}\eta^I\eta^J\eta^K\psi_-^L+\eta^1\eta^2\eta^3\eta^4A_-\,.
\ee
The four-dimensional chiral superspace coordinates $\eta^I$
carry SU(4) R-symmetry indices $\{I,J,\ldots\}$.
For later use, let us remark that this multiplet is CPT self conjugate
and can equally well be written
in terms of anti-chiral superspace coordinates $\bar{\eta}_I$ using
\be
   \cV_{\cN=4}(\bar{\eta}_I)
    = \int\!\d^4\eta\,e^{\eta^I\bar{\eta}_I}\cV_{\cN=4} (\eta^I)
    = A_- + \bar{\eta}_I\psi_-^I + \dots
%    + \frac{1}{2} \bar{\eta}_I \bar{\eta}_J \varphi^{IJ}
%    + \frac{1}{3!} \epsilon^{IJKL} \bar{\eta}_I \bar{\eta}_J \bar{\eta}_K \psi_{L}^+
    + \bar{\eta}_1\bar{\eta}_2\bar{\eta}_3\bar{\eta}_4A^+\,,
\label{eq:antichiralSS}
\ee
where the measure is $\d^4\eta=\d\eta^1\d\eta^2\d\eta^3\d\eta^4$.
%The anti-chiral superspace coordinates will become useful later when considering $\MHVb$ amplitudes.

The $\cN=4$ multiplet naturally decomposes on $\eta^3$ and $\eta^4$
into $\cN=2$ multiplets:
\be\label{eq:neq4decomp}
   \cV_{\cN=4}
    = V^+_{\cN=2} + \eta^3\Phi_{\cN=2} + \eta^4\wPhi_{\cN=2}
    + \eta^3\eta^4V^-_{\cN=2}\,.
\ee
Here the $\cN=2$ vector multiplets are
\be\label{eq:neq2multiplet}
   V^+_{\cN=2}(\eta^I) = A^+ + \eta^I\psi^+_I + \eta^1\eta^2\varphi_{12}\,,\qquad \quad
   V^-_{\cN=2}(\eta^I) = \varphi_{34} + \eps_{I34J}\eta^I\psi_-^J
    + \eta^1\eta^2A_-\,,
\ee
where the SU(2) indices $I,J=1,2$ are inherited from SU(4);
the hypermultiplets (hypers) are
\be
   \Phi_{\cN=2}(\eta^I) = \psi^+_3 - \eta^I\varphi_{I3}
    + \eta^1\eta^2\psi_-^4\,,\qquad\,
   \wPhi_{\cN=2}(\eta^I) = \psi^+_4 - \eta^I\varphi_{I4}
    - \eta^1\eta^2\psi_-^3\,.\quad\,
\ee
All four have $\cN=2$ supersymmetries
represented by the remaining Grassmann variables $\eta^1$ and $\eta^2$.
Moreover, $V^+_{\cN=2}$ is related to $V^-_{\cN=2}$ by CPT conjugation,
and likewise for $\Phi_{\cN=2}$ and $\wPhi_{\cN=2}$.

%%%%%%%%% Figure %%%%%%%%%%%%%%%
\begin{table}[t]
\centering
\vspace{-7pt}
\includegraphics[scale=1.25]{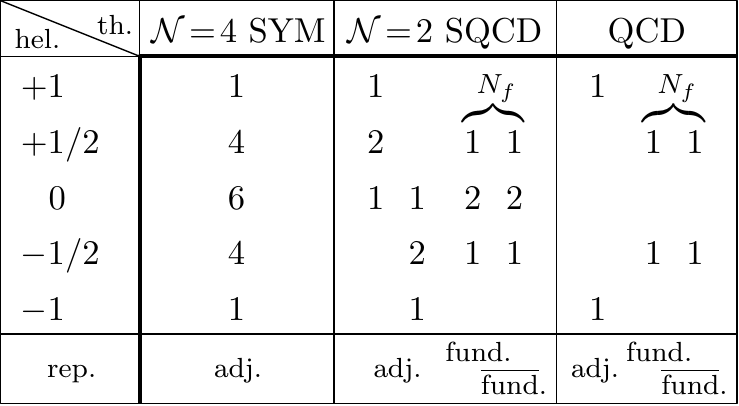}
\caption{\small Helicity content of $\cN=2$ supersymmetric QCD
in comparison to $\cN=4$ super-Yang-Mills theory and conventional QCD.
For these theories,
the helicities and the representations of the particles
are listed in the left column and the lower row, respectively.}
\label{tab:SQCD}
\end{table}
%%%%%%%%%%%%%%%%%%%%%%%%%%%%%%%%

% Now the maximally supersymmetric character of the $\cN=4$ multiplet
% means that it may only belong to the adjoint representation of the gauge group.
The on-shell content of $\cN=2$ SQCD is obtained by switching
the representation of the hypers $\Phi_{\cN=2}$ and $\wPhi_{\cN=2}$
from the adjoint to the fundamental and anti-fundamental representations,
respectively.
This explicitly breaks supersymmetry on $\eta^3$ and $\eta^4$,
but not on $\eta^1$ and~$\eta^2$.
Furthermore, the hypers can be generalized
to an arbitrary number $N_f=\delta^\alpha_\alpha$ of flavors
by attaching flavor indices $\{\alpha,\beta,\ldots\}$ to them,
as in $(\Phi_{\cN=2})^\alpha$ and $(\wPhi_{\cN=2})_\alpha$.
Alternatively, the flavor indices can be conflated with the color indices,
implying reducible gauge-group representations for the matter multiplets.
In this paper will use the tree amplitudes for $N_f=1$ to construct unitarity cuts;
the diagrammatic form of the resulting loop integrands
will allow for an arbitrary $N_f$.

By analogy to QCD, $V^+_{\cN=2}$ and $V^-_{\cN=2}$ act as positive- and negative-helicity gluons, $A^+$ and $A_-$,
and can be regarded as their respective on-shell supersymmetrizations.
The hypermultiplets $(\Phi_{\cN=2})^\alpha$ and $(\wPhi_{\cN=2})_\alpha$ play the roles of massless quarks and anti-quarks.
In this way, $\cN=2$ SQCD can be viewed as the middle ground between $\cN=4$ SYM
and the actual QCD, as illustrated in \tab{tab:SQCD}.
Although it is less well studied than the other two,
its one-loop MHV and NMHV amplitudes are known for any number of external gluons~\cite{Bern:1994cg,BjerrumBohr:2007vu,Ochirov:2013oca}
via their relation to those in $\cN=1,2,4$ SYM~\cite{Bern:1994zx}.

%%%%%%%%%%%%%%%%%%%%%%%%%%%%%%%%%%%%%%%%%%%%%%%%%%
\subsection{Tree-level amplitudes}
\label{sec:tree}

Tree-level $\cN=2$ SQCD amplitudes are simply related to those of $\cN=4$ SYM. In the maximally helicity-violating (MHV) sector, to which we specialize in this paper, planar tree-level $\cN=4$ SYM amplitudes are given by the famous Parke-Taylor formula~\cite{Parke:1986gb,Nair:1988bq}:\footnote{We adopt the usual spinor-helicity notation --- see for example \rcites{Dixon:1996wi,Elvang:2015rqa}.}
\be\label{eqn:PT}
   A_n^{(0),\text{MHV}}\big(\cV_{\cN=4},\cV_{\cN=4},\ldots,\cV_{\cN=4}\big)
    = \frac{i\delta^8(Q)}{\braket{12}\braket{23}\cdots\braket{n1}}\,.
\ee
Here the Grassmann delta function imposes conservation of supercharges;
for $\cN$ supersymmetries it is
\be\label{eq:deltasusy}
  \delta^{2\cN}\!(Q)
    = \delta^{2\cN}\!\bigg( \sum_{i=1}^n \ket{i} \eta_i \bigg)
    = \prod_{I=1}^\cN\sum_{i<j}^n\eta_i^I\braket{i\,j}\eta_j^I \,.
\ee
Tree-level $\cN=2$ SQCD amplitudes with $N_f=1$ massless hypermultiplet flavors are obtained by projecting out the relevant multiplets using the decomposition given in \eqn{eq:neq4decomp}. The two extra superspace coordinates $\eta^3$ and $\eta^4$, associated with the two broken supersymmetries, serve to identify the multiplets:
$\Phi_{\cN=2}$ carries $\eta^3$, $\wPhi_{\cN=2}$ carries $\eta^4$,
$V^+_{\cN=2}$ carries neither, and $V^-_{\cN=2}$ carries both.
Several $n$-point examples are given in \rcite{Johansson:2017bfl}.

In this paper we will mostly use four-point amplitudes.
In order to track the ${\cal N}=2$ multiplets on their external legs,
we introduce the superspace combination
\be\label{eq:kappa}
  \kappa_{(ab)(cd)}(1,2,3,4)\equiv\frac{[12][34]}{\braket{12}\braket{34}}\delta^4(Q)
  \eta_a^3\braket{a\,b}\eta_b^3\eta_c^4\braket{c\,d}\eta_d^4\,,
\ee
where $\{a,b,c,d\}\in\{1,2,3,4\}$.
Here the spinor-helicity prefactor is permutation invariant
and is familiar from the commonly used $\cN=4$ amplitude prefactor
\begin{equation}
  \kappa(1,2,3,4) \equiv \frac{[12][34]}{\braket{12}\braket{34}}\delta^8(Q) \,.
\label{eq:kappaN4}
\end{equation}
In \rcites{Johansson:2014zca,Johansson:2017bfl} a similar notation was used to label the two anti-chiral $V^-_{\cN=2}$ vector multiplets in the MHV sector:
$\kappa_{ab}\equiv\kappa_{(ab)(ab)}$. Our updated notation is more flexible, as it allows us to also track the hypermultiplets $\Phi_{\cN=2}$ and $\wPhi_{\cN=2}$ on external legs. For instance, we can now compactly write
\begin{subequations}
\begin{align}
   A_4^{(0),\text{MHV}}(V^-_{\cN=2},V^+_{\cN=2},V^-_{\cN=2},V^+_{\cN=2}) &
    = -\frac{i}{st}\kappa_{(13)(13)}=-\frac{i}{st}\kappa_{13}\,,\\
   A_4^{(0),\text{MHV}}(V^-_{\cN=2},\Phi_{\cN=2},\wPhi_{\cN=2},V^+_{\cN=2}) &
    = -\frac{i}{st}\kappa_{(12)(13)}\,,\\
   A_4^{(0),\text{MHV}}(\Phi_{\cN=2},\wPhi_{\cN=2},\Phi_{\cN=2},\wPhi_{\cN=2}) &
    = -\frac{i}{st}\kappa_{(13)(24)}\,,\\*
   A_4^{(0),\text{MHV}}(\Phi_{\cN=2},\Phi_{\cN=2},\wPhi_{\cN=2},\wPhi_{\cN=2}) &
    = -\frac{i}{st}\kappa_{(12)(34)}\,,
\end{align}
\end{subequations}
where $s=(p_1+p_2)^2$ and $t=(p_2+p_3)^2$ are the usual Mandelstam variables.

Using CPT invariance of the theory we can equally well study $\MHVb$ amplitudes.
These are related to the MHV by exchanging $|i\rangle\leftrightarrow|i]$ and $\eta_i^I\leftrightarrow\bar{\eta}_{i,I}$.
For instance, the $n$-point tree-level $\MHVb$ amplitude is also given by a Parke-Taylor formula:
\be\label{eqn:PTb}
   A_n^{(0),\MHVb}\big(\cV_{\cN=4},\cV_{\cN=4},\ldots,\cV_{\cN=4}\big)
    = \frac{i\delta^8(\bar{Q})}{[12][23]\cdots[n1]}\,,
\ee
where the anti-chiral supermomentum-conserving delta function is defined as
\be
   \delta^{2\cN}(\bar{Q})
    = \prod_{I=1}^\cN\sum_{i<j}^n\bar{\eta}_i^I[i\,j]\bar{\eta}_j^I\,.
\ee
To compare amplitudes formulated in different superspaces we switch between chiral and anti-chiral superspace coordinates using Fourier transforms:
\be\label{eqn:FT}
  A_n(\eta_i^I) = \int\!\d^4\bar{\eta}_1 \cdots \d^4\bar{\eta}_n e^{\bar{\eta}_{1,I}\eta_1^I}\cdots e^{\bar{\eta}_{n,I}\eta_n^I}A_n(\bar{\eta}_{i,I})\,.
\ee
Note that even with $\cN=2$ supersymmetries we continue to use
the four-dimensional Grassmann integration $\d^4\eta_i$ ---
the superspace variables of the broken supersymmetry are retained
in order to discern between the $\cN=2$ multiplets.

At four points the $\text{MHV}$ and $\MHVb$ amplitudes are equivalent, so related by the Fourier transform~(\ref{eqn:FT}).
We can therefore also track the external state configuration using
\be\label{eq:kappab}
  \bar{\kappa}_{(ab)(cd)}(1,2,3,4) \equiv
      \frac{\braket{12}\braket{34}}{[12][34]}\delta^4(\bar{Q})
      \bar{\eta}_{a,3}[a\,b]\bar{\eta}_{b,3}\bar{\eta}_{c,4}[c\,d]\bar{\eta}_{d,4}\,.
\ee
Under the Fourier transform~(\ref{eqn:FT}) this maps into $\kappa_{(\o{ab})(\o{cd})}(1,2,3,4)$, where a barred pair of indices $\{\o{a,b}\}\equiv\{1,2,3,4\}\setminus\{a,b\}$ denotes the complement with respect to the set of external labels.

\subsection{Loop-level amplitudes}
\label{sec:bcj}

Proceeding now to consider loop-level amplitudes, we adopt diagrammatic representations from the outset. In a general Yang-Mills theory,
the trivalent nature of the gauge group generators
allows us to write any $L$-loop amplitude as a sum of cubic graphs:
\begin{equation}\label{eq:colorKinDualAmp}
\cA_n^{(L)}=i^{L-1}g^{n+2L-2}\sum_{\text{cubic graphs }\Gamma_i}
\int\!\frac{\d^{LD}\ell}{(2\pi)^{LD}}\frac1{S_i} \frac{n_ic_i}{D_i}\,.
\end{equation}
Here $g$ is the coupling, $S_i$ are the symmetry factors,
$D_i$ are the usual products of massless propagators,
and $n_i$ are the kinematic numerators associated with each graph,
depending on both external and loop momenta.
To regulate potentially divergent integrals,
we use dimensional regularization in $D=4-2\eps$ dimensions.
Finally, we assume $N_f=1$ hypermultiplet flavors;
one can generalize to $N_f\neq1$ by assigning flavor-conserving delta functions to each diagram,
with $N_f=\delta^\alpha_\alpha$ for closed matter loops (see~\rcites{Dixon:2010ik,Melia:2013epa,Johansson:2015oia} for more details).

One of the main advantages of such a cubic representation is that
the color factors~$c_i$ are unambiguously assigned to the graphs.
There are two kinds of trivalent vertices:
pure-adjoint, and those with a particle in each of the adjoint, fundamental and anti-fundamental representations.
These are associated with the structure constant $\tf^{abc}=\ttr([T^a,T^b]T^c)$ and generator $T^a_{i\bar\jmath}$, respectively:\footnote{In this paper,
we label cubic diagrams by their graphical representations.
Their explicit layout encodes a sign due to the antisymmetry of the vertices.
}
\be
   %\tikzset{external/force remake}%
   \tf^{abc} =
      c\!\left(\!\!\gTreeTri[eLA=$b$,eLB=$c$,eLC=$a$]{}\right) , \qquad \quad
   T^a_{i\bar\jmath} =
      c\!\left(\!\!\gTreeTri[eB=aquark,eC=quark,eLA=$a$,eLC=$i$,eLB=$\bar\jmath$]{}
         \right) .
%    =-c\!\left(\!\!\gTreeTri[eB=quark,eC=aquark,eLA=$a$,eLB=$i$,eLC=$\bar\jmath$]{}
%         \right) = -T^a_{\bar\imath j}\,.
\ee
Both are antisymmetric:
$\tf^{abc}=-\tf^{acb}$, $T^a_{\bar\imath j}\equiv-T^a_{j\bar\imath}$
(the latter relationship defines $T^a_{\bar\imath j}$).
Fundamental structure constants are normalized such that
$\ttr(T^aT^b) = \delta^{ab}$.

We seek loop-level amplitude representations obeying color-kinematics duality~\cite{Bern:2008qj,Bern:2010ue}. In this case, the same linear identities satisfied by the color factors $c_i$ should also be satisfied by the kinematic numerators $n_i$, which we refer to as color-dual. Such relationships include commutation relations
\beal\label{eq:commutation}
\tf^{ba_3a_4}T^b_{i_1\bar\imath_2}&=
T^{a_3}_{i_1\bar\jmath}T^{a_4}_{j\bar\imath_2}-
T^{a_4}_{i_1\bar\jmath}T^{a_3}_{j\bar\imath_2}=[T^{a_3},T^{a_4}]_{i_1\bar\imath_2}\,,\\
c\!\left(\gTreeS[eLA=$1$,eLB=$2$,eLC=$3$,eLD=$4$,eA=quark,eB=aquark]{}\right)&=
c\!\left(\gTreeT[eLA=$1$,eLB=$2$,eLC=$3$,eLD=$4$,eA=quark,eB=aquark,iA=aquark]{}\right)-
c\!\left(\gTreeT[eLA=$1$,eLB=$2$,eLC=$4$,eLD=$3$,eA=quark,eB=aquark,iA=aquark]{}\right) ,
\eeal
and their adjoint-representation counterparts --- the Jacobi identities
\beal\label{eq:jacobi}
\tf^{a_1a_2b}\tf^{ba_3a_4}&=\tf^{a_4a_1b}\tf^{ba_2a_3}-\tf^{a_2a_4b}\tf^{ba_3a_1}\,,\\
c\!\left(\gTreeS[eLA=$1$,eLB=$2$,eLC=$3$,eLD=$4$]{}\right)&=
c\!\left(\gTreeT[eLA=$1$,eLB=$2$,eLC=$3$,eLD=$4$]{}\right)-
c\!\left(\gTreeT[eLA=$1$,eLB=$2$,eLC=$4$,eLD=$3$]{}\right) .
\eeal
Color-kinematics duality requires that
\beal
    c_i = c_j - c_k & \qquad \Leftrightarrow \qquad n_i = n_j - n_k\,, %\\
%    c_i \rightarrow -c_i & \quad \Leftrightarrow \quad n_i \rightarrow -n_i\,.
\eeal
%The second set of identities are symmetries, arising from antisymmetry of the vertices as described above. They incorporate a sign change for a swap of two legs at a cubic vertex.

The usual motivation for finding so-called color-dual representations is to enable use of the double copy~\cite{Bern:2008qj,Bern:2010ue}, which allows supergravity amplitudes to be obtained by replacing the color factors $c_i$ with a second copy of the kinematic numerators $n_i$ in the amplitude~\eqref{eq:colorKinDualAmp}.
Fundamental-representation hypers play an important role when dealing with $\cN<4$ supergravities, for example as they allow unwanted additional vector multiplets to be subtracted from the resulting supergravity multiplet~\cite{Johansson:2014zca}.
For instance, \rcites{Johansson:2014zca,Johansson:2017bfl} described how pure $\cN=4$ supergravity amplitudes could be obtained from a double copy of $\cN=2$ SYM with itself, the hypermultiplets being used internally to remove unwanted $\cN=4$ SYM multiplets from the supergravity theory.

There are, however, considerable advantages to finding color-dual representations even if the goal is merely efficient computation of gauge-theory amplitudes.
First, such representations are cubic, so the assignment of color factors to diagrams is trivial
(for alternative non-cubic constructions of full-color integrands from unitarity cuts
see \eg \rcites{Badger:2015lda,Ochirov:2016ewn,Kalin:2017oqr}).
Moreover, the kinematic numerators being interlinked by commutation and Jacobi relations implies that only a limited subset of the numerators need to be calculated directly.
The corresponding graphs, which are referred to as masters, are chosen to ensure that the numerators of all other graphs can be obtained using commutation and Jacobi identities.
For instance, using the commutation relation~\eqref{eq:commutation}
implies
\begin{equation}\label{eq:egMaster}
%\tikzset{external/force remake}
n\!\left(\gTriC[eLA=$1$,eLB=$2$,eLC=$3$,eLD=$4$,eA=quark,eB=aquark]{}\right)=
n\!\left(\gBox[eLA=$1$,eLB=$2$,eLC=$3$,eLD=$4$,eA=quark,eB=aquark,iA=aquark]{}\right)-
n\!\left(\gBox[eLA=$2$,eLB=$1$,eLC=$3$,eLD=$4$,eA=aquark,eB=quark,iA=quark]{}\right).
\end{equation}
In this case, the triangle is uniquely determined by the two boxes;
these, in turn, are related by a symmetry through the horizontal axis (after relabeling $p_3\leftrightarrow p_4$).
The box is in this case a master:
it is on these masters that we will focus our attention in later sections.

In general, the existence of a consistent set of color-dual numerators is not trivial. At tree level, it is proven~\cite{Kiermaier,BjerrumBohr:2010hn}
for gauge theories, in which color-ordered amplitudes satisfy
the BCJ relations~\cite{Bern:2008qj,BjerrumBohr:2009rd,Stieberger:2009hq,Feng:2010my}.
For (super-)Yang-Mills theories with arbitrary fundamental matter, which is less studied, the corresponding BCJ relations~\cite{Johansson:2014zca,Johansson:2015oia} have been proven in the case of QCD~\cite{delaCruz:2015dpa}.

The task of finding color-dual representations is further complicated by the fact that, for a given amplitude, such representations are generally not unique.
For this reason, a large part of \rcite{Johansson:2017bfl} was devoted to finding additional constraints to be imposed on the numerators, with the intention of shrinking the space of allowed solutions while manifesting certain desirable properties.
These constraints also reduce the number of masters to be computed.
In the remainder of this section we discuss the constraints that we have found helpful. 
Note that not all the integrands presented here have all the properties presented below;
\app{sec:allRes} summarizes the various representations and their properties.

%%%%%%%%%%%%%%%%%%%%%%%%%%%%%%%%%%%%%%%%%%%%%%%%%%
\subsubsection{Two-term identities}
\label{sec:2term}

If imposed, the two-term identities require that, for indistinguishable matter multiplets,
\begin{align}\label{eq:2term}
%\tikzset{external/force remake}
c\!\left(\gTreeS[eLA=$1$,eLB=$2$,eLC=$3$,eLD=$4$,eA=quark,eB=aquark,eC=quark,eD=aquark]{}\right)
\stackrel{?}{=}
c\!\left(\gTreeT[eLA=$1$,eLB=$2$,eLC=$3$,eLD=$4$,eA=quark,eB=aquark,eC=quark,eD=aquark]{}\right).
\end{align}
Although this is not true in general,
it holds if the gauge group is chosen as $G=\,$U(1),
or for specific tensor representations of U($N_c$)~\cite{Johansson:2014zca}.
Like the commutation and Jacobi relations, we impose these identities on the numerators whose graphs contain internal subgraphs of the above form.
For instance, these one-loop box and triangle numerators are equated:
\begin{equation}\label{eq:eg2term}
%\tikzset{external/force remake}
n\!\left(\gBox[eLA=$1$,eLB=$2$,eLC=$3$,eLD=$4$,eA=quark,eB=aquark,eC=quark,eD=aquark,iB=quark,iD=quark]{}\right)=
n\!\left(\gTriC[eLA=$1$,eLB=$2$,eLC=$3$,eLD=$4$,eA=quark,eB=aquark,eC=quark,eD=aquark,iB=quark,iD=quark]{}\right).
\end{equation}
One can also regard the two-term identities as their own kind of commutation relations, the difference being that the $u$-channel graphs are excluded as their routing of fundamental matter lines is not sensible.

As we shall see in \sec{sec:N2rules}, the two-term identities have their origin in the structure of the $\cN=2$ cuts, and the diagrammatic rules will help to clarify this.
In \rcite{Johansson:2017bfl}, these identities allowed all numerators with two matter loops to be reduced to those with one matter loop.
The two-term identities are equally useful in restricting the set of masters when matter is taken on external legs.

%%%%%%%%%%%%%%%%%%%%%%%%%%%%%%%%%%%%%%%%%%%%%%%%%%
\subsubsection{CPT conjugation}
\label{sec:CPT}

Amplitudes respect the CPT invariance of the theory,
and we can extend this to a manifest off-shell symmetry acting on individual numerators.
CPT conjugation acts by transforming $|i\rangle\leftrightarrow|i]$ and $\eta_i^I\leftrightarrow\bar{\eta}_{i,I}$;
graphically, this corresponds to flipping the helicity of external vectors and reversing arrow directions on hypermultiplets.
The transformation should correspond to a replacement of $\kappa_{(ab)(cd)}$ by its complement and an additional sign flip of parity-odd terms:
\begin{equation}\label{eqn:CPT}
   n_i(1,2,3,4;\ell_1,\ell_2) = \bar{n}_i(1,2,3,4;\ell_1,\ell_2)
      |_{ \kappa_{(ab)(cd)}\rightarrow
          \kappa_{(\o{ab})(\o{cd})},
          |i\rangle\leftrightarrow|i]}\,,
\end{equation}
where $\bar{n}_i$ stands for the numerator of the graph with flipped arrows on matter lines;
$\{\o{a,b}\}\equiv\{1,2,3,4\}\setminus\{a,b\}$.
For instance, we could equate
\begin{equation}\label{eq:egCPT}
%\tikzset{external/force remake}
n\!\left(\gBox[eLA=$1$,eLB=$2$,eLC=$3$,eLD=$4$,
eA=quark,eB=aquark,eC=quark,eD=aquark,iB=quark,iD=quark]{}\right)=
\left.n\!\left(\gBox[eLA=$1$,eLB=$2$,eLC=$3$,eLD=$4$,
eA=aquark,eB=quark,eC=aquark,eD=quark,iB=aquark,iD=aquark]{}\right)
\right|_{\kappa_{(ab)(cd)}\rightarrow
          \kappa_{(\o{ab})(\o{cd})},
          |i\rangle\leftrightarrow|i]}\,.
\end{equation}
Note that a change of direction of external hyper lines together with the conjugation of the indices of $\kappa$ lands us back on the same external state configuration as we started from.

%%%%%%%%%%%%%%%%%%%%%%%%%%%%%%%%%%%%%%%%%%%%%%%%%%
\subsubsection{Matter-reversal symmetry}
\label{sec:matterReverse}

The matter multiplets $\Phi_{\cN=2}$ and $\wPhi_{\cN=2}$ are identical up to R-symmetry indices and the gauge-group representation,
which leads to another potential off-shell symmetry of the numerators.
In \rcite{Johansson:2017bfl} only vector multiplets were allowed on external legs,
so the symmetry was invariance under arrow reversal for all numerators containing matter loops.
The symmetry held for each matter loop individually.

With hypermultiplets on external legs the situation is more subtle.
One cannot simply equate numerators with reversed hypermultiplets,
as they carry different external states.
But this is easily remedied:
by inspection of the $\cN=4$ state decomposition~\eqref{eq:neq4decomp} the symmetry clearly exchanges $\eta^3\leftrightarrow\eta^4$ (with no effect on $V^+_{\cN=2}$ or $V^-_{\cN=2}$).
With this additional exchange imposed, we can implement the same identity,
for instance
\begin{equation}\label{eq:egMatterReverse}
%\tikzset{external/force remake}
n\!\left(\gBox[eLA=$1$,eLB=$2$,eLC=$3$,eLD=$4$,
eA=quark,eB=quark,eC=aquark,eD=aquark,iB=aquark,iD=quark]{}\right)=
-\left.n\!\left(\gBox[eLA=$1$,eLB=$2$,eLC=$3$,eLD=$4$,
eA=quark,eB=aquark,eC=quark,eD=aquark,iB=quark,iD=quark]{}\right)
\right|_{\eta_2^4\to\eta_2^3,\eta_3^3\to\eta_3^4}.
\end{equation}
This symmetry is required if one considers matter multiplets in a pseudo-real representation~\cite{Chiodaroli:2015wal,Chiodaroli:2016jqw}.

%%%%%%%%%%%%%%%%%%%%%%%%%%%%%%%%%%%%%%%%%%%%%%%%%%
\subsubsection{Matching with \texorpdfstring{$\cN=4$}{N=4} SYM}
\label{sec:Neq4ident}

As we have already seen in \eqn{eq:neq4decomp}, the vector multiplet $\cV_{\cN=4}$'s $2^4=16$ states can be distributed between $V^+_{\cN=2}$, $V^-_{\cN=2}$, $\Phi_{\cN=2}$ and $\wPhi_{\cN=2}$.
This offers another constraint on the color-dual $\cN=2$ SQCD numerators: that summing them over the internal multiplets corresponding to the on-shell content of $\cN=4$ SYM should reproduce those same numerators.
For instance, we can demand that
\beal\label{eq:egNeq4ident}
n^{[\cN=4]}\!\left(\gBoxBox[scale=.6,eLA=$1$,eLB=$2$,eLC=$3$,eLD=$4$]{}\right) =
n\!\left(\gBoxBox[scale=.6,eLA=$1$,eLB=$2$,eLC=$3$,eLD=$4$]{}\right)+
n\!\left(\gBoxBox[scale=.6,eLA=$1$,eLB=$2$,eLC=$3$,eLD=$4$,
                  iA=aquark,iB=aquark,iG=quark,iF=aquark]{}\right)+
n\!\left(\gBoxBox[scale=.6,eLA=$1$,eLB=$2$,eLC=$3$,eLD=$4$,
                  iG=aquark,iE=aquark,iD=aquark,iC=aquark]{}\right)+
n\!\left(\gBoxBox[scale=.6,eLA=$1$,eLB=$2$,eLC=$3$,eLD=$4$,
   iA=aquark,iB=aquark,iF=aquark,iE=aquark,iD=aquark,iC=aquark]{}\right)&\\+\,
n\!\left(\gBoxBox[scale=.6,eLA=$1$,eLB=$2$,eLC=$3$,eLD=$4$,
                  iA=quark,iB=quark,iG=aquark,iF=quark]{}\right)+
n\!\left(\gBoxBox[scale=.6,eLA=$1$,eLB=$2$,eLC=$3$,eLD=$4$,
                  iG=quark,iE=quark,iD=quark,iC=quark]{}\right)+
n\!\left(\gBoxBox[scale=.6,eLA=$1$,eLB=$2$,eLC=$3$,eLD=$4$,
                  iA=quark,iB=quark,iF=quark,iE=quark,iD=quark,iC=quark]{}\right)&,
\eeal
where in this case a suitable expression for the $\cN=4$ double-box numerator is simply
\begin{align}\label{eq:egNeq4ident2}
n^{[\cN=4]}\left(\gBoxBox[scale=.8,eLA=$1$,eLB=$2$,eLC=$3$,eLD=$4$]{}\right)=
s(\kappa_{12}+\kappa_{13}+\kappa_{14}+\kappa_{23}+\kappa_{24}+\kappa_{34})\,.
\end{align}
Here we have projected the $\cN=2$ vector multiplets out of the supersymmetric delta function $\delta^8(Q)$.
This statement must be true for kinematic configurations where the propagators are taken on-shell, \ie on the maximal cut, because summing over these diagrams then simply amounts to taking $\eta^3$ and $\eta^4$ integrals on the right-hand side.
This logic also holds when hypers are taken on external legs. The non-trivial observation is that we can demand it also be true for off-shell loop momenta.

%%%%%%%%%%%%%%%%%%%%%%%%%%%%%%%%%%%%%%%%%%%%%%%%%%
%%%%%%%%%%%%%%%%%%%%%%%%%%%%%%%%%%%%%%%%%%%%%%%%%%
\section{Iterated two-particle cuts}
\label{sec:rungrule}

In this section we describe the iterative two-particle cut construction, as it applies to both $\cN=4$ and $\cN=2$ SYM in strictly four dimensions (in section~\ref{sec:oneloop} we will explain how to find higher-dimensional corrections for the purpose of dimensional regularization). The construction is underpinned by that fact that when two four-point amplitudes are glued together to form a cut, the result is proportional to another four-point tree amplitude. This allows the gluing procedure to be iterated, leading to diagrammatic rules for cut assembly without the need to perform intermediate supersums. As we shall see, in $\cN=4$ SYM this construction leads to the so-called ``rung rule'' for assembling Mondrian-type diagrams~\cite{Bern:1997nh,Bern:1998ug,Bern:2004kq}.

%%%%%%%%%%%%%%%%%%%%%%%%%%%%%%%%%%%%%%%%%%%%%%%%%%
\subsection{\texorpdfstring{$\cN=4$}{N=4} SYM}

To understand the iterated nature of two-particle cuts in $\cN=4$ SYM, we begin with the four-point one-loop $s$-channel cut
\beal
   %\tikzset{external/force remake}
   \cBoxA[eLA=3,eLB=4,eLC=1,eLD=2,
          iLA=$\oset{\rightarrow}{l_2}$,iLB=$\uset{\rightarrow}{l_1}$]{} &
   \begin{aligned}[t]
     &= \int\!\d^4\eta_{l_1}\d^4\eta_{l_2} A_4^{(0),\text{MHV}}(1,2,l_1,l_2)
                                         A_4^{(0),\text{MHV}}(3,4,-l_2,-l_1) \\ &
                                         =-i\frac{st}{(l_1+p_2)^2(l_1-p_3)^2}A_4^{(0),\text{MHV}}(1,2,3,4)\,.
   \end{aligned}
\eeal
The fact that this cut is proportional to the tree amplitude is a well-known result, following from Green, Schwarz and Brink's original computation of the one-loop amplitude~\cite{Green:1982sw}. If we extract the physical poles from the tree amplitudes using the MHV formula
\begin{equation}\label{eq:treeAmpN4}
  A_4^{(0),\text{MHV}}(1,2,3,4) = -\frac{i}{st}\kappa(1,2,3,4) \,,
\end{equation}
then using $s=s_{l_1l_2}$ (in this case) --- where $s_{ij}=(p_i+p_j)^2$, which we extend to include loop momenta --- the cut identity can be more compactly written as
\begin{equation}\label{eq:supersumN4}
   \int\!\d^4\eta_{l_1}\d^4\eta_{l_2} \kappa(1,2,l_1,l_2) \kappa(3,4,-l_2,-l_1)
    = s_{l_1 l_2}^2 \kappa(1,2,3,4)\,.
\end{equation}
A similar construction was presented in \rcite{Bern:2010tq};
for the sake of completeness we give a proof of this relation in \app{app:superspace}.

\subsubsection{\texorpdfstring{$\cN=4$}{N=4} diagrammatic rules}

The two-particle cut being proportional to the tree-level amplitude allows for an iterated construction.
We can attach more four-point tree-level amplitudes to the cut
and glue two pairs of legs at a time by using \eqn{eq:supersumN4}.
So any iterated two-particle cut is a product of terms coming from its four-point tree-level amplitudes
and two-particle supersums.
The following diagrammatic rules summarize the construction:
\be
  %\tikzset{external/force remake}
  \begin{tikzpicture}
    [line width=1pt,
    baseline={([yshift=-0.5ex]current bounding box.center)},
    font=\scriptsize]
    \fill[blob] (0,0) circle (0.25);
    \draw[gluon] (45:0.25) -- (45:0.5) node[right] {$a$};
    \draw[gluon] (-45:0.25) -- (-45:0.5) node[right] {$b$};
    \draw[gluon] (-135:0.25) -- (-135:0.5) node[left] {$c$};
    \draw[gluon] (135:0.25) -- (135:0.5) node[left] {$d$};
  \end{tikzpicture}
  \rightarrow -\frac{i}{s_{ab}s_{ac}}\,,
  \qquad \quad
  %\tikzset{external/force remake}
  \begin{tikzpicture}
    [line width=1pt,
    baseline={([yshift=-0.5ex]current bounding box.center)},
    font=\scriptsize]
    \fill[pattern=north west lines] (-0.5,0) -- (-0.5,-0.25) arc (-90:90:0.25) -- cycle;
    \draw (-0.5,-0.25) arc (-90:90:0.25);
    \fill[pattern=north west lines] (0.5,0) -- (0.5,0.25) arc (90:270:0.25) -- cycle;
    \draw (0.5,0.25) arc (90:270:0.25);
    \draw[gluon] ($(-0.5,0) + (30:0.25)$) -- node[above] {$\underset{\rightarrow}{l_1}$} ($(0.5,0) + (150:0.25)$);
    \draw[gluon] ($(-0.5,0) + (-30:0.25)$) -- node[below=0.05] {$\overset{\rightarrow}{l_2}$} ($(0.5,0) + (-150:0.25)$);
  \end{tikzpicture}
  \rightarrow s_{l_1l_2}^2\,,
  \qquad \quad
  %\tikzset{external/force remake}
  \begin{tikzpicture}
    [line width=1pt,
    baseline={([yshift=-0.5ex]current bounding box.center)},
    font=\scriptsize]
    \draw[dotted] (0,0) circle (0.2);
    \draw[gluon] (45:0.2) -- (45:0.5) node[right] {$q$};
    \draw[gluon] (-45:0.2) -- (-45:0.5) node[right] {$r$};
    \draw[gluon] (-135:0.2) -- (-135:0.5) node[left] {$s$};
    \draw[gluon] (135:0.2) -- (135:0.5) node[left] {$t$};
  \end{tikzpicture}
  \rightarrow\kappa(q,r,s,t)\,.
\label{eq:rungruleN4cuts}
\ee
The first rule comes from \eqn{eq:treeAmpN4}; for each tree-amplitude constituent, we must insert its physical poles.
The second rule is the result of \eqn{eq:supersumN4};
it tells us that a factor $s_{l_1l_2}^2$ is obtained whenever two tree-level amplitudes are glued together.
Finally, the last ``external'' rule tells us that, once all partons have been assembled, we should multiply by an overall $\kappa$ factor to encode the configuration of external states.

The $s$-channel cut from before is now easily assembled:
\begin{equation}\label{eq:twoCutNeq4}
   %\tikzset{external/force remake}
   \cBoxA[eLA=3,eLB=4,eLC=1,eLD=2,
          iLA=$\oset{\rightarrow}{l_2}$,iLB=$\uset{\rightarrow}{l_1}$]{}
    = \frac{-i}{s_{12}s_{1l_2}}\times s_{l_1l_2}^2\times\frac{-i}{s_{34}s_{4(-l_2)}}\times\kappa
    = -\frac{s_{l_1 l_2}^2\kappa}{s^2 s_{1l_2} s_{4(-l_2)}}\,.
\end{equation}
The poles are the physical poles of both tree-level constituents. Re-using $s=s_{l_1l_2}$ to cancel unwanted poles, it becomes clear that the only ones leftover are $s_{1l_2}$ and $s_{4(-l_2)}$. In terms of off-shell numerators,
\begin{equation}\label{eq:N4box}
%\tikzset{external/force remake}
n^{[\cN=4]}\left(\gBox[eLA=$1$,eLB=$2$,eLC=$3$,eLD=$4$]{}\right)=\kappa
\end{equation}
is the only contributor~\cite{Green:1982sw}.

%%%%%%%%%%%%%%%%%%%%%%%%%%%%%%%%%%%%%%%%%%%%%%%%%%
\subsubsection{The rung rule}
\label{sec:rungruleN4}

The rung rule takes this construction one step further, directly giving off-shell expressions for box-like Mondrian diagrams without the need for cut assembly~\cite{Bern:1997nh,Bern:2004kq}. In the $s$-channel cut given above, we noticed a cancellation between kinematic factors $s_{l_1l_2}$, coming from the gluing rule~\eqref{eq:supersumN4}, with physical poles coming from the physical tree amplitudes~\eqref{eq:treeAmpN4}. The cancellation is completely general, as gluing a four-point tree amplitude to an arbitrary four-point MHV amplitude gives
%(up to an overall $\kappa$)
\be\label{eq:rungruleN4onshell}
  %\tikzset{external/force remake}
  \begin{tikzpicture}
    [line width=1pt,
    baseline={([yshift=-0.5ex]current bounding box.center)},
    font=\scriptsize]
    \fill[blob] (-0.5,0) ellipse (0.2 and 0.3);
    \fill[blob] (0.5,0) ellipse (0.3 and 0.4);
    \filldraw[fill=white,draw=black] (0.55,0.15) circle (0.1);
    \filldraw[fill=white,draw=black] (0.45,-0.1) circle (0.1);
    \draw ($(-0.5,0) + (135:0.23)$) -- ($(-0.5,0) + (135:0.5)$) node[left] {$2$};
    \draw ($(-0.5,0) + (-135:0.23)$) -- ($(-0.5,0) + (-135:0.5)$) node[left] {$1$};
    \draw ($(-0.5,0) + (30:0.23)$) -- node[above] {$\uset{\leftarrow}{l_1}$} ($(0.5,0) + (157:0.3)$);
    \draw ($(-0.5,0) + (-30:0.23)$) -- node[below] {$\oset{\leftarrow}{l_2}$} ($(0.5,0) + (-157:0.3)$);
    \draw ($(0.5,0) + (45:0.35)$) -- ($(0.5,0) + (45:0.7)$) node[right] {$3$};
    \draw ($(0.5,0) + (-45:0.35)$) -- ($(0.5,0) + (-45:0.7)$) node[right] {$4$};
  \end{tikzpicture}
  = \frac{-i}{s_{12}s_{1(-l_2)}}\times s_{l_1l_2}^2\times
  %\tikzset{external/force remake}
  \begin{tikzpicture}
    [line width=1pt,
    baseline={([yshift=-0.5ex]current bounding box.center)},
    font=\scriptsize]
    \fill[blob] (0.5,0) ellipse (0.3 and 0.4);
    \filldraw[fill=white,draw=black] (0.55,0.15) circle (0.1);
    \filldraw[fill=white,draw=black] (0.45,-0.1) circle (0.1);
    \draw ($(0.5,0) + (157:0.3)$) -- ($(-0.3,0) + (30:0.23)$) node[left] {$l_1$};
    \draw ($(0.5,0) + (-157:0.3)$) -- ($(-0.3,0) + (-30:0.23)$)  node[left] {$l_2$};
    \draw ($(0.5,0) + (45:0.35)$) -- ($(0.5,0) + (45:0.7)$) node[right] {$3$};
    \draw ($(0.5,0) + (-45:0.35)$) -- ($(0.5,0) + (-45:0.7)$) node[right] {$4$};
  \end{tikzpicture}
  = -\frac{is_{l_1l_2}}{s_{1(-l_2)}}\times
  %\tikzset{external/force remake}
  \begin{tikzpicture}
    [line width=1pt,
    baseline={([yshift=-0.5ex]current bounding box.center)},
    font=\scriptsize]
    \fill[blob] (0.5,0) ellipse (0.3 and 0.4);
    \filldraw[fill=white,draw=black] (0.55,0.15) circle (0.1);
    \filldraw[fill=white,draw=black] (0.45,-0.1) circle (0.1);
    \draw ($(0.5,0) + (157:0.3)$) -- ($(-0.3,0) + (30:0.23)$) node[left] {$l_1$};
    \draw ($(0.5,0) + (-157:0.3)$) -- ($(-0.3,0) + (-30:0.23)$)  node[left] {$l_2$};
    \draw ($(0.5,0) + (45:0.35)$) -- ($(0.5,0) + (45:0.7)$) node[right] {$3$};
    \draw ($(0.5,0) + (-45:0.35)$) -- ($(0.5,0) + (-45:0.7)$) node[right] {$4$};
  \end{tikzpicture}\,.
\ee
This suggests that a triangle-like diagram should not contribute as the $s_{12}$ pole is absent. By further cutting into the left tree-level amplitude we obtain an on-shell rung rule:
\be
  %\tikzset{external/force remake}
  \begin{tikzpicture}
    [line width=1pt,
    baseline={([yshift=-0.5ex]current bounding box.center)},
    font=\scriptsize]
    \fill[blob] (0.5,0) ellipse (0.3 and 0.4);
    \filldraw[fill=white,draw=black] (0.55,0.15) circle (0.1);
    \filldraw[fill=white,draw=black] (0.45,-0.1) circle (0.1);
    \draw ($(-0.5,0) + (30:0.23)$) -- ($(-0.2,0) + (135:0.5)$) node[left] {$2$};
    \draw ($(-0.5,0) + (-30:0.23)$) -- ($(-0.2,0) + (-135:0.5)$) node[left] {$1$};
    \draw ($(-0.5,0) + (30:0.23)$) -- ($(-0.5,0) + (-30:0.23)$);
    \draw ($(-0.5,0) + (30:0.23)$) -- node[above] {$\uset{\leftarrow}{l_1}$} ($(0.5,0) + (157:0.3)$);
    \draw ($(-0.5,0) + (-30:0.23)$) -- node[below] {$\oset{\leftarrow}{l_2}$} ($(0.5,0) + (-157:0.3)$);
    \draw ($(0.5,0) + (45:0.35)$) -- ($(0.5,0) + (45:0.7)$) node[right] {$3$};
    \draw ($(0.5,0) + (-45:0.35)$) -- ($(0.5,0) + (-45:0.7)$) node[right] {$4$};
  \end{tikzpicture}
  = -is_{l_1l_2}\times
  %\tikzset{external/force remake}
  \begin{tikzpicture}
    [line width=1pt,
    baseline={([yshift=-0.5ex]current bounding box.center)},
    font=\scriptsize]
    \fill[blob] (0.5,0) ellipse (0.3 and 0.4);
    \filldraw[fill=white,draw=black] (0.55,0.15) circle (0.1);
    \filldraw[fill=white,draw=black] (0.45,-0.1) circle (0.1);
    \draw ($(0.5,0) + (157:0.3)$) -- ($(-0.3,0) + (30:0.23)$) node[left] {$l_1$};
    \draw ($(0.5,0) + (-157:0.3)$) -- ($(-0.3,0) + (-30:0.23)$)  node[left] {$l_2$};
    \draw ($(0.5,0) + (45:0.35)$) -- ($(0.5,0) + (45:0.7)$) node[right] {$3$};
    \draw ($(0.5,0) + (-45:0.35)$) -- ($(0.5,0) + (-45:0.7)$) node[right] {$4$};
  \end{tikzpicture}\,.
\ee
In other words, attaching an on-shell rung to an existing cut amounts to
multiplication by $-is_{l_1l_2}$. The off-shell continuation of this statement for the amplitude numerators is typically written as
\be\label{eq:rungruleN4offshell}
  %\tikzset{external/force remake}
  \begin{tikzpicture}
    [line width=1pt,
    baseline={([yshift=-0.5ex]current bounding box.center)},
    font=\scriptsize]
    \fill[blob] (0.5,0) ellipse (0.3 and 0.4);
    \filldraw[fill=white,draw=black] (0.55,0.15) circle (0.1);
    \filldraw[fill=white,draw=black] (0.45,-0.1) circle (0.1);
    \draw ($(-0.5,0) + (30:0.23)$) -- ($(-0.2,0) + (135:0.5)$) node[left] {$2$};
    \draw ($(-0.5,0) + (-30:0.23)$) -- ($(-0.2,0) + (-135:0.5)$) node[left] {$1$};
    \draw ($(-0.5,0) + (30:0.23)$) -- ($(-0.5,0) + (-30:0.23)$);
    \draw ($(-0.5,0) + (30:0.23)$) -- node[above] {$\uset{\leftarrow}{\ell_1}$} ($(0.5,0) + (157:0.3)$);
    \draw ($(-0.5,0) + (-30:0.23)$) -- node[below] {$\oset{\leftarrow}{\ell_2}$} ($(0.5,0) + (-157:0.3)$);
    \draw ($(0.5,0) + (45:0.35)$) -- ($(0.5,0) + (45:0.7)$) node[right] {$3$};
    \draw ($(0.5,0) + (-45:0.35)$) -- ($(0.5,0) + (-45:0.7)$) node[right] {$4$};
  \end{tikzpicture}
  = -i(\ell_1 + \ell_2)^2\times
  %\tikzset{external/force remake}
  \begin{tikzpicture}
    [line width=1pt,
    baseline={([yshift=-0.5ex]current bounding box.center)},
    font=\scriptsize]
    \fill[blob] (0.5,0) ellipse (0.3 and 0.4);
    \filldraw[fill=white,draw=black] (0.55,0.15) circle (0.1);
    \filldraw[fill=white,draw=black] (0.45,-0.1) circle (0.1);
    \draw ($(0.5,0) + (157:0.3)$) -- ($(-0.3,0) + (30:0.23)$) node[left] {$\ell_1$};
    \draw ($(0.5,0) + (-157:0.3)$) -- ($(-0.3,0) + (-30:0.23)$)  node[left] {$\ell_2$};
    \draw ($(0.5,0) + (45:0.35)$) -- ($(0.5,0) + (45:0.7)$) node[right] {$3$};
    \draw ($(0.5,0) + (-45:0.35)$) -- ($(0.5,0) + (-45:0.7)$) node[right] {$4$};
  \end{tikzpicture}\,,
\ee
where the legs $\ell_1$ and $\ell_2$ are now understood
to carry unconstrained loop momenta.\footnote{In principle,
the off-shell continuation of $s_{l_1l_2}$ to $(\ell_1+\ell_2)^2$
in \eqn{eq:rungruleN4offshell} is not unique.
One may, for example, choose $2(\ell_1\!\cdot \ell_2)$ instead,
thereby ignoring the terms $\ell_1^2$ and $\ell_2^2$ that vanish on the cut.}

For instance, beginning with the box numerator given in \eqn{eq:N4box}, attaching a first rung gives the two-loop double box numerator~\cite{Bern:1997nh},
\begin{equation}
  %\tikzset{external/force remake}
  n^{[\cN=4]}\!\left(\gBoxBox[scale=0.8,eLA=$1$,eLB=$2$,eLC=$3$,eLD=$4$]{}\right)
    = s\,\kappa\,.
\end{equation}
The two possible ways of attaching a second rung give the 3-loop triple-box and ``tennis-court'' numerators:
\begin{equation}
  %\tikzset{external/force remake}
  n^{[\cN=4]}\!\left(\gBoxBoxBoxA[scale=0.8,eLA=$1$,eLB=$2$,eLC=$3$,eLD=$4$]{}\right)
    = s^2\,\kappa\,, \qquad \quad
    n^{[\cN=4]}\!\left(\gBoxBoxBoxB[scale=.8,yshift=-1,eLA=$1$,eLB=$2$,eLC=$3$,eLD=$4$,
      iLG=$\uset{\rightarrow}{\ell}$]{}\right)
    = s(\ell+p_4)^2\,\kappa\,.
\end{equation}
This pattern agrees with the three-loop amplitude~\cite{Bern:1997nh,ArkaniHamed:2010kv,ArkaniHamed:2010gh}.

However, we should recognize the circumstances under which the rung rule is too na\"{i}ve. While cuts are unique, off-shell numerators are not:
we are free to shift terms between numerators by adding terms that vanish on support of the on-shell conditions.
So the numerators may require modification, for instance, if
\begin{itemize}
\item the same diagram contributes to cuts that suggest different on-shell forms;
\item additional off-shell constraints --- like color duality --- are demanded.
\end{itemize}
A classic example where the rung rule fails to provide color-dual numerators is the 4-point, 3-loop MHV amplitude~\cite{Bern:2010ue,Bern:2012cd} --- the numerators given above are not color-dual.
Therefore, as we now proceed to consider $\cN=2$ SQCD, we will use the iteration only to construct cuts, and remember that off-shell numerators may require modification.

%%%%%%%%%%%%%%%%%%%%%%%%%%%%%%%%%%%%%%%%%%%%%%%%%%
\subsection{\texorpdfstring{$\cN=2$}{N=2} SQCD}

The iterative cut construction works in $\cN=2$ SYM for the same reason as in $\cN=4$ SYM:
because the result of gluing together a pair of tree-level amplitudes is proportional to another tree-level amplitude.
The generalization of the supersum~\eqref{eq:supersumN4}, written in terms of $\kappa_{(ab)(cd)}$ as introduced in~\sec{sec:tree}, is
\begin{equation}\label{eq:supersumN2}
\int\d^4\eta_{l_1} \d^4\eta_{l_2} \kappa_{(ab)(cd)}^{\text{(L)}}\kappa_{(ef)(gh)}^{\text{(R)}}=
s_{l_1l_2}\braket{ab}[\o{cd}]\braket{ef}[\o{gh}]
[qr]\braket{\o{st}}\frac{\kappa_{(qr)(st)}}{s_{qr}s_{st}}\,,
\end{equation}
which we will prove below; we denote
\beal
   \kappa_{(ab)(cd)}^{\text{(L)}} & = \kappa_{(ab)(cd)}(1,2,l_1,l_2)\,, \qquad \quad
   \kappa_{(ef)(gh)}^{\text{(R)}} = \kappa_{(ef)(gh)}(3,4,-l_2,-l_1)\,, \\
   \kappa_{(qr)(st)} & = \kappa_{(qr)(st)}(1,2,3,4)\,,
\eeal
where $(qr)(st)$ denotes the overall state configuration. Bars denote the complement with respect to external legs on a tree amplitude; for instance, $\{\o{c,d}\}=\{1,2,l_1,l_2\}\setminus\{c,d\}$. We omit the overall sign since it depends on the ordering of the complement, which affects the spinor-helicity brackets.

As we shall discuss in \sec{sec:multiparticlecuts}, similar relations work for higher-point tree amplitudes.
In the conclusions, we will also discuss possible generalizations to $\cN=0$ and $\cN=1$ SYM amplitudes.

\subsubsection{\texorpdfstring{$\cN=2$}{N=2} diagrammatic rules}
\label{sec:N2rules}
The relationship~\eqref{eq:supersumN2} leads to simple rules for assembling any iterated two-particle cut.
For each four-point constituent, regardless of the configuration of external or intermediate states, we include the same physical poles as we did for $\cN=4$ SYM:
\begin{align}\label{eqn:physPolesRule}
  %\tikzset{external/force remake}
  \begin{tikzpicture}
    [line width=1pt,
    baseline={([yshift=-0.5ex]current bounding box.center)},
    font=\scriptsize]
    \fill[blob] (0,0) circle (0.2);
    \draw[line] (45:0.2) -- (45:0.5) node[right] {$a$};
    \draw[line] (-45:0.2) -- (-45:0.5) node[right] {$b$};
    \draw[line] (-135:0.2) -- (-135:0.5) node[left] {$c$};
    \draw[line] (135:0.2) -- (135:0.5) node[left] {$d$};
  \end{tikzpicture}
  &\rightarrow -\frac{i}{s_{ab}s_{bc}}\,.
\end{align}
The solid lines are used to indicate that it does not matter whether the states are vectors or hypermultiplets. For each amplitude there is now an additional factor: $\braket{ab}[\o{cd}]$ for the left-hand side of the cut, and $\braket{ef}[\o{gh}]$ for the right-hand side. This gives a numerator contribution depending on the particle content:
\begin{subequations}\label{eq:N2blobRules}
  \begin{align}
  \label{eq:cutRule1}
  %\tikzset{external/force remake}
  \begin{tikzpicture}
    [line width=1pt,
    baseline={([yshift=-0.5ex]current bounding box.center)},
    font=\scriptsize]
    \fill[blob] (0,0) circle (0.2);
    \draw[gluon] (45:0.2) -- (45:0.5) node[right] {$a^-$};
    \draw[gluon] (-45:0.2) -- (-45:0.5) node[right] {$b^-$};
    \draw[gluon] (-135:0.2) -- (-135:0.5) node[left] {$c^+$};
    \draw[gluon] (135:0.2) -- (135:0.5) node[left] {$d^+$};
  \end{tikzpicture}
  &\rightarrow \braket{ab}[cd]\,,\\
  \label{eq:cutRule2}
  %\tikzset{external/force remake}
  \begin{tikzpicture}
    [line width=1pt,
    baseline={([yshift=-0.5ex]current bounding box.center)},
    font=\scriptsize]
    \fill[blob] (0,0) circle (0.2);
    \draw[gluon] (45:0.2) -- (45:0.5) node[right] {$a^-$};
    \draw[gluon] (-45:0.2) -- (-45:0.5) node[right] {$b^+$};
    \draw[quark] (-135:0.2) -- (-135:0.5) node[left] {$c$};
    \draw[aquark] (135:0.2) -- (135:0.5) node[left] {$d$};
  \end{tikzpicture}
  &\rightarrow \langle a|c|b]\,,\\ %= -\langle a|p_d+p_a|b]\,,\\
  \label{eq:cutRule3}
  %\tikzset{external/force remake}
  \begin{tikzpicture}
    [line width=1pt,
    baseline={([yshift=-0.5ex]current bounding box.center)},
    font=\scriptsize]
    \fill[blob] (0,0) circle (0.2);
    \draw[quark] (45:0.2) -- (45:0.5) node[right] {$a$};
    \draw[aquark] (-45:0.2) -- (-45:0.5) node[right] {$b$};
    \draw[quark] (-135:0.2) -- (-135:0.5) node[left] {$c$};
    \draw[aquark] (135:0.2) -- (135:0.5) node[left] {$d$};
  \end{tikzpicture}~\,
  &\rightarrow s_{ac} = s_{bd}\,,
\end{align}
\end{subequations}
where the ordering of legs is irrelevant. When gluing two tree-level amplitudes, we multiply by
\begin{align}\label{eq:N2gluingrule}
  %\tikzset{external/force remake}
  \begin{tikzpicture}
    [line width=1pt,
    baseline={([yshift=-0.5ex]current bounding box.center)},
    font=\scriptsize]
    \fill[pattern=north west lines] (-0.5,0) -- (-0.5,-0.25) arc (-90:90:0.25) -- cycle;
    \draw (-0.5,-0.25) arc (-90:90:0.25);
    \fill[pattern=north west lines] (0.5,0) -- (0.5,0.25) arc (90:270:0.25) -- cycle;
    \draw (0.5,0.25) arc (90:270:0.25);
    \draw ($(-0.5,0) + (30:0.25)$) -- node[above] {$\underset{\rightarrow}{l_1}$} ($(0.5,0) + (150:0.25)$);
    \draw ($(-0.5,0) + (-30:0.25)$) -- node[below] {$\overset{\rightarrow}{l_2}$} ($(0.5,0) + (-150:0.25)$);
  \end{tikzpicture}
  &\rightarrow s_{l_1l_2}\,,
\end{align}
where again the type of particles is irrelevant. Finally, the leftover $[qr]\braket{\o{st}}$, as well as the poles in $s_{qr}$ and $s_{st}$,
yield an overall factor depending on the configuration of the four external legs:
\begin{subequations}\label{eq:N2extRules}
\begin{align}
  %\tikzset{external/force remake}
  \label{eq:cutRule4}
  \begin{tikzpicture}
    [line width=1pt,
    baseline={([yshift=-0.5ex]current bounding box.center)},
    font=\scriptsize]
    \draw[dotted] (0,0) circle (0.2);
    \draw[gluon] (45:0.2) -- (45:0.5) node[right] {$q^-$};
    \draw[gluon] (-45:0.2) -- (-45:0.5) node[right] {$r^-$};
    \draw[gluon] (-135:0.2) -- (-135:0.5) node[left] {$s^+$};
    \draw[gluon] (135:0.2) -- (135:0.5) node[left] {$t^+$};
  \end{tikzpicture}
  &\rightarrow [qr]\braket{st}\kT_{(qr)(qr)}\,,\\
  \label{eq:cutRule5}
  %\tikzset{external/force remake}
  \begin{tikzpicture}
    [line width=1pt,
    baseline={([yshift=-0.5ex]current bounding box.center)},
    font=\scriptsize]
    \draw[dotted] (0,0) circle (0.2);
    \draw[gluon] (45:0.2) -- (45:0.5) node[right] {$q^-$};
    \draw[gluon] (-45:0.2) -- (-45:0.5) node[right] {$r^+$};
    \draw[quark] (-135:0.2) -- (-135:0.5) node[left] {$s$};
    \draw[aquark] (135:0.2) -- (135:0.5) node[left] {$t$};
  \end{tikzpicture}
  &\rightarrow  [q|s|r\rangle\kT_{(qs)(qt)}\,,\\ %= -[a|p_c+p_b|b\rangle\kT_{(ac)(ad)}\,,\\
  \label{eq:cutRule6}
  %\tikzset{external/force remake}
  \begin{tikzpicture}
    [line width=1pt,
    baseline={([yshift=-0.5ex]current bounding box.center)},
    font=\scriptsize]
    \draw[dotted] (0,0) circle (0.2);
    \draw[quark] (45:0.2) -- (45:0.5) node[right] {$q$};
    \draw[aquark] (-45:0.2) -- (-45:0.5) node[right] {$r$};
    \draw[quark] (-135:0.2) -- (-135:0.5) node[left] {$s$};
    \draw[aquark] (135:0.2) -- (135:0.5) node[left] {$t$};
  \end{tikzpicture}~\,
  &\rightarrow s_{rt}\kT_{(qs)(rt)} = s_{qs}\kT_{(qs)(rt)}\,,
\end{align}
\end{subequations}
where we have introduced
\be
    \kT_{(qr)(st)} \equiv \frac{\kappa_{(qr)(st)}}{s_{qr}s_{st}}\,.
\ee
This completes the set of rules required to assemble any cut obtained by gluing four-point amplitudes.

\subsubsection{Derivation of \texorpdfstring{$\cN=2$}{N=2} diagrammatic rules}

A convenient way to derive the recursion formula given in \eqn{eq:supersumN2} is by treating chiral and anti-chiral superspace coordinates democratically. We begin with
\begin{subequations}
\begin{align}
	\label{eq:supersumN2A}
   \int\!\d^4\eta_{l_1} \d^4\eta_{l_2} \kappa_{(ab)(cd)}^{\text{(L)}}
                            \kappa_{(ef)(gh)}^{\text{(R)}}
    &= [l_1 l_2]^2 \braket{ab}\braket{cd}\braket{ef}\braket{gh}[qr][st]
      \frac{\kappa_{(qr)(st)}}{s_{qr}s_{st}}\,,\\
\label{eqn:supersumN2B}
   \int\!\d^4\bar{\eta}_{l_1} \d^4\bar{\eta}_{l_2} \bar{\kappa}_{(\o{ab})(\o{cd})}^{\text{(L)}}\bar{\kappa}_{(\o{ef})(\o{gh})}^{\text{(R)}}
    &= \braket{l_1 l_2}^2 [\o{ab}][\o{cd}][\o{ef}][\o{gh}]\braket{\o{qr}}\braket{\o{st}}
      \frac{\bar{\kappa}_{(\o{qr})(\o{st})}}{s_{\o{qr}}s_{\o{st}}}\,,
\end{align}
\end{subequations}
which are CPT conjugates of each other; the former is proved in \app{app:superspace}. The Fourier transform~\eqref{eqn:FT} that brings the second expression back into the chiral superspace amounts to replacing
\begin{equation}
  \bar{\kappa}_{(\o{qr})(\o{st})}
  \rightarrow\kappa_{(qr)(st)}\,,
\end{equation}
which reflects the equivalence of four-point MHV and $\MHVb$ amplitudes. As $s_{qr}=s_{\o{qr}}$, $s_{st}=s_{\o{st}}$ by momentum conservation, this implies that the strings of spinor-helicity brackets are equal across the two formulas.

We can therefore treat chiral and anti-chiral spinor variables democratically. This is naturally accomplished using a square root:
\begin{equation}\label{eq:supersumSR}
  \begin{aligned}
   \MoveEqLeft[1]\int\!\d^4\eta_{l_1} \d^4\eta_{l_2} \kappa_{(ab)(cd)}^{\text{(L)}}\kappa_{(ef)(gh)}^{\text{(R)}}\\
   &= s_{l_1l_2}\left(\braket{ab}[\o{ab}]
   \braket{cd}[\o{cd}]\braket{ef}[\o{f}]
   \braket{gh}[\o{gh}][qr]\braket{\o{qr}}
   [st]\braket{\o{st}}\right)^{\frac12}
    \frac{\kappa_{(qr)(st)}}{s_{qr}s_{st}}\,,
  \end{aligned}
\end{equation}
where the sign is left ambiguous.
The last step is to show that, for the left-hand tree amplitude, $\braket{ab}[\o{cd}]=\pm[\o{ab}]\braket{cd}$;
this follows by examination for the three possible external helicity configurations listed in the diagrammatic rules~\eqref{eq:N2blobRules}.
A similar identity holds for the right-hand side, $\braket{ef}[\o{gh}]=\pm[\o{ef}]\braket{gh}$, and for the external configuration, $\braket{\o{qr}}[st]=\pm[qr]\braket{\o{st}}$.
Using these identities we can eliminate the square root, yielding the iterative formula~\eqref{eq:supersumN2}.

%%%%%%%%%%%%%%%%%%%%%%%%%%%%%%%%%%%%%%%%%%%%%%%%%%
\subsubsection{Locality}
\label{sec:locality}

An important property of the recursive formula~\eqref{eq:supersumN2} is that unphysical poles cancel between successive iterations.
Suppose we glued a third tree amplitude, with external states encoded by $\kappa_{(ij)(kl)}^{\text{(X)}}$.
The additional rung carries two new loop momenta $l_3$ and~$l_4$,
where the orientation is irrelevant.
Using $\kappa_{(qr)(st)}$ to denote the new external state configuration, we can write
\begin{equation}
  \begin{aligned}
   \MoveEqLeft[1]\int\!\d^4\eta_{l_1}\d^4\eta_{l_2}\d^4\eta_{l_3}\d^4\eta_{l_4}
   \kappa_{(ab)(cd)}^{\text{(L)}}\kappa_{(ef)(gh)}^{\text{(R)}}\kappa_{(ij)(kl)}^{\text{(X)}}\\
   &=s_{l_1l_2}\frac{\braket{ab}[\o{cd}]\braket{ef}[\o{gh}]
   [wx]\braket{\o{yz}}}{s_{wx}s_{yz}}\int\!\d^4\eta_{l_3}\d^4\eta_{l_4}
   \kappa^{\text{(L}+\text{R)}}_{(wx)(yz)}\kappa_{(ij)(kl)}^{\text{(X)}}\\
   &=s_{l_1l_2}\frac{\braket{ab}[\o{cd}]\braket{ef}[\o{gh}]
   [wx]\braket{\o{yz}}}{s_{wx}s_{yz}}
   s_{l_3l_4}\braket{wx}[\o{yz}]\braket{ij}[\o{kl}][qr]\braket{\o{st}}\frac{\kappa_{(qr)(st)}}{s_{qr}s_{st}}\\
   &=s_{l_1l_2}s_{l_3l_4}\braket{ab}[\o{cd}]\braket{ef}
   [\o{gh}]\braket{ij}[\o{kl}][qr]
   \braket{\o{st}}\frac{\kappa_{(qr)(st)}}{s_{qr}s_{st}}\,,
  \end{aligned}
\end{equation}
having used $s_{wx}=s_{\o{yz}}$. We are left only with the poles of the external state configuration, $s_{qr}$ and $s_{st}$.
The above expression is consistent with the diagrammatic rules given in \sec{sec:N2rules}.

The upshot is that all poles are handled transparently.
Physical poles coming from the trees are directly accommodated for by the first rule~\eqref{eqn:physPolesRule};
the only other poles are those associated with the external rule~\eqref{eq:N2extRules}, and which we absorb into $\kT_{(qr)(st)} \equiv \kappa_{(qr)(st)}/(s_{qr}s_{st})$.
They are relics of the spinor-helicity notation, arising from the representation of four-dimensional gluon polarization vectors in terms of spinor-helicity brackets (see \eg \rcites{Dixon:1996wi,Elvang:2015rqa}):
\be
  \varepsilon_{+}^\mu(p;q) = \frac{[p|\sigma^\mu|q\rangle}{\sqrt{2}\braket{qp}}\,,
  \qquad \quad
  \varepsilon_{-}^\mu(p;q) = \frac{[q|\sigma^\mu|p\rangle}{\sqrt{2}[pq]}\,.
\ee
As we will see in the generalization to three-particle cuts (and higher), at two loops these are the only non-physical poles.
Consequently, they are the only poles allowed in a local representation of the integrand.

%%%%%%%%%%%%%%%%%%%%%%%%%%%%%%%%%%%%%%%%%%%%%%%%%%
\subsubsection{Off-shell continuation}
\label{sec:offShell}

Our ability to control the physical pole structure of cuts
allows us to guess expressions for individual numerators:
we lift Lorentz-invariant terms off shell,
then attach them to numerators of graphs with the corresponding pole structure.
Strings of spinor-helicity brackets arrange themselves into Lorentz inner products and Dirac traces:
\begin{equation}
\begin{aligned} \!\!\!
   [i_1i_2]\braket{i_2i_3}\cdots[i_{k-1}i_k]\braket{i_ki_1} &
    = p_{i_1}^{\mu_1} p_{i_2}^{\mu_2} \cdots p_{i_k}^{\mu_k}
      \ttr(\bar{\sigma}_{\mu_1}\sigma_{\mu_2}\cdots\sigma_{\mu_k})%\\&
    = \trp(i_1i_2\cdots i_k)\,,\! \\ \!\!\!
   \braket{i_1i_2}[i_2i_3]\cdots\braket{i_{k-1}i_k}[i_ki_1] &
    = p_{i_1}^{\mu_1} p_{i_2}^{\mu_2} \cdots p_{i_k}^{\mu_k}
      \ttr(\sigma_{\mu_1}\bar{\sigma}_{\mu_2}\cdots\bar\sigma_{\mu_k})%\\&
    = \trm(i_1i_2\cdots i_k)\,,\!
\end{aligned}
\end{equation}
where $\trpm(i_1i_2\cdots i_k)=\frac12\ttr((1\pm\gamma_5)i_1i_2\cdots i_k)$,
and $\trpm(ij)=s_{ij}$.
In contrast to the rung rule for $\cN=4$ SYM, the cut structure often leads to triangular subgraphs, enabled by the cut structure allowing both $s$- and $t$-channel poles to cancel.
The off-shell continuation is not unique and will in some cases lead us to several different representations of the same integrand.

Now let us point out some general features of the $\cN=2$ diagrammatic rules
that will be reflected in the explicit off-shell numerators in the next sections.
\begin{itemize}
\item In case of two adjacent fundamental hypermultiplets
      (with aligned matter arrows) on one side of a unitarity cut,
      the four-hyper rule~\eqref{eq:cutRule3} implies a pole cancellation
      that makes it equivalent to
      the $\cN=4$ rung rule~\eqref{eq:rungruleN4offshell}:
\be
  %\tikzset{external/force remake}
  \begin{tikzpicture}
    [line width=1pt,
    baseline={([yshift=-0.5ex]current bounding box.center)},
    font=\scriptsize]
    \fill[blob] (-0.5,0) ellipse (0.2 and 0.3);
    \fill[blob] (0.5,0) ellipse (0.3 and 0.4);
    \filldraw[fill=white,draw=black] (0.55,0.15) circle (0.1);
    \filldraw[fill=white,draw=black] (0.45,-0.1) circle (0.1);
    \draw[quark] ($(-0.5,0) + (135:0.23)$) -- ($(-0.5,0) + (135:0.5)$) node[left] {$2$};
    \draw[quark] ($(-0.5,0) + (-135:0.23)$) -- ($(-0.5,0) + (-135:0.5)$) node[left] {$1$};
    \draw[aquark] ($(-0.5,0) + (30:0.23)$) -- node[above] {$\uset{\leftarrow}{l_1}$} ($(0.5,0) + (157:0.3)$);
    \draw[aquark] ($(-0.5,0) + (-30:0.23)$) -- node[below] {$\oset{\leftarrow}{l_2}$} ($(0.5,0) + (-157:0.3)$);
    \draw ($(0.5,0) + (45:0.35)$) -- ($(0.5,0) + (45:0.7)$) node[right] {$3$};
    \draw ($(0.5,0) + (-45:0.35)$) -- ($(0.5,0) + (-45:0.7)$) node[right] {$4$};
  \end{tikzpicture}
  = \frac{-is_{12}}{s_{12}s_{1(-l_2)}}\times s_{l_1l_2}\times
  %\tikzset{external/force remake}
  \begin{tikzpicture}
    [line width=1pt,
    baseline={([yshift=-0.5ex]current bounding box.center)},
    font=\scriptsize]
    \fill[blob] (0.5,0) ellipse (0.3 and 0.4);
    \filldraw[fill=white,draw=black] (0.55,0.15) circle (0.1);
    \filldraw[fill=white,draw=black] (0.45,-0.1) circle (0.1);
    \draw[quark] ($(0.5,0) + (157:0.3)$) -- ($(-0.3,0) + (30:0.23)$) node[left] {$l_1$};
    \draw[quark] ($(0.5,0) + (-157:0.3)$) -- ($(-0.3,0) + (-30:0.23)$)  node[left] {$l_2$};
    \draw ($(0.5,0) + (45:0.35)$) -- ($(0.5,0) + (45:0.7)$) node[right] {$3$};
    \draw ($(0.5,0) + (-45:0.35)$) -- ($(0.5,0) + (-45:0.7)$) node[right] {$4$};
  \end{tikzpicture}
  = -\frac{is_{12}}{s_{1(-l_2)}}\times
  %\tikzset{external/force remake}
  \begin{tikzpicture}
    [line width=1pt,
    baseline={([yshift=-0.5ex]current bounding box.center)},
    font=\scriptsize]
    \fill[blob] (0.5,0) ellipse (0.3 and 0.4);
    \filldraw[fill=white,draw=black] (0.55,0.15) circle (0.1);
    \filldraw[fill=white,draw=black] (0.45,-0.1) circle (0.1);
    \draw[quark] ($(0.5,0) + (157:0.3)$) -- ($(-0.3,0) + (30:0.23)$) node[left] {$l_1$};
    \draw[quark] ($(0.5,0) + (-157:0.3)$) -- ($(-0.3,0) + (-30:0.23)$)  node[left] {$l_2$};
    \draw ($(0.5,0) + (45:0.35)$) -- ($(0.5,0) + (45:0.7)$) node[right] {$3$};
    \draw ($(0.5,0) + (-45:0.35)$) -- ($(0.5,0) + (-45:0.7)$) node[right] {$4$};
  \end{tikzpicture}\,,
\ee
      where we have ignored the external states to expose the similarity
      to \eqn{eq:rungruleN4onshell}.
      The resulting absence of the triangle-like term
      happens because the above four-hyper tree amplitude
      contains a single $t$-channel diagram.

\item In case of adjacent fundamental and anti-fundamental hypermultiplets
      on one side of a cut,
      the four-hyper rule~\eqref{eq:cutRule3} gives two contributors:
\be
  %\tikzset{external/force remake}
  \begin{tikzpicture}
    [line width=1pt,
    baseline={([yshift=-0.5ex]current bounding box.center)},
    font=\scriptsize]
    \fill[blob] (-0.5,0) ellipse (0.2 and 0.3);
    \fill[blob] (0.5,0) ellipse (0.3 and 0.4);
    \filldraw[fill=white,draw=black] (0.55,0.15) circle (0.1);
    \filldraw[fill=white,draw=black] (0.45,-0.1) circle (0.1);
    \draw[aquark] ($(-0.5,0) + (135:0.23)$) -- ($(-0.5,0) + (135:0.5)$) node[left] {$2$};
    \draw[quark] ($(-0.5,0) + (-135:0.23)$) -- ($(-0.5,0) + (-135:0.5)$) node[left] {$1$};
    \draw[quark] ($(-0.5,0) + (30:0.23)$) -- node[above] {$\uset{\leftarrow}{l_1}$} ($(0.5,0) + (157:0.3)$);
    \draw[aquark] ($(-0.5,0) + (-30:0.23)$) -- node[below] {$\oset{\leftarrow}{l_2}$} ($(0.5,0) + (-157:0.3)$);
    \draw ($(0.5,0) + (45:0.35)$) -- ($(0.5,0) + (45:0.7)$) node[right] {$3$};
    \draw ($(0.5,0) + (-45:0.35)$) -- ($(0.5,0) + (-45:0.7)$) node[right] {$4$};
  \end{tikzpicture}
  = i\left(\frac1{s_{12}}+\frac1{s_{1(-l_2)}}\right)s_{12}\times
  %\tikzset{external/force remake}
  \begin{tikzpicture}
    [line width=1pt,
    baseline={([yshift=-0.5ex]current bounding box.center)},
    font=\scriptsize]
    \fill[blob] (0.5,0) ellipse (0.3 and 0.4);
    \filldraw[fill=white,draw=black] (0.55,0.15) circle (0.1);
    \filldraw[fill=white,draw=black] (0.45,-0.1) circle (0.1);
    \draw[aquark] ($(0.5,0) + (157:0.3)$) -- ($(-0.3,0) + (30:0.23)$) node[left] {$l_1$};
    \draw[quark] ($(0.5,0) + (-157:0.3)$) -- ($(-0.3,0) + (-30:0.23)$)  node[left] {$l_2$};
    \draw ($(0.5,0) + (45:0.35)$) -- ($(0.5,0) + (45:0.7)$) node[right] {$3$};
    \draw ($(0.5,0) + (-45:0.35)$) -- ($(0.5,0) + (-45:0.7)$) node[right] {$4$};
  \end{tikzpicture}\,,
\ee
      which correspond to a box-like and a triangle-like diagrams,
      with different matter line routings.
      The fact that they both have equal numerators
      justifies our use of the two-term identity
      for off-shell numerators, as introduced in \sec{sec:2term}.

\item Finally, our approach implies explicit IR structure
      for the kinematic numerators.
      The numerator rule~\eqref{eq:cutRule2} is especially important
      in this regard, as it can be re-expressed as
\begin{equation}
%\tikzset{external/force remake}
  \begin{tikzpicture}
    [line width=1pt,
    baseline={([yshift=-0.5ex]current bounding box.center)},
    font=\scriptsize]
    \fill[blob] (0,0) circle (0.2);
    \draw[gluon] (45:0.2) -- (45:0.5) node[right] {$a^-$};
    \draw[gluon] (-45:0.2) -- (-45:0.5) node[right] {$b^+$};
    \draw[quark] (-135:0.2) -- (-135:0.5) node[left] {};
    \draw[aquark] (135:0.2) -- (135:0.5) node[left] {};
    \draw[line width=0.45pt,-to] (160:0.35) -- node[left] {$\ell$} (200:0.35);
  \end{tikzpicture}
  \rightarrow \langle a|\ell|b]\,,
\end{equation}
      The resulting numerators will evidently vanish
      whenever $\ell\to0$, $\ell\to -p_a$, or $\ell\to p_b$,
      which correspond to the soft regions of loop integration.
      The numerators will also vanish in the collinear regions
      where $\ell$ becomes collinear to $p_a$ or $p_b$.
      Such behavior makes a lot of sense,
      as only vectors should give rise to soft and collinear divergences
      in the loop integrals,
      not hypermultiplets (which in our case are also massless) ---
      see \eg \rcites{Catani:1996jh,Catani:1996vz,Catani:1998bh}.
      As we shall see (here and in future work currently in progress),
      this vanishing of the numerators in specific regions
      serves to block potentially singular regions arising for the hypers.
\end{itemize}

%%%%%%%%%%%%%%%%%%%%%%%%%%%%%%%%%%%%%%%%%%%%%%%%%%
%%%%%%%%%%%%%%%%%%%%%%%%%%%%%%%%%%%%%%%%%%%%%%%%%%
\section{One-loop examples}
\label{sec:oneloop}

To illustrate the iterative method of calculating cuts in $\cN=2$ SYM, we begin by considering all three four-point one-loop amplitudes in the MHV sector: with zero, one, and two external hypermultiplet pairs.
In each case, the formation of the four-dimensional cuts tells us what structure we should expect in the off-shell numerators, and we find it unnecessary to use ans\"atze.
A complete listing of the non-zero numerators for all three solutions is provided in \app{sec:allRes}.

Given our desire to regulate integrals in $D=4-2\eps$ dimensions, it is also necessary for us to obtain unitarity cuts from higher-dimensional trees.
As explained in \rcite{Johansson:2017bfl}, a convenient method is to calculate cuts arising from the tree amplitudes of six-dimensional $\cN=(1,0)$ SYM --- the dimensional uplift of four-dimensional $\cN=2$ --- using the six-dimensional spinor-helicity formalism~\cite{deWit:2002vz,Cheung:2009dc,Boels:2009bv,Dennen:2009vk,Bern:2010qa,Huang:2011um,Elvang:2011fx}.
One restricts the six-dimensional external momenta to a four-dimensional subspace, and re-interprets the extra two loop momentum components as complex masses: $\mu^2=\bar{\ell}^2-\ell^2$, where $\bar{\ell}$ is the four-dimensional part of $\ell$.

From these six-dimensional cuts, terms proportional to $\mu^2$ are found by subtracting the previously obtained four-dimensional cuts.
These terms are sufficiently simple that they can be lifted off shell without interference to the color-kinematic structure of the four-dimensional numerators.
We use a $D$-dimensional Clifford algebra to write Dirac traces involving $\ell$~\cite{Collins:1984xc,tHooft:1972tcz}; $\trpm$ are defined in terms of $\gamma_5=i\gamma^0\gamma^1\gamma^2\gamma^3$, which anticommutes with elements of the four-dimensional subalgebra but commutes with the rest.\footnote{A recent
  review of dimensional-regularization schemes was given in \rcite{Gnendiger:2017pys}.
}
Computing six-dimensional cuts also provides a check on all of the four-dimensional cuts computed in this paper.

%%%%%%%%%%%%%%%%%%%%%%%%%%%%%%%%%%%%%%%%%%%%%%%%%%
\subsection{External vectors}

The one-loop amplitudes with four external vector multiplets,
previous versions of which have been obtained
in \rcites{Carrasco:2012ca,Bern:2013yya,Nohle:2013bfa,
Ochirov:2013xba,Chiodaroli:2013upa,Johansson:2014zca},
are particularly simple to determine.
%The one-loop amplitude with four external vector multiplets, previously obtained by Johansson and one of the present authors~\cite{Johansson:2014zca}, is particularly simple to determine.
An $\cN=4$ matching identity (see \sec{sec:Neq4ident}) relates the pure-adjoint box numerator to the fundamental:
\begin{align}
%\tikzset{external/force remake}
n^{[\cN=4]}\!\left(\gBox[eLA=$1$,eLB=$2$,eLC=$3$,eLD=$4$]{}\right)&=
%\tikzset{external/force remake}
n\!\left(\gBox[eLA=$1$,eLB=$2$,eLC=$3$,eLD=$4$]{}\right)+
n\!\left(\gBox[eLA=$1$,eLB=$2$,eLC=$3$,eLD=$4$,iA=aquark,iB=aquark,iD=aquark,iC=aquark]{}\right)+
n\!\left(\gBox[eLA=$1$,eLB=$2$,eLC=$3$,eLD=$4$,iA=quark,iB=quark,iD=quark,iC=quark]{}\right)\,,
\end{align}
where the $\cN=4$ box numerator is given by $\kappa$ and can be rewritten as
\begin{align}
%\tikzset{external/force remake}
n^{[\cN=4]}\!\left(\gBox[eLA=$1$,eLB=$2$,eLC=$3$,eLD=$4$]{}\right)=
s^2(\kT_{12}+\kT_{34})+t^2(\kT_{23}+\kT_{14})+u^2(\kT_{13}+\kT_{24})\,.
\end{align}
All possible combinations of external $\cN=2$ vector multiplets are projected out of $\kappa$ as coefficients of $\kT_{ij}=\kappa_{ij}/s_{ij}^2$.
The fundamental box is the only master --- from it we can uniquely fix all other numerators.

We isolate this box numerator from the following family of four-dimensional cuts, determined using the diagrammatic rules given in \sec{sec:N2rules}.
Negative helicities are placed on different external legs to yield coefficients of different $\kT_{ij}$:
\begin{subequations}\label{eq:1lboxbarcuts}
%\tikzset{external/force remake}
\begin{align}
\cBoxA[eLA=$1^-$,eLB=$2^-$,eLC=$3^+$,eLD=$4^+$,iLC=$\downarrow l_1$,iLD=$l_2\uparrow$,
iA=aquark,iB=aquark,iC=aquark,iD=aquark]{}&=0\,,\\
%%%
\cBoxA[eLA=$1^-$,eLB=$2^+$,eLC=$3^-$,eLD=$4^+$,iLC=$\downarrow l_1$,iLD=$l_2\uparrow$,
iA=aquark,iB=aquark,iC=aquark,iD=aquark]{}&=
-\frac{\langle3|l_2|4]}{s\,l_2^2}\times s\times\frac{\langle1|l_1|2]}{s\,l_1^2}\times[13]\braket{24}\kT_{13}=
\frac{\trm(1l_124l_23)}{s\,l_1^2l_2^2}\kT_{13}\,,\\
%%%
\cBoxA[eLA=$1^-$,eLB=$2^+$,eLC=$3^+$,eLD=$4^-$,iLC=$\downarrow l_1$,iLD=$l_2\uparrow$,
iA=aquark,iB=aquark,iC=aquark,iD=aquark]{}&=
-\frac{[3|l_2|4\rangle}{s\,l_2^2}\times s\times\frac{\langle1|l_1|2]}{s\,l_1^2}\times[14]\braket{23}\kT_{14}=
\frac{\trm(1l_123l_24)}{s\,l_1^2l_2^2}\kT_{14}\,.
\end{align}
\end{subequations}
The box is separated from the triangles and bubbles which also contribute by further cutting into the $l_1$ and $l_2$ propagators; in that case $\trpm(1l_124l_23)=-s\trpm(1l_1l_23)$ and $\trpm(1l_123l_24)=0$.
% \footnote{
%   The first identity can be derived by simple use of the Dirac algebra on support of the on-shell conditions $l_1^2=l_2^2=(l_1-p_2)^2=(l_2-p_4)^2=0$ and momentum conservation $l_1-l_2=p_{23}$:
%   \begin{align*}
%     \trm(1l_124l_23)&=-\trm(1l_1(l_1\!-\!p_2)4l_23)=
%     \trm(1(l_2\!-\!p_4)(l_1\!-\!p_2)(l_2\!-\!p_4)l_23)\\
%     &=2(l_2\!-\!p_4)\!\cdot\!(l_1\!-\!p_2)\trm(1(l_2\!-\!p_4)l_23)=
%     -s\trm(1l_1l_23)\,.
%   \end{align*}
% }
The $\kT_{23}$, $\kT_{24}$, and $\kT_{34}$ coefficients are related by CPT conjugation;
the descendants are determined using commutation relations:
\begin{subequations}
%\tikzset{external/force remake}
\begin{align}
   n\!\left(\gBox[yshift=-1,eLA=$1$,eLB=$2$,eLC=$3$,eLD=$4$,
                  iA=aquark,iB=aquark,iD=aquark,iC=aquark,
                  iLD=$\uset{\rightarrow}{\ell}$]{}\right) &
    = \begin{aligned}[t]
         \kT_{13}\trm(1(\ell-p_1)(\ell+p_4)3)
       + \kT_{24}\trp(1(\ell-p_1)(\ell+p_4)3) & \\
       + \mu^2 \big( s(\kT_{12}\!+\!\kT_{34}) + t(\kT_{23}\!+\!\kT_{14})
                   + u(\kT_{13}\!+\!\kT_{24}) \big) & \,,
      \end{aligned} \\
   n\!\left(\gTriC[eLA=$2$,eLB=$3$,eLC=$4$,eLD=$1$,iB=aquark,iD=aquark,iC=aquark,
                   iLC=$\ell\uparrow$]{}\right) &
    = \begin{aligned}[t] &
         (\kT_{13}+\kT_{34})\trm(1(\ell-p_1)(\ell+p_4)3) \\ &
       + (\kT_{12}+\kT_{24})\trp(1(\ell-p_1)(\ell+p_4)3)
       + (\kT_{12}+\kT_{34})t\ell^2\,,
      \end{aligned} \\
   n\bigg(\gBubB[yshift=-3,eLA=$1$,eLB=$2$,eLC=$3$,eLD=$4$,iB=aquark,iC=aquark,
                 iLC=$\uset{\rightarrow}{\ell}$]{}\bigg) &
    = 2\ell\!\cdot\!(p_{12}-\ell) [t(\kT_{23}+\kT_{14})-u(\kT_{13}+\kT_{24})]\,,
\end{align} \label{eq:1Lchiralbox}%
\end{subequations}
%where the shorthand $\trpm=\trpm(1(\ell-p_1)(\ell+p_4)3)$ is used.
By resubstituting back into the cuts given in \eqn{eq:1lboxbarcuts},
we confirm the color duality of these kinematic numerators.\footnote{There are
other descendants with bubbles on external legs or tadpoles,
listed in \app{sec:allRes}. However, as explained in \rcite{Johansson:2017bfl},
these contributions vanish upon integration as they lack a proper mass scale.
}

Box numerators involving four-term traces of this kind have previously appeared in the context of $D$-dimensional local integrands~\cite{Badger:2016ozq,Badger:2016egz}.
A box integral with numerator $\trpm[1(\ell-p_1)(\ell+p_4)3]$ was there found to be free of both UV and IR divergences --- in fact, it is proportional to the $(D+2)$-dimensional scalar box integral, which does not diverge.
The mechanism blocking these divergences was discussed at the end of section~\ref{sec:offShell}:
when $\ell$ enters any potentially soft or collinear region,
vanishing of the trace blocks the divergence.
The descendants have similar properties,
so these are our first examples of numerators with an IR structure
that manifests vanishing of soft and collinear limits of matter-line momenta.

%%%%%%%%%%%%%%%%%%%%%%%%%%%%%%%%%%%%%%%%%%%%%%%%%%
\subsection{External vectors + matter}

\begin{figure}
  %\tikzset{external/force remake}
  \centering
  \begin{subfigure}[a]{0.25\textwidth}
    \centering
    \gBox[scale=1.5,eB=quark,eC=aquark,iA=quark,iC=quark,iD=quark]{}
  \end{subfigure}
  \begin{subfigure}[b]{0.25\textwidth}
    \centering
    \gBox[scale=1.5,eD=aquark,eB=quark,iA=quark,iD=quark]{}
  \end{subfigure}
  \begin{subfigure}[c]{0.25\textwidth}
    \centering
    \gBox[scale=1.5,eA=quark,eD=aquark,iD=quark]{}
  \end{subfigure}
  \caption{\small The three one-loop masters with mixed external particle content.}
  \label{fig:1LMixedMasters}
\end{figure}

There are three masters, drawn in \fig{fig:1LMixedMasters}.
We examine the first on its $s$-channel cut:
\begin{equation}\label{eq:1LMixedCutA}
%\tikzset{external/force remake}
\cBoxA[eLA=$1^-$,eLB=$2$,eLC=$3$,eLD=$4^+$,iLB=$\uset{\rightarrow}{l_1}$,
iLA=$\oset{\leftarrow}{l_2}$,
eB=quark,eC=aquark,iB=quark]{}=
\frac{\trp(4l_1l_2l_112)}{s\,s_{4l_1}s_{2l_2}}\kT_{(12)(13)}=
-\frac{\trp(4l_112)}{s_{4l_1}s_{2l_2}}\kT_{(12)(13)}\,,
\end{equation}
where the Dirac algebra is used to cancel the $s$-channel pole.
This implies that, similar to the rung rule in $\cN=4$ SYM, there are no triangles; only the first box drawn in~\fig{fig:1LMixedMasters} contributes.
The second and third masters are isolated on near-identical cuts: the former by exchanging $p_3\leftrightarrow p_4$, and the latter by also exchanging $p_1\leftrightarrow p_2$.
As the diagrammatic rules for the numerators do not depend on the ordering of the tree amplitudes (except for relabeling) the same result is obtained.
The three masters are
\begin{subequations}\label{eq:1LMixedNums}
%\tikzset{external/force remake}
\begin{align}
n\!\left(\gBox[yshift=-1,eLA=$1$,eLB=$2$,eLC=$3$,eLD=$4$,iA=quark,iD=quark,iC=quark,
eB=quark,eC=aquark,iLD=$\uset{\rightarrow}{\ell}$]{}\right)&=
\trp(4\ell12)\kT_{(12)(13)}+\trm(4\ell12)\kT_{(24)(34)}\,,\\
n\!\left(\gBox[yshift=-1,eLA=$1$,eLB=$2$,eLC=$3$,eLD=$4$,iA=quark,iD=quark,
eB=quark,eD=aquark,iLD=$\uset{\rightarrow}{\ell}$]{}\right)&=
\trp(3\ell12)\kT_{(12)(14)}+\trm(3\ell12)\kT_{(23)(34)}\,,\\
n\!\left(\gBox[yshift=-1,eLA=$1$,eLB=$2$,eLC=$3$,eLD=$4$,iD=quark,
eA=quark,eD=aquark,iLD=$\uset{\rightarrow}{\ell}$]{}\right)&=
\trp(3\ell21)\kT_{(12)(24)}+\trm(3\ell21)\kT_{(13)(34)}\,,
\end{align}
\end{subequations}
where in all three cases the coefficient of $\trm$ is related by CPT conjugation.
It can be checked using six-dimensional cuts that continuation to $D=4-2\eps$
does not introduce terms proportional to $\mu^2$;
this can also be argued from the absence of quadratic $\ell$ terms in the
numerators.

As further confirmation of these expressions we can also examine the $t$-channel cuts. The first master contributes to
\begin{equation}
%\tikzset{external/force remake}
\cBoxB[eLA=$1^-$,eLB=$2$,eLC=$3$,eLD=$4^+$,
iLD=$\uset{\rightarrow}{l_1}$,iLC=$\oset{\leftarrow}{l_2}$,
eB=quark,eC=aquark,iB=quark,iA=quark]{}=
-\left(\frac1t+\frac1{l_2^2}\right)\frac{\trp(4l_112)}{l_1^2}\kT_{(12)(13)}\,.
\end{equation}
In this case there are two contributors;
as explained in \sec{sec:offShell}, the four-hyper amplitude naturally separates this cut into two contributions with equal numerators:
\begin{equation}\label{eq:1LMixedTwoTerm}
%\tikzset{external/force remake}
n\!\left(\gBox[eLA=$2$,eLB=$3$,eLC=$4$,eLD=$1$,iD=quark,iC=quark,iB=quark,
eA=quark,eB=aquark,iLC=$\ell\!\uparrow$]{}\right)=
n\!\left(\gTriC[eLA=$2$,eLB=$3$,eLC=$4$,eLD=$1$,iB=quark,iD=quark,iC=quark,
eA=quark,eB=aquark,iLC=$\ell\!\uparrow$]{}\right)\,,
\end{equation}
which is a two-term identity (see \sec{sec:2term}).
The $t$-channel cut for the second master is redundant as it related
by symmetry to the $s$-channel one.
For the third master, it gives
\be\label{eq:1LMixedCutB}
%\tikzset{external/force remake}
   \cBoxB[eLA=$1$,eLB=$2^-$,eLC=$3^+$,eLD=$4$,
          iLD=$\uset{\rightarrow}{l_1}$,iLC=$\oset{\leftarrow}{l_2}$,eA=quark,eD=aquark]{}\!
 = \cBoxB[eLA=$1$,eLB=$2^-$,eLC=$3^+$,eLD=$4$,eA=quark,eD=aquark,
          iLA=$\substack{+\\-}$,iLB=$\substack{-\\+}$]{}
   \cBoxB[eLA=$1$,eLB=$2^-$,eLC=$3^+$,eLD=$4$,eA=quark,eD=aquark,
          iLA=$\substack{-\\+}$,iLB=$\substack{+\\-}$]{}\!
 =-\left(\frac{\trp(12l_13)}{l_1^2l_2^2}-\frac{2\trp(12l_23)}{t\,l_2^2}\right)
   \kT_{(12)(24)}\,.
\ee
We are required to sum over the two possible helicity configurations, and the Dirac algebra is used to expose the two contributors: the box and another triangle.
They are automatically related by a commutation relation:
\begin{equation}\label{eq:1LMixedCommutation}
%\tikzset{external/force remake}
\begin{aligned}
n\!\left(\gTriC[eLA=$2$,eLB=$3$,eLC=$4$,eLD=$1$,eA=quark,eB=aquark,iLC=$\ell\!\uparrow$]{}\right)&=
n\!\left(\gBox[eLA=$2$,eLB=$3$,eLC=$4$,eLD=$1$,iA=aquark,eA=quark,eB=aquark,iLC=$\ell\!\uparrow$]{}\right)-
n\!\left(\gBox[eLA=$3$,eLB=$2$,eLC=$4$,eLD=$1$,iA=quark,eA=aquark,eB=quark,iLC=$\ell\!\uparrow$]{}\right)\\
&=-2\trp(4\ell12)\kT_{(12)(13)}-2\trm(4\ell12)\kT_{(24)(34)}\,.
\end{aligned}
\end{equation}
This completes the set of non-zero numerators, which are also listed in~\app{eqn:solComplMixed}; all of the extra off-shell identities described in \sec{sec:bcj} are satisfied by these numerators including $\cN=4$ matching identities:
\begin{subequations}
%\tikzset{external/force remake}
\begin{align}
n\!\left(\gBox[yshift=-.13cm,eLA=$1$,eLB=$2$,eLC=$3$,eLD=$4$,iA=quark,iD=quark,iC=quark,
eB=quark,eC=aquark,iLD=$\uset{\rightarrow}{\ell}$]{}\right)+
n\!\left(\gBox[yshift=-.13cm,eLA=$1$,eLB=$2$,eLC=$3$,eLD=$4$,iB=aquark,
eB=quark,eC=aquark,iLD=$\uset{\rightarrow}{\ell}$]{}\right)
&=su(\kT_{(12)(13)}+\kT_{(24)(34)})\,,\\
n\!\left(\gBox[yshift=-.13cm,eLA=$1$,eLB=$2$,eLC=$3$,eLD=$4$,iA=quark,iD=quark,
eB=quark,eD=aquark,iLD=$\uset{\rightarrow}{\ell}$]{}\right)+
n\!\left(\gBox[yshift=-.13cm,eLA=$1$,eLB=$2$,eLC=$3$,eLD=$4$,iB=aquark,iC=aquark,
eB=quark,eD=aquark,iLD=$\uset{\rightarrow}{\ell}$]{}\right)
&=st(\kT_{(12)(14)}+\kT_{(23)(34)})\,,\\
n\!\left(\gTriC[eLA=$2$,eLB=$3$,eLC=$4$,eLD=$1$,eA=quark,eB=aquark,iLC=$\ell\!\uparrow$]{}\right)+
2n\!\left(\gTriC[eLA=$2$,eLB=$3$,eLC=$4$,eLD=$1$,iB=quark,iD=quark,iC=quark,
eA=quark,eB=aquark,iLC=$\ell\!\uparrow$]{}\right)
&=0\,.
\end{align}
\end{subequations}
These numerators have good IR behavior ---
taking the loop momentum associated with an edge carrying hypermultiplets in any one of them to zero, the numerator does indeed vanish;
this does not happen for internal gluon lines.
This indicates that soft divergences can indeed develop, but only as a result of soft vectors being exchanged, not soft hypers.
Collinear divergences can also develop, but only at vertices connecting an internal gluon line.

%%%%%%%%%%%%%%%%%%%%%%%%%%%%%%%%%%%%%%%%%%%%%%%%%%
\subsection{External matter}
\label{sec:matter1loop}

% \begin{figure}
%   %\tikzset{external/force remake}
%   \centering
%   \begin{subfigure}[a]{0.25\textwidth}
%     \centering
%     \gBox[scale=1.5,eA=quark,eB=quark,eC=aquark,eD=aquark,iB=aquark,iD=quark]{}
%   \end{subfigure}
%   \begin{subfigure}[b]{0.25\textwidth}
%     \centering
%     \gBox[scale=1.5,eA=quark,eB=aquark,eC=quark,eD=aquark,iB=quark,iD=quark]{}
%   \end{subfigure}
%   \caption{\small Two one-loop masters with four external matter multiplets.}
%   \label{fig:1LHyperMasters}
% \end{figure}

A color-dual representation for four external matter multiplets has previously been obtained in~\rcite{Chiodaroli:2013upa} via an orbifold construction.\footnote{We are especially thankful to Marco Chiodaroli for sharing unpublished material containing the explicit orbifold construction for hypermultiplets.
}
A single master is sufficient;
we choose the box numerator contributing to
\begin{equation}\label{eq:1LHyperCutA}
%\tikzset{external/force remake}
\cBoxA[eLA=$1$,eLB=$2$,eLC=$3$,eLD=$4$,iLD=$l_2\!\uparrow$,iLC=$\downarrow\!l_1$,
eA=quark,eB=quark,eC=aquark,eD=aquark,iA=aquark,iB=quark]{}=
-\frac{s^2}{l_1^2l_2^2}\kT_{(12)(34)}\,.
\end{equation}
The box numerator is easily read off as the only contributor,
and the full set of numerators for this amplitude is
\begin{subequations}\label{eq:1LHyperSoln}
%\tikzset{external/force remake}
\begin{align}
n\!\left(\gBox[eLA=$1$,eLB=$2$,eLC=$3$,eLD=$4$,
eA=quark,eB=quark,eC=aquark,eD=aquark,iB=aquark,iD=quark]{}\right)&=
s^2\kT_{(12)(34)}\,,\\
n\!\left(\gBox[eLA=$1$,eLB=$2$,eLC=$3$,eLD=$4$,
eA=quark,eB=aquark,eC=quark,eD=aquark,iB=quark,iD=quark]{}\right)&=
n\!\left(\gTriC[eLA=$1$,eLB=$2$,eLC=$3$,eLD=$4$,
eA=quark,eB=aquark,eC=quark,eD=aquark,iB=quark,iD=quark]{}\right)=
-n\!\left(\gTriC[eLA=$1$,eLB=$2$,eLC=$3$,eLD=$4$,
eA=quark,eB=aquark,eC=quark,eD=aquark,iC=aquark]{}\right)\\
%%%
&=n\!\left(\gBubB[eLA=$1$,eLB=$2$,eLC=$3$,eLD=$4$,
eA=quark,eB=aquark,eC=quark,eD=aquark,iB=quark,iC=quark]{}\right)=
-\frac{1}{2}n\!\left(\gBubB[eLA=$1$,eLB=$2$,eLC=$3$,eLD=$4$,iB=agluon,iC=agluon,
eA=quark,eB=aquark,eC=quark,eD=aquark]{}\right)=
-su\kT_{(13)(24)}\,. \nn
\end{align}
\end{subequations}
In particular, the other box numerator is related by the matter-reversal symmetry identity given in \eqn{eq:egMatterReverse};
the first triangle and first bubble are equated to the box using two-term identities.
Once again, by determining the relevant cuts in $D=6$ dimensions we have confirmed that these expressions are unmodified by terms proportional to $\mu^2$.

An incongruous feature of these numerators is their soft behavior.
Until now, there has been a well-established pattern: when loop momenta carried on hypermultiplet edges go to zero the numerators vanish.
This is consistent with the fact that only vector multiplets should give rise to soft regions, not hypermultiplets, as can be seen from the IR factorization formulae \cite{Catani:1996jh,Catani:1996vz,Catani:1998bh}.
Without any loop-momentum dependence in these numerators, this clearly cannot happen. 

We therefore question whether one should add loop-momentum-dependent terms vanishing on the $s$-channel cuts in order to restore this behavior.
The statement clarifies further upon examination of a $t$-channel cut:
\begin{equation}
%\tikzset{external/force remake}
\cBoxB[eLA=$1$,eLB=$2$,eLC=$3$,eLD=$4$,
iLD=$\uset{\rightarrow}{l_1}$,
iLC=$\oset{\leftarrow}{l_2}$,
eA=quark,eB=quark,eC=aquark,eD=aquark]{}=
\cBoxB[eLA=$1$,eLB=$2$,eLC=$3$,eLD=$4$,iLA=$\overset+-$,iLB=$\overset-+$,
eA=quark,eB=quark,eC=aquark,eD=aquark]{}+
\cBoxB[eLA=$1$,eLB=$2$,eLC=$3$,eLD=$4$,iLA=$\overset-+$,iLB=$\overset+-$,
eA=quark,eB=quark,eC=aquark,eD=aquark]{}=
\frac{2l_1^2l_2^2+2tl_1\!\cdot\!l_2}{t\,l_1^2l_2^2}s\,\kT_{(12)(34)}\,.
\end{equation}
Now from this perspective, a natural suggestion for the two box numerators is
\begin{subequations}
%\tikzset{external/force remake}
\begin{align}
n\!\left(\gBox[yshift=-1,eLA=$1$,eLB=$2$,eLC=$3$,eLD=$4$,iLD=$\uset{\rightarrow}{\ell}$,
eA=quark,eB=quark,eC=aquark,eD=aquark,iB=aquark,iD=quark]{}\right)&=
2s\,\ell\cdot(p_{12}-\ell)\kT_{(12)(34)}\,,\\
%%%
n\!\left(\gBox[yshift=-1,eLA=$1$,eLB=$2$,eLC=$3$,eLD=$4$,iLD=$\uset{\rightarrow}{\ell}$,
eA=quark,eB=aquark,eC=quark,eD=aquark,iB=quark,iD=quark]{}\right)&=
2u\,\ell\cdot(\ell-p_{12})\kT_{(13)(24)}\,,
\end{align}
\end{subequations}
where the latter is again related to the former by matter-reversal symmetry~\eqref{eq:egMatterReverse}.
These numerators vanish when $\ell\to0$ or $\ell\to p_{12}$
and still reduce to the previous expressions~\eqref{eq:1LHyperSoln}
on the $s$-channel cuts due to $2\ell\cdot(p_{12}-\ell)=s-\ell^2-(\ell-p_{12})^2$.

The resulting set of numerators satisfies the cuts conditions, color-kinematics duality and the two-term identities but violates the $\cN=4$ matching conditions.
Moreover, an unfortunate consequence of the above numerator rearrangement is that the descendants become both numerous and more complicated
(for instance, there are now non-vanishing bubbles and tadpoles).
This suggests that, should one wish to expose the IR behavior in this way,
maintaining manifest color-kinematics duality is not always the best approach.
Nevertheless, when scattering four hypermultiplets at two loops we will find that such rearrangements become necessary in order to obtain valid color-dual numerators.

%%%%%%%%%%%%%%%%%%%%%%%%%%%%%%%%%%%%%%%%%%%%%%%%%%
%%%%%%%%%%%%%%%%%%%%%%%%%%%%%%%%%%%%%%%%%%%%%%%%%%
\section{Two-loop examples}
\label{sec:twoloop}

We now proceed to the main result of this paper: the complete set of two-loop four-point MHV amplitudes in $\cN=2$ SQCD.
A color-dual representation of the amplitude with four external vector multiplets was already found in \rcite{Johansson:2017bfl} ---  in fact, two solutions were found which emphasized different physical properties.
We begin by re-deriving one of these two solutions from iterative two-particle cuts; when expressed in terms of Dirac traces, the resulting solution is written far more compactly than as originally presented.
We then proceed to calculate the two-loop amplitudes with one and two external hypermultiplet pairs.

As we explained at one loop, using cuts in strictly four dimensions misses extra-dimensional terms $\mu_{ij}=\bar{\ell}_i\cdot\bar{\ell}_j-\ell_i\cdot\ell_j$ needed for dimensional regularization ($\bar{\ell}_i$ is the four-dimensional part of $\ell_i$).
By evaluating cuts in six-dimensional $\cN=(1,0)$ SYM we recover the missing terms, which are simple enough not to interfere with the color duality of the four-dimensional numerators (again, see \rcite{Johansson:2017bfl} for details).
A new feature is the antisymmetric object $\eps(\mu_1,\mu_2)$, which is an extra-dimensional echo of the six-dimensional Levi-Civita tensor:
\begin{equation}
\eps(\mu_1,\mu_2)=\frac{\eps^{(6)}(v_1,v_2,v_3,v_4,\ell_1,\ell_2)}{\eps(v_1,v_2,v_3,v_4)}\,,
\end{equation}
where $v_i$ are four-dimensional vectors. Its appearance is due to the unavoidably chiral nature of certain six-dimensional internal states; although it vanishes upon integration, we keep it here as it gives rise to non-chiral contributions after the double copy: $\eps(\mu_1,\mu_2)^2=\mu_{11}\mu_{22}-\mu_{12}^2$.

%%%%%%%%%%%%%%%%%%%%%%%%%%%%%%%%%%%%%%%%%%%%%%%%%%
\subsection{External vectors}

\begin{figure}
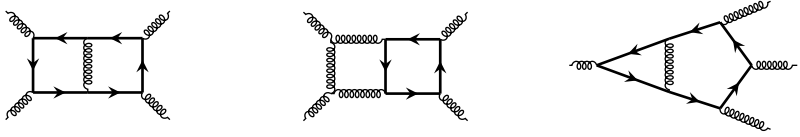

  %\tikzset{external/force remake}
  \centering
  \begin{subfigure}[a]{0.25\textwidth}
    \centering
    \gBoxBox[scale=1.2,iA=aquark,iB=aquark,iF=aquark,iE=aquark,iD=aquark,iC=aquark]{}
  \end{subfigure}
  \begin{subfigure}[b]{0.25\textwidth}
    \centering
    \gBoxBox[scale=1.2,iA=aquark,iB=aquark,iF=aquark,iG=quark]{}
  \end{subfigure}
  \begin{subfigure}[c]{0.25\textwidth}
    \centering
    \gTriPenta[scale=1.2,iA=aquark,iB=aquark,iC=aquark,iD=aquark,iE=aquark,iF=aquark]{}
  \end{subfigure}
  \caption{\small Three two-loop masters with four external vector multiplets.}
  \label{fig:twoLoopExtVectorMasters}
\end{figure}

First we summarize the main result of \rcite{Johansson:2017bfl}, updating the notation as necessary to make use of Dirac traces. The solution we are interested in was chosen to satisfy two-term and $\cN=4$ identities, as well as matter-reversal symmetry and CPT conjugation.
A suitable choice of three masters is displayed in \fig{fig:twoLoopExtVectorMasters}.
The pentagon triangle vanishes on its maximal cut; by setting it to zero, while demanding locality of all numerators, it was found that the other two masters are uniquely fixed. This left a total of 19 non-zero numerators.

The $\cN=4$ identities are particularly useful, as they ensure that all numerators with pure-adjoint content (no hyper loops) can be uniquely written in terms of those with internal hyper loops using $\cN=4$ identities. There are only two non-zero two-loop four-point $\cN=4$ SYM numerators in the MHV sector:
\begin{equation}
%\tikzset{external/force remake}
n^{[\cN=4]}\!\left(\gBoxBox[scale=.8,eLA=$1$,eLB=$2$,eLC=$3$,eLD=$4$]{}\right)=
n^{[\cN=4]}\!\left(\gBoxBoxNP[scale=.6,eLA=$1$,eLB=$2$,eLC=$4$,eLD=$3$]{}\right)=
s\sum_{i<j}\kappa_{ij}\,,
\end{equation}
where the $\cN=2$ content has been projected out. Using the two-term identities, any numerator with two hypermultiplet loops can be uniquely specified in terms of one with a single loop. So only numerators with a single hypermultiplet loop need to be specified --- there are 10 of these.

Up to relabeling of loop momenta and overall constants,
four are equal:\footnote{
In this section we use both $\kappa_{ab}$
and $\kT_{ab}\equiv\kT_{(ab)(ab)}=\kappa_{(ab)}/s_{ab}^2$
to our convenience.
}
\begin{align}\label{eq:2LVectorEasyNums}
&
%\tikzset{external/force remake}
n\!\left(\gBoxBox[scale=.8,eLA=$1$,eLB=$2$,eLC=$3$,eLD=$4$,iA=aquark,iB=aquark,iF=aquark,
iE=aquark,iD=aquark,iC=aquark,iLA=$\downarrow\!\ell_1$,iLD=$\ell_2\!\downarrow$]{}\right)=
n\!\left(\gBoxBox[scale=.8,eLA=$1$,eLB=$2$,eLC=$3$,eLD=$4$,iA=aquark,iB=aquark,iF=aquark,iG=quark,
  iLA=$\downarrow\!\ell_1$,iLG=$\!\downarrow\!\!\ell_2$]{}\right)=
n\!\left(\gBoxBoxNP[scale=.6,eLA=$1$,eLB=$2$,eLC=$4$,eLD=$3$,iA=aquark,iB=aquark,iC=aquark,iD=aquark,
  iE=aquark,iLA=$\downarrow\!\ell_1$,iLD=$\ell_2\,$]{%
  \draw[line width=0.45pt,-to] (0.6,1.2) -- (0.25,0.9);
}\right)=
-\frac{1}{2}n\!\left(\gTriTri[scale=.6,eLA=$1$,eLB=$2$,eLC=$3$,eLD=$4$,iA=aquark,iB=aquark,iC=aquark,
iLA=$\downarrow\!\ell_1$,iLF=$\ell_2\!\downarrow$]{}\right)\nn\\
&\qquad=\kT_{13}\trm(1\ell_124\ell_23)+\kT_{14}\trm(1\ell_123\ell_24)+
\kT_{23}\trp(1\ell_123\ell_24)+\kT_{24}\trp(1\ell_124\ell_23)\nn\\
&\qquad\qquad-
s\mu_{12}\left(s(\kT_{12}+\kT_{34})+t(\kT_{23}+\kT_{14})+u(\kT_{13}+\kT_{24})\right)\nn\\
&\qquad\qquad+
i\eps(\mu_1,\mu_2)s^2(\kT_{12}-\kT_{34})\,.
\end{align}
There is another non-planar double box with a similar structure, given by
\begin{equation}
  \begin{aligned}
%\tikzset{external/force remake}
    n\!\left(\gBoxBoxNP[scale=.6,eLA=$1$,eLB=$2$,eLC=$4$,eLD=$3$,iC=aquark,iD=aquark,
      iF=aquark,iG=aquark,iLE=$\uset{\rightarrow}{\ell_1}$,iLD=$\ell_2$]{%
    \draw[line width=0.45pt,-to] (0.6,1.2) -- (0.25,0.9);
  }\right)
=
\begin{aligned}[t]
&s(\kT_{12}\trp(3\ell_{12}\ell_24)+\kT_{34}\trm(3\ell_{12}\ell_24))\\
&+\kT_{13}\trm(1\ell_242\ell_{12}3)+\kT_{23}\trp(1\ell_{12}32\ell_24)\\
&+\kT_{14}\trm(1\ell_{12}32\ell_24)+\kT_{24}\trp(1\ell_242\ell_{12}3)\\
&+s(\mu_{12}+\mu_{22})[s(\kT_{12}+\kT_{34})+t(\kT_{23}+\kT_{14})+u(\kT_{13}+\kT_{24})]\\
&+i\eps(\mu_1,\mu_2)[t^2(\kT_{23}-\kT_{14})+u^2(\kT_{13}-\kT_{24})]\,,
\end{aligned}
\end{aligned}
\end{equation}
where $\ell_{12}=\ell_1+\ell_2$.
The only non-zero pentagon triangle with internal matter is
\begin{equation}
%\tikzset{external/force remake}
\begin{aligned}
n\!\!\left(\!\!\gTriPenta[scale=.6,eLA=$1$,eLB=$2$,eLC=$3$,eLD=$4$,iD=aquark,iG=aquark,iE=aquark,
  iLF=$\ell_1\,\,\,\,$,iLE=$\ell_2\,\,$]{%
  \draw[line width=0.45pt,-to] (-0.6,0.4) -- (-1.2,0.15);
  \draw[line width=0.45pt,-to] (-0.4,0.45) -- (0,0.6);
}\!\right)\!\!&=\!\!\!
\begin{aligned}[t]
&-s(\kT_{12}\trp(3\ell_{12}\ell_24)+\kT_{34}\trm(3\ell_{12}\ell_24))\\
&-t(\kT_{23}\trp(1\ell_{12}\ell_24)+\kT_{14}\trm(1\ell_{12}\ell_24))\\
&-u(\kT_{13}\trp(2\ell_{12}\ell_24)+\kT_{24}\trm(2\ell_{12}\ell_24))\\
&-i\eps(\mu_1,\mu_2)[s^2 (\kT_{12}-\kT_{34}) + t^2 (\kT_{23}-\kT_{14}) + u^2 (\kT_{13}-\kT_{24})]\,,
\end{aligned}
\end{aligned}
\end{equation}
Finally, the other four non-zero numerators are
\beal\label{eq:2LExtVectorsLastNums}
  %\tikzset{external/force remake}
  n\Bigg(\gBoxBubA[scale=.8,yshift=-5,eLA=$1$,eLB=$2$,eLC=$3$,eLD=$4$,iE=aquark,iF=aquark,
    iLF={$\oset[0.1ex]{\leftarrow}{\ell_2}$},iLE=$\uset{\rightarrow}{\ell_1}$]{}\Bigg)&=
  -2\ell_1\cdot\ell_2\sum_{i<j}\kappa_{ij}\,, \qquad \quad\,
  %\tikzset{external/force remake}
  n\!\left(\!\!\gPentaTad[eLA=$1$,eLB=$2$,eLC=$3$,eLD=$4$,iG=aquark,
    iLG=$\ell_2\!\!\downarrow\!$,iLE=$\ell_1\,\,$]{%
    \draw[line width=0.4pt,->,>=to] (160:0.45) -- (128:0.45);}\right)=
  -4\ell_1\cdot\ell_2\sum_{i<j}\kappa_{ij}\,, \!\!\!\!\!\! \\ \!\!\!\!
  %\tikzset{external/force remake}
  n\!\left(\!\!\gBoxBubB[scale=.8,eLA=$1$,eLB=$2$,eLC=$3$,eLD=$4$,iG=aquark,iF=aquark,iLG=$\uset{\leftarrow}{\ell_2}$]{}\!\!\right)&=
  2\ell_2\cdot(p_4-\ell_2)\sum_{i<j}\kappa_{ij}\,, \quad
  %\tikzset{external/force remake}
  n\!\left(\!\!\gBoxTadA[eLA=$1$,eLB=$2$,eLC=$3$,eLD=$4$,iG=aquark,iLG=$\ell_2\!\!\downarrow\!$]{}\!\!\right)=
  4\ell_2\cdot p_4\sum_{i<j}\kappa_{ij}\,.\!\!\!\!\!\!
\eeal
As explained in \rcite{Johansson:2017bfl}, the three diagrams with bubbles on external legs or tadpoles~\eqref{eq:2LExtVectorsLastNums} are of no concern as they vanish upon integration.

This solution has good soft behavior: setting the loop momentum of any internal edge carrying hypermultiplets to zero, the corresponding numerator vanishes.
In fact, double-box integrals involving the six-term traces appearing in \eqn{eq:2LVectorEasyNums} have already been calculated by Caron-Huot and Larsen~\cite{CaronHuot:2012ab};
they were suggested as forming part of a basis of IR-finite integrals.
From their results, we conclude that the integral of the first double box with hypermultiplets circulating the outside edge is both UV and IR finite to all orders in $\eps$.

To see how the solution arises, we strategically choose cuts to yield information about the non-vanishing masters. Beginning with the first double box, we consider
\begin{subequations}
%\tikzset{external/force remake}
\begin{align}
\cBoxBoxA[eLA=$1^-$,eLB=$2^-$,eLC=$3^+$,eLD=$4^+$,iA=aquark,iB=aquark,iC=aquark,iD=aquark,
iLE=$\downarrow\!l_1$,iLG=$l_2\!\downarrow$,iLF=$\!\uparrow\!\!l_3$]{}&=0\,,\\
%%%
\cBoxBoxA[eLA=$1^-$,eLB=$2^+$,eLC=$3^-$,eLD=$4^+$,iA=aquark,iB=aquark,iC=aquark,iD=aquark,
iLE=$\downarrow\!l_1$,iLG=$l_2\!\downarrow$,iLF=$\!\uparrow\!\! l_3$]{}&=
\left(\frac1s+\frac1{l_3^2}\right)\frac{\trm(1l_124l_23)}{l_1^2l_2^2}\kT_{13}\,,\\
%%%
\cBoxBoxA[eLA=$1^-$,eLB=$2^+$,eLC=$3^+$,eLD=$4^-$,iA=aquark,iB=aquark,iC=aquark,iD=aquark,
iLE=$\downarrow\!l_1$,iLG=$l_2\!\downarrow$,iLF=$\!\uparrow\!\! l_3$]{}&=
\left(\frac1s+\frac1{l_3^2}\right)\frac{\trm(1l_123l_24)}{l_1^2l_2^2}\kT_{14}\,.
\end{align}
\end{subequations}
These are almost identical to the one-loop cuts with four external vectors given in \eqn{eq:1lboxbarcuts};
the new feature is the central tree amplitude insertion which, as explained in \sec{sec:offShell},
naturally implies two numerators equated by a two-term identity:
\begin{align}\label{eq:dbExtVectors}
  %\tikzset{external/force remake}
  n\!\left(\gBoxBox[scale=.8,eLA=$1$,eLB=$2$,eLC=$3$,eLD=$4$,iA=aquark,iB=aquark,iF=aquark,iE=aquark,
  iD=aquark,iC=aquark,iLA=$\downarrow\!\ell_1$,iLD=$\ell_2\!\downarrow$]{}\right)=
  %\tikzset{external/force remake}
  n\!\left(\gTriTri[scale=.6,eLA=$1$,eLB=$2$,eLC=$3$,eLD=$4$,iA=aquark,iB=aquark,iF=aquark,iE=aquark,
iG=aquark,iC=aquark,iLA=$\downarrow\!\ell_1$,iLF=$\ell_2\!\downarrow$]{}\right)\,.
\end{align}
The double box can also be isolated from the double triangle on
\begin{subequations}
%\tikzset{external/force remake}
\begin{align}
\begin{split}
\cBoxBoxB[eLA=$1^-$,eLB=$2^-$,eLC=$3^+$,eLD=$4^+$,iA=quark,iB=aquark,iC=aquark,
iLE=$\downarrow\!l_1$,iLB=$l_2\!\downarrow$,iLD=$\!\uparrow\!\!l_3$]{}&=0\,,
\end{split}\\
\begin{split}
\cBoxBoxB[eLA=$1^-$,eLB=$2^+$,eLC=$3^-$,eLD=$4^+$,iA=quark,iB=aquark,iC=aquark,
  iLE=$\downarrow\!l_1$,iLB=$l_2\!\downarrow$,iLD=$\!\uparrow\!\!l_3$]{}&
\begin{aligned}[t]
  &=-\frac{\langle3|l_2|4]}{(l_2+p_4)^2(l_2-p_3)^2}\times s\times\frac{\langle1|l_1|2]}{s\,l_1^2}\times
[13]\braket{24}\kT_{13}\\
&=\frac{\trm(1l_124l_23)}{l_1^2(l_2+p_4)^2(l_2-p_3)^2}\kT_{13}\,,
\end{aligned}
\end{split}\\
\begin{split}
\cBoxBoxB[eLA=$1^-$,eLB=$2^+$,eLC=$3^+$,eLD=$4^-$,iA=quark,iB=aquark,iC=aquark,
  iLE=$\downarrow\!l_1$,iLB=$l_2\!\downarrow$,iLD=$\!\uparrow\!\!l_3$]{}&
\begin{aligned}[t]
&=-\frac{\langle4|l_2|3]}{(l_2+p_4)^2(l_2-p_3)^2}\times s\times\frac{\langle1|l_1|2]}{s\,l_1^2}\times
[14]\braket{23}\kT_{14}\\
&=\frac{\trm(1l_123l_24)}{l_1^2(l_2+p_4)^2(l_2-p_3)^2}\kT_{14}\,.
\end{aligned}
\end{split}
\end{align}
\end{subequations}
In the second two cases, we have re-used the one-loop cut given in \eqn{eq:1LMixedCutA} for the two amplitudes on the left-hand side ---
we simply stripped away the part of the expression given by the external rules.
So only the double box contributes, and we reproduce the same expression.
We are also reminded that the pentagon triangle can safely be set to zero.

Finally, the other double box can be determined from
\begin{subequations}
%\tikzset{external/force remake}
\begin{align}
\begin{split}
\cBoxBoxA[eLA=$1^-$,eLB=$2^-$,eLC=$3^+$,eLD=$4^+$,iA=aquark,iD=aquark,
iLE=$\downarrow\!l_1$,iLG=$l_2\!\downarrow$,iLF=$\!\uparrow\!\!l_3$]{}&=0\,,
\end{split}\\
\begin{split}
\cBoxBoxA[eLA=$1^-$,eLB=$2^+$,eLC=$3^-$,eLD=$4^+$,iA=aquark,iD=aquark,
iLE=$\downarrow\!l_1$,iLG=$l_2\!\downarrow$,iLF=$\!\uparrow\!\!l_3$]{}&=
-\frac{\trm(1l_124l_33)}{l_1^2l_2^2l_3^2}\kT_{13}-
\frac{2\trm(1l_124l_23)}{s\,l_1^2l_2^2}\kT_{13}\,,
\end{split}\\
\begin{split}
\cBoxBoxA[eLA=$1^-$,eLB=$2^+$,eLC=$3^+$,eLD=$4^-$,iA=aquark,iD=aquark,
iLE=$\downarrow\!l_1$,iLG=$l_2\!\downarrow$,iLF=$\!\uparrow\!\!l_3$]{}&=
-\frac{\trm(1l_123l_34)}{l_1^2l_2^2l_3^2}\kT_{14}-
\frac{2\trm(1l_123l_24)}{s\,l_1^2l_2^2}\kT_{14}\,,
\end{split}
\end{align}
\end{subequations}
This time we re-used the one-loop cut given in \eqn{eq:1LMixedCutB}, saving us the need to sum over helicity configurations of the vector multiplets.
The two contributing numerators, a double box and double triangle, are related by a commutation relation:
\begin{align}
%\tikzset{external/force remake}
n\!\left(\gTriTri[scale=.6,eLA=$1$,eLB=$2$,eLC=$3$,eLD=$4$,iA=aquark,iB=aquark,iC=aquark,
iLA=$\downarrow\!\ell_1$,iLF=$\ell_2\!\downarrow$]{}\right)=
n\!\left(\gBoxBox[scale=.8,eLA=$1$,eLB=$2$,eLC=$3$,eLD=$4$,iA=aquark,iB=aquark,iF=aquark,iG=quark,
iLA=$\downarrow\!\ell_1$,iLD=$\ell_2\!\downarrow$]{}\right)-
n\!\left(\gBoxBox[scale=.8,eLA=$1$,eLB=$2$,eLC=$4$,eLD=$3$,iA=aquark,iB=aquark,iF=aquark,iG=quark,
iLA=$\downarrow\!\ell_1$,iLD=$\ell_2\!\uparrow$]{}\right)\,.
\end{align}
This works by precise analogy to the one-loop relation given in \eqn{eq:1LMixedCommutation}.

%%%%%%%%%%%%%%%%%%%%%%%%%%%%%%%%%%%%%%%%%%%%%%%%%%
\subsection{External matter}

\begin{figure}
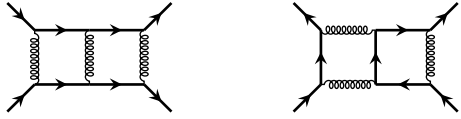

  %\tikzset{external/force remake}
  \centering
  \begin{subfigure}[a]{0.24\textwidth}
    \centering
    \gBoxBox[scale=1.2,eA=quark,eB=quark,eC=aquark,eD=aquark,iB=aquark,iC=aquark,iE=quark,iF=quark]{}
  \end{subfigure}
  %% \begin{subfigure}[b]{0.24\textwidth}
  %%   \centering
  %%   \gBoxBox[scale=1.2,eA=quark,eB=aquark,eC=quark,eD=aquark,iB=quark,iC=quark,iE=quark,iF=quark]{}
  %% \end{subfigure}
  \begin{subfigure}[c]{0.24\textwidth}
    \centering
    \gBoxBox[scale=1.2,eA=quark,eB=aquark,eC=aquark,eD=quark,iB=quark,iD=quark,iF=quark,iG=aquark]{}
  \end{subfigure}
  %% \begin{subfigure}[d]{0.24\textwidth}
  %%   \centering
  %%   \gBoxBox[scale=1.2,eA=quark,eB=aquark,eC=quark,eD=aquark,iB=quark,iD=aquark,iF=quark,iG=aquark]{}
  %% \end{subfigure}
  \caption{\small The two two-loop masters with external hypermultiplets.}
  \label{fig:twoLoopExtHyperMasters}
\end{figure}

In this case, assuming that the matter-reversal symmetry discussed in \sec{sec:matterReverse} holds for non-tadpole diagrams,
there are two masters displayed in \fig{fig:twoLoopExtHyperMasters}.
The first is most naturally isolated from
\begin{equation}\label{eqn:cutMatterLeq21}
%\tikzset{external/force remake}
%% \begin{align}\label{eqn:cutMatterLeq21}
\cBoxBoxA[eLA=$1$,eLB=$2$,eLC=$3$,eLD=$4$,eA=quark,eB=quark,eC=aquark,eD=aquark,iA=aquark,iB=aquark,
iC=quark,iD=quark,iLE=$\downarrow\!l_1$,iLG=$ l_2\!\downarrow$,iLF=$\!\uparrow\!\!l_3$]{}=
\frac{s^3}{l_1^2l_2^2l_3^2}\kT_{(12)(34)}\,.
%%%
%% \label{eqn:cutMatterLeq22}%
%% \cBoxBoxA[eLA=$1$,eLB=$2$,eLC=$3$,eLD=$4$,eA=quark,eB=aquark,eC=quark,eD=aquark,iA=quark,iB=quark,
%% iC=quark,iD=quark,iLE=$\downarrow\!l_1$,iLG=$ l_2\!\downarrow$,iLF=$\!\uparrow\!\! l_3$]{}&=
%% \frac{u(s+l_1^2)(s+l_2^2)(s+l_3^2)}{s\,l_1^2l_2^2l_3^2}\kT_{(13)(24)}\,.
%% \end{align}
\end{equation}
We encountered a similar pattern when dealing with four external hypermultiplets at one loop~\eqref{eq:1LHyperCutA}, and there it was simple to read off a color-dual box numerator.
As explained in \sec{sec:offShell}, the new tree amplitude insertion implies that the double box should be given by a rung rule from the first box numerator in \eqn{eq:1LHyperSoln},
so in this case simply $s^3\kT_{(12)(34)}$.

Unfortunately, we have confirmed by direct calculation that such a choice is incompatible with a color-dual representation of the complete amplitude assuming two-term identities.
From the one-loop discussion in \sec{sec:matter1loop} we learned that there is a second solution if we change the off-shell continuation to carry loop momenta.
Converting one factor of $s$ into $2\,(l_1+p_1)\cdot(p_2-l_1)$
or $2\,(l_2+p_4)\cdot(p_3-l_2)$ inside~\eqn{eqn:cutMatterLeq21},
we make the following simple ansatz for the numerator:
\begin{equation}\label{eqn:ansatzMatter}
  \begin{aligned}
    %\tikzset{external/force remake}
    n\!\left(\!\!\gBoxBox[scale=0.8,eLA=$1$,eLB=$2$,eLC=$3$,eLD=$4$,eA=quark,eB=quark,eC=aquark,eD=aquark,iB=aquark,iC=aquark,iE=quark,iF=quark,iLA=$\downarrow\!\ell_1$,iLD=$\ell_2\!\downarrow$]{}\!\!\right)
    &= s^2\left[c_1 s + c_2 (\ell_1+p_1)\cdot(\ell_1-p_2) + c_3 (\ell_2+p_4)\cdot(\ell_2-p_3)\right]\kT_{(12)(34)}\,,%\\
  \end{aligned}
\end{equation}
where $c_i$ are numerical coefficients to be determined. The same terms are also suggested by cuts of the form
\begin{align}
%\tikzset{external/force remake}
\cBoxBoxB[eLA=$1$,eLB=$2$,eLC=$3$,eLD=$4$,eA=quark,eB=quark,eC=aquark,eD=aquark,iA=aquark,iC=quark]{}\,, & &
%\tikzset{external/force remake}
\cBoxBoxB[rotate=180,eLA=$3$,eLB=$4$,eLC=$1$,eLD=$2$,eA=aquark,eB=aquark,eC=quark,eD=quark,iA=quark,
  iC=aquark]{}\,,
\end{align}
which can again be obtained by recycling the one-loop results.

The situation is even more difficult for the second master
in \fig{fig:twoLoopExtHyperMasters}.
The three cuts above naturally propose the same off-shell continuation,
which does not conform to a color-dual representation
(assuming two-term identities), as we explicitly checked.
We therefore use an ansatz construction for this master
along the lines of~\rcite{Johansson:2017bfl}.
It is, however, simplified by the diagrammatic rule~\eqref{eq:cutRule6},
which suggests an overall factor of $s_{ab}$,
where $a$ and $b$ denote the two external hyper legs.
Combining with the ansatz~\eqref{eqn:ansatzMatter}
and applying the constraints described in \sec{sec:bcj},
we arrive at a solution with a single free parameter.
In our final numerators
\begin{subequations}
\begin{align} \!\!
    %\tikzset{external/force remake}
    n\!\left(\gBoxBox[scale=0.8,eLA=$1$,eLB=$2$,eLC=$3$,eLD=$4$,eA=quark,eB=quark,eC=aquark,eD=aquark,iB=aquark,iC=aquark,iE=quark,iF=quark,iLA=$\downarrow\!\ell_1$,iLD=$\ell_2\!\downarrow$]{}\right)
    &= -s^2\left[(\ell_1+p_1)\cdot(\ell_1-p_2) + (\ell_2+p_4)\cdot(\ell_2-p_3)\right]\kT_{(12)(34)}\,,\\ \!\!
    %% %\tikzset{external/force remake}
    %% n\!\left(\gBoxBox[scale=0.8,eLA=$1$,eLB=$2$,eLC=$3$,eLD=$4$,eA=quark,eB=aquark,eC=quark,eD=aquark,iB=quark,iC=quark,iE=quark,iF=quark,iLA=$\downarrow\!\ell_1$,iLD=$\ell_2\!\downarrow$]{}\right)
    %% &= -us\left((\ell_1+p_1)\cdot(\ell_1-p_2) + (\ell_2+p_4)\cdot(\ell_2-p_3)\right)\kT_{(13)(24)}\,,\\
    %\tikzset{external/force remake}
    n\!\left(\gBoxBox[scale=0.8,eLA=$1$,eLB=$2$,eLC=$3$,eLD=$4$,eA=quark,eB=aquark,eC=aquark,eD=quark,iB=quark,iD=quark,iF=quark,iG=aquark,iLA=$\downarrow\!\ell_1$,iLD=$\ell_2\!\downarrow$,iLG=$\!\uparrow\!\!\ell_3$]{}\right)
    &= \frac{1}{2}st\left[(\ell_1-p_1-p_3)^2-2\,(\ell_2\cdot\ell_3)\right]\kT_{(14)(23)}\,,%,\\
    %% %\tikzset{external/force remake}
    %% n\!\left(\gBoxBox[scale=0.8,eLA=$1$,eLB=$2$,eLC=$3$,eLD=$4$,eA=quark,eB=aquark,eC=quark,eD=aquark,iB=quark,iD=aquark,iF=quark,iG=aquark,iLA=$\downarrow\!\ell_1$,iLD=$\ell_2\!\downarrow$,iLG=$\!\uparrow\!\!\ell_3$]{}\right)
    %% &= -\frac{1}{2}su\left((\ell_1-p_1-p_3)^2-2\,\ell_2\cdot\ell_3\right)\kT_{(13)(24)}\,.
\end{align}
\end{subequations}
we fixed it to have the shortest possible expression for the second master.

%%%%%%%%%%%%%%%%%%%%%%%%%%%%%%%%%%%%%%%%%%%%%%%%%%
\subsection{External vectors + matter}
\begin{figure}
  %\tikzset{external/force remake}
   \centering
   \begin{subfigure}[a]{0.2\textwidth} \centering
      \gBoxBox[eA=quark,eB=aquark,iB=aquark,iC=quark,iD=quark,iE=quark,iF=quark]{}
   \end{subfigure}
   \begin{subfigure}[b]{0.2\textwidth} \centering
      \gBoxBox[eA=quark,eC=aquark,iA=aquark,iB=aquark,iC=aquark]{}
   \end{subfigure}
   \begin{subfigure}[c]{0.2\textwidth} \centering
      \gBoxBox[eA=quark,eD=aquark,iE=quark,iF=quark]{}
   \end{subfigure}
   \begin{subfigure}[d]{0.2\textwidth} \centering
      \gBoxBox[eA=quark,eC=aquark,iC=aquark,iF=quark,iG=aquark]{}
   \end{subfigure}
   \begin{subfigure}[e]{0.2\textwidth} \centering
      \gBoxBox[eA=quark,eB=aquark,iB=quark,iF=quark,iG=aquark]{}
   \end{subfigure}
   \begin{subfigure}[f]{0.2\textwidth} \centering
      \gBoxBox[eA=quark,eB=aquark,
               iA=aquark,iC=aquark,iD=aquark,iE=aquark,iG=aquark]{}
   \end{subfigure}
   \begin{subfigure}[g]{0.2\textwidth} \centering
      \gBoxBox[eA=quark,eD=aquark,iA=aquark,iB=aquark,iC=aquark,iD=aquark]{}
   \end{subfigure}
   \begin{subfigure}[h]{0.2\textwidth} \centering
      \gBoxBoxNP[scale=0.9,eA=quark,eB=aquark,
                 iA=aquark,iC=aquark,iD=aquark,iF=aquark,iG=aquark]{}
   \end{subfigure}
\caption{\small Two-loop masters with two external vector and matter multiplets.}
\label{fig:2LMixedMasters}
\end{figure}

The last --- and most difficult --- two-loop external MHV state configuration is with a single hypermultiplet pair and two vectors of opposite chirality.
The number of masters given in \fig{fig:2LMixedMasters} is significantly larger than for all the other cases.
Similar to what we have seen with four external matter states, a color-dual representation does not always agree with the off-shell continuation suggested by the cuts.
Nevertheless, we are able to find a valid color-dual representation for two of the master numerators in \fig{fig:2LMixedMasters} directly from the cuts.
Re-using computations from the corresponding one-loop example,
we bring the cuts for these two masters into the form
\begin{subequations}
\begin{align} \!\!\!\!
    %\tikzset{external/force remake}
    \cBoxBoxA[eA=quark,eB=aquark,eLA=$1$,eLB=$2$,eLC=$3$,eLD=$4$,iA=quark,iB=quark,
      iC=quark,iD=quark,iLE=$\downarrow\!l_1$,iLF=$\!\uparrow\!\!l_3$,iLG=$l_2\!\downarrow$]{}\!
    &=\!\!\begin{aligned}[t] &
      -\frac{\trm(4l_231)(s+l_3^2)(s+l_1^2)}{sl_1^2l_2^2l_3^2}\kT_{(41)(42)} \\ &
      -\frac{\trp(4l_231)(s+l_3^2)(s+l_1^2)}{sl_1^2l_2^2l_3^2}\kT_{(31)(32)}\,,
    \end{aligned} \\
    %\tikzset{external/force remake}
    \cBoxBoxB[eA=quark,eB=aquark,eLA=$1$,eLB=$2$,eLC=$3$,eLD=$4$,iA=quark,iB=quark,iC=quark,iLB=$l_2\!\downarrow$,iLD=$\uparrow\!l_3$,iLE=$\downarrow\!l_1$]{}\!
    &=-\frac{\trm(4l_231)(s+l_1^2)}{l_1^2l_2^2l_3^2}\kT_{(41)(42)}
      -\frac{\trp(4l_231)(s+l_1^2)}{l_1^2l_2^2l_3^2}\kT_{(31)(32)}\,,
      \!\!\! \\ \!\!\!\!
    %\tikzset{external/force remake}
    \cBoxBoxA[eA=quark,eB=aquark,eLA=$1$,eLB=$2$,eLC=$3$,eLD=$4$,iA=quark,iD=quark,iLE=$\downarrow\!l_1$,iLF=$\!\uparrow\!\!l_3$,iLG=$l_2\!\downarrow$]{}\!
    &= \begin{aligned}[t] &
      \frac{[s\trp(3l_341)+2l_3^2\trp(3l_241)](s+l_1^2)}{sl_1^2l_2^2l_3^2}\kT_{(41)(42)} \\ &
      +\frac{[s\trm(3l_341) +2l_3^2\trm(3l_241)](s+l_1^2)}{sl_1^2l_2^2l_3^2}\kT_{(31)(32)}\,,
    \end{aligned}\\
    %\tikzset{external/force remake}
    \cBoxBoxB[eA=quark,eB=aquark,eLA=$1$,eLB=$2$,eLC=$3$,eLD=$4$,iA=quark,iC=quark,iD=aquark,iLB=$l_2\!\downarrow$,iLD=$\uparrow\!l_3$,iLE=$\downarrow\!l_1$]{}\!
    &= \frac{\trp(3l_341)(s+l_1^2)}{l_1^2l_2^2l_3^2}\kT_{(41)(42)}
      +\frac{\trm(3l_341)(s+l_1^2)}{l_1^2l_2^2l_3^2}\kT_{(31)(32)}\,.
\end{align}
\end{subequations}
This leads us to simple expressions for the corresponding double-box numerators:
\begin{subequations}
\begin{align}
    %\tikzset{external/force remake}
    n\!\left(\gBoxBox[scale=0.8,eA=quark,eB=aquark,eLA=$1$,eLB=$2$,eLC=$3$,eLD=$4$,iB=quark,iC=quark,iD=quark,
      iE=quark,iF=quark,iLD=$\ell_2\!\downarrow$]{}\right)
    &= -s\trm(4\ell_231)\kT_{(41)(42)} -s\trp(4\ell_231)\kT_{(31)(32)}\,,\\
    %\tikzset{external/force remake}%
    n\!\left(\gBoxBox[scale=0.8,eA=quark,eB=aquark,eLA=$1$,eLB=$2$,eLC=$3$,eLD=$4$,iB=quark,iF=quark,
      iG=aquark,iLG=$\!\uparrow\!\!\ell_3$]{}\right)
    &= s\trp(3\ell_341)\kT_{(41)(42)} +s\trm(3\ell_341)\kT_{(31)(32)}\,,
\end{align}
\end{subequations}
and these do form a valid representation.

The other seven masters fall out of the pattern and had to be computed through an ansatz construction.
After implementing as many constraints as possible from \sec{sec:bcj} the solution contains one free parameter, that we fix by hand to obtain the shortest possible representation.
%To obtain a presentable representation we fixed these parameters to a solution of the constraint equations with a minimal numerical denominator using an algorithm presented in \rcite{Chen:2005:BBC}.
The expressions for all numerators are attached in an ancillary file to the arXiv submission of this paper as discussed in \app{sec:allRes}.

%%%%%%%%%%%%%%%%%%%%%%%%%%%%%%%%%%%%%%%%%%%%%%%%%%
%%%%%%%%%%%%%%%%%%%%%%%%%%%%%%%%%%%%%%%%%%%%%%%%%%
\section{Multi-particle cuts}
\label{sec:multiparticlecuts}

We generalize the recursion to multi-particle cuts,
deriving general formulas for cuts of MHV or $\MHVb$ amplitudes built from MHV and $\MHVb$ trees.
The structure of supersums for less than maximal supersymmetries has previously been studied in \rcite{Bern:2009xq},
using similar on-shell superspace techniques as we will employ here.
As there is no natural generalization of $\kappa$ at higher points,
our generalizations of the $\cN=4$ and $\cN=2$ supersymmetric recursion formulas,
given in \eqns{eq:supersumN4}{eq:supersumN2},
are built out of full tree-level amplitudes containing physical poles in their Parke-Taylor factors;
the resulting cut formulas are therefore less compact.
Nevertheless, if the final result is a four-point cut then there is a simple mechanism to reintroduce $\kappa$ and cancel all unphysical poles.
These expressions can be used for an iteration to any loop order.

In order to determine higher-loop or higher-point amplitudes one generally also requires cuts containing non-MHV (and non-$\MHVb$) amplitudes.
However, to obtain two-loop MHV amplitudes this is not necessary;
the techniques described here have been used to check the two-loop representations detailed in the previous section.

%%%%%%%%%%%%%%%%%%%%%%%%%%%%%%%%%%%%%%%%%%%%%%%%%%
\subsection{\texorpdfstring{$\cN=4$}{N=4} SYM}

Consider a cut of the form
\begin{equation}\label{eqn:cutGeneral}
  %\tikzset{external/force remake}
  \begin{tikzpicture}
    [line width=1pt,
    baseline={([yshift=-0.5ex]current bounding box.center)},
    font=\scriptsize]
    \fill[blob] (-0.4,0) ellipse (0.2 and 0.3);
    \fill[blob] (0.4,0) ellipse (0.2 and 0.3);
    \draw ($(-0.4,0) + (-135:0.23)$) -- ($(-0.4,0) + (-135:0.5)$) node[left] {$1$};
    \draw[dotted] ($(-0.4,0) + (-155:0.55)$) .. controls ($(-0.4,0) + (-160:0.66)$) and ($(-0.4,0) + (160:0.65)$) .. ($(-0.4,0) + (155:0.55)$);
    \draw ($(-0.4,0) + (135:0.23)$) -- ($(-0.4,0) + (135:0.5)$) node[left] {$k$};
    \draw ($(-0.4,0) + (45:0.23)$) -- node[above] {$\uset{\rightarrow}{l_1}$} ($(0.4,0) + (135:0.23)$);
    \draw[dotted] (0,0.13) -- (0,-0.13);
    \draw ($(-0.4,0) + (-45:0.23)$) -- node[below] {$\oset{\rightarrow}{l_m}$} ($(0.4,0) + (-135:0.23)$);
    \draw ($(0.4,0) + (45:0.23)$) -- ($(0.4,0) + (45:0.5)$) node[right] {$k+1$};
    \draw[dotted] ($(0.4,0) + (25:0.55)$) .. controls ($(0.4,0) + (20:0.65)$) and ($(0.4,0) + (-20:0.65)$) .. ($(0.4,0) + (-25:0.55)$);
    \draw ($(0.4,0) + (-45:0.23)$) -- ($(0.4,0) + (-45:0.5)$) node[right] {$n$};
  \end{tikzpicture}
 = \int\!\d^4\eta_{l_1}\!\cdots
   \begin{aligned}[t]
      \d^4\eta_{l_r} & A_{k+m}^{(0)}(1,\ldots,k,l_1,\ldots,l_r) \\
             \times\,& A_{n-k+m}^{(0)}(k+1,\ldots,n,-l_m,\ldots,-l_1)\,.
   \end{aligned}
\end{equation}
We are interested in cuts for which the individual trees and the full external state configuration live in either the MHV or $\MHVb$ sector.
There are two ways this can happen:
(i) one of the trees is MHV and the other is $\MHVb$, in which case we require $k=2$ or $k=n-2$;
(ii) both trees are MHV or $\MHVb$, in which case we require $m=2$ (a two-particle cut).
When $k=m=2$ both cases should reduce to the existing iterated two-particle cuts;
for more than a three-particle cut ($m>3$) this will give only specific contributions to the full cut.
We consider the two possibilities in turn.

\subsubsection{\texorpdfstring{$\MHVb\times\text{MHV}$}{MHV-bar x MHV}}

In this configuration the cut~\eqref{eqn:cutGeneral} is given by the superspace integration
\begin{align}
  %\tikzset{external/force remake}
  \begin{tikzpicture}
    [line width=1pt,
    baseline={([yshift=-0.5ex]current bounding box.center)},
    font=\scriptsize]
    \draw[] (-0.4,0) ellipse (0.2 and 0.3);
    \fill[black] (0.4,0) ellipse (0.2 and 0.3);
    %\node[fill=white,rotate=90,inner sep=0.1] at (-0.4,0) {\bf\scalebox{.6}{$\MHVb$}};
    %\node[fill=white,rotate=90,inner sep=0.1] at (0.4,0) {\bf\scalebox{.6}{MHV}};
    \draw ($(-0.4,0) + (-135:0.23)$) -- ($(-0.4,0) + (-135:0.5)$) node[left] {$1$};
    \draw ($(-0.4,0) + (135:0.23)$) -- ($(-0.4,0) + (135:0.5)$) node[left] {$2$};
    \draw ($(-0.4,0) + (45:0.23)$) -- node[above] {$\uset{\rightarrow}{l_1}$} ($(0.4,0) + (135:0.23)$);
    \draw[dotted] (0,0.13) -- (0,-0.13);
    \draw ($(-0.4,0) + (-45:0.23)$) -- node[below] {$\oset{\rightarrow}{l_m}$} ($(0.4,0) + (-135:0.23)$);
    \draw ($(0.4,0) + (45:0.23)$) -- ($(0.4,0) + (45:0.5)$) node[right] {$3$};
    \draw[dotted] ($(0.4,0) + (25:0.55)$) .. controls ($(0.4,0) + (20:0.65)$) and ($(0.4,0) + (-20:0.65)$) .. ($(0.4,0) + (-25:0.55)$);
    \draw ($(0.4,0) + (-45:0.23)$) -- ($(0.4,0) + (-45:0.5)$) node[right] {$n$};
  \end{tikzpicture}
  = -\int\!\d^4\eta_{l_1}\cdots \d^4\eta_{l_m} \frac{\delta^8(\bar{Q}_{\text{L}})}{[12][2\,l_1]\cdots[l_m1]}\frac{\delta^8(Q_{\text{R}})}{\braket{34}\cdots\braket{n\,l_m}\cdots\braket{l_13}}\,,
\end{align}
where we have inserted the Parke-Taylor formulas~\eqref{eqn:PT} and~\eqref{eqn:PTb} for right-hand MHV and left-hand $\MHVb$ tree amplitudes respectively;
using the symmetry between chiral and anti-chiral superspace we can specialize to $k=2$ without loss of generality.
We also implicitly assume a Fourier transform~\eqref{eqn:FT} of the first delta function to bring it into the chiral superspace.

To see the iterated structure some manipulation is required.
The product of the two Parke-Taylor denominators is brought into the form
\beal
    \frac{1}{[12][2l_1]\cdots[l_m1]} &
    \frac{1}{\braket{34}\cdots\braket{n\,l_m}\cdots\braket{l_13}} \\
    %&=\frac{1}{[12][2r][s1]\braket{3,4}\cdots\braket{n\,s}\braket{r,3}s_{r,r+1}\cdots s_{s-1,s}}\\
    %&=\frac{\braket{12}\braket{23}\braket{n1}}{[12][2r][s1]\braket{ns}\braket{r3}\braket{12}\cdots\braket{n1}s_{r,r+1}\cdots s_{s-1,s}}\\
    %&=\frac{\braket{12}\braket{23}\braket{n1}\braket{2r}\braket{s1}[ns][r3]}{[12]\braket{12}\cdots\braket{n1}s_{2r}s_{s1}s_{ns}s_{r3}s_{r,r+1}\cdots s_{s-1,s}}\\
     = \frac{\braket{12}}{[12]} &
       \underbrace{\frac{\braket{2|3|l_1|2}\braket{1|l_m|n|1}}{s_{2l_1}s_{l_m1}s_{n(-l_m)}s_{(-l_1)3}s_{l_1l_2}\cdots s_{l_{m-1},l_m}}}_{\text{phys. poles}}
       \underbrace{\frac{1}{\braket{12}\cdots\braket{n1}}}_{\text{Parke-Taylor}}\,,
\eeal
which exposes another Parke-Taylor factor.
The overall supersum is given by
\begin{equation}\label{eq:supersumMHVbMHV}
  \int\!\d^\cN\eta_{l_1}\cdots\d^\cN\eta_{l_m}
  \delta^{2\cN}(\bar{Q}_{\text{L}})\delta^{2\cN}(Q_{\text{R}}) = [12]^\cN\delta^{2\cN}(Q)\,,
\end{equation}
where $Q$ is the overall supermomentum.
Putting the pieces together,
\begin{align}
  %\tikzset{external/force remake}
  \begin{tikzpicture}
    [line width=1pt,
    baseline={([yshift=-0.5ex]current bounding box.center)},
    font=\scriptsize]
    \draw[] (-0.4,0) ellipse (0.2 and 0.3);
    \fill[black] (0.4,0) ellipse (0.2 and 0.3);
    %\node[fill=white,rotate=90,inner sep=0.1] at (-0.4,0) {\bf\scalebox{.6}{$\MHVb$}};
    %\node[fill=white,rotate=90,inner sep=0.1] at (0.4,0) {\bf\scalebox{.6}{MHV}};
    \draw ($(-0.4,0) + (-135:0.23)$) -- ($(-0.4,0) + (-135:0.5)$) node[left] {$1$};
    \draw ($(-0.4,0) + (135:0.23)$) -- ($(-0.4,0) + (135:0.5)$) node[left] {$2$};
    \draw ($(-0.4,0) + (45:0.23)$) -- node[above] {$\uset{\rightarrow}{l_1}$} ($(0.4,0) + (135:0.23)$);
    \draw[dotted] (0,0.13) -- (0,-0.13);
    \draw ($(-0.4,0) + (-45:0.23)$) -- node[below] {$\oset{\rightarrow}{l_m}$} ($(0.4,0) + (-135:0.23)$);
    \draw ($(0.4,0) + (45:0.23)$) -- ($(0.4,0) + (45:0.5)$) node[right] {$3$};
    \draw[dotted] ($(0.4,0) + (25:0.55)$) .. controls ($(0.4,0) + (20:0.65)$) and ($(0.4,0) + (-20:0.65)$) .. ($(0.4,0) + (-25:0.55)$);
    \draw ($(0.4,0) + (-45:0.23)$) -- ($(0.4,0) + (-45:0.5)$) node[right] {$n$};
  \end{tikzpicture}
  = \frac{i\,s_{12}\trm(23l_121l_mn1)}{s_{2l_1}s_{(-l_1)3}s_{1l_m}s_{(-l_m)n}s_{l_1l_2}\cdots s_{l_{m-1}\,l_m}}A^{(0),\text{MHV}}_n(1,\dots,n)\,.
\end{align}
Via CPT conjugation the cut with $\text{MHV}\leftrightarrow\MHVb$ is given by replacing $|i\rangle\leftrightarrow|i]$,
which exchanges $\trp\leftrightarrow\trm$.

If, at the end of several iteration steps, we are left with the cut of a four-point amplitude the result is simplified by reinstating $\kappa$:
\begin{equation}
  \begin{aligned}
    %\tikzset{external/force remake}
    \begin{tikzpicture}
      [line width=1pt,
        baseline={([yshift=-0.5ex]current bounding box.center)},
        font=\scriptsize]
      \draw[] (-0.4,0) ellipse (0.2 and 0.3);
      \fill[black] (0.4,0) ellipse (0.2 and 0.3);
      %\node[fill=white,rotate=90,inner sep=0.1] at (-0.4,0) {\bf\scalebox{.6}{$\MHVb$}};
      %\node[fill=white,rotate=90,inner sep=0.1] at (0.4,0) {\bf\scalebox{.6}{MHV}};
      \draw ($(-0.4,0) + (-135:0.23)$) -- ($(-0.4,0) + (-135:0.5)$) node[left] {$1$};
      \draw ($(-0.4,0) + (135:0.23)$) -- ($(-0.4,0) + (135:0.5)$) node[left] {$2$};
      \draw ($(-0.4,0) + (45:0.23)$) -- node[above] {$\uset{\rightarrow}{l_1}$} ($(0.4,0) + (135:0.23)$);
      \draw[dotted] (0,0.13) -- (0,-0.13);
      \draw ($(-0.4,0) + (-45:0.23)$) -- node[below] {$\oset{\rightarrow}{l_m}$} ($(0.4,0) + (-135:0.23)$);
      \draw ($(0.4,0) + (45:0.23)$) -- ($(0.4,0) + (45:0.5)$) node[right] {$3$};
      \draw ($(0.4,0) + (-45:0.23)$) -- ($(0.4,0) + (-45:0.5)$) node[right] {$4$};
    \end{tikzpicture}
    % &= \frac{i\,s_{12}\trm(23l_121l_r41)}{s_{2l_1}s_{(-l_1)3}s_{1l_r}s_{(-l_r)4}s_{l_1l_2}\cdots s_{l_{r-1}l_r}}A^{(0),\text{MHV}}_n(1,\dots,n)\\
    % &= \frac{s_{12}\braket{23}[3|l_1|2|1|l_r|4]\braket{41}[12]}{s_{2l_1}s_{(-l_1)3}s_{1l_r}s_{(-l_r)4}s_{l_1l_2}\cdots s_{l_{r-1}l_r}}\frac{\kappa}{s_{12} s_{23}}\\
    %&= \frac{}{s_{2l_1}s_{l_13}s_{1l_r}s_{l_r4}s_{l_1l_2}\cdots s_{l_{r-1}l_r}}\frac{\kappa}{\braket{23}[32]}\\
    %&= -\frac{[3|l_1|2|1|l_r|4]\braket{43}}{s_{2l_1}s_{l_13}s_{1l_r}s_{l_r4}s_{l_1l_2}\cdots s_{l_{r-1}l_r}}\kappa\\
    &= -\frac{\trm(43l_121l_m)}{s_{2l_1}s_{(-l_1)3}s_{1l_m}s_{(-l_m)4}s_{l_1l_2}\cdots s_{l_{m-1}l_m}}\kappa\,.
  \end{aligned}
\end{equation}
We recover the two-particle cut~\eqref{eq:twoCutNeq4} for $m=2$ using
$\trm(43l_121l_2) = -s_{1l_2}s_{4(-l_2)}s_{l_1l_2}$.
For $m=3$ this construction determines the full cut, given by
\begin{equation}
  \begin{aligned}
    %\tikzset{external/force remake}%
    \begin{tikzpicture}
      [line width=1pt,
        baseline={([yshift=-0.5ex]current bounding box.center)},
        font=\scriptsize]
      \draw[] (-0.4,0) ellipse (0.2 and 0.3);
      \fill[black] (0.4,0) ellipse (0.2 and 0.3);
      %\node[fill=white,rotate=90,inner sep=0.1] at (-0.4,0) {\bf\scalebox{.6}{$\MHVb$}};
      %\node[fill=white,rotate=90,inner sep=0.1] at (0.4,0) {\bf\scalebox{.6}{MHV}};
      \draw ($(-0.4,0) + (-135:0.23)$) -- ($(-0.4,0) + (-135:0.5)$) node[left] {$1$};
      \draw ($(-0.4,0) + (135:0.23)$) -- ($(-0.4,0) + (135:0.5)$) node[left] {$2$};
      \draw ($(-0.4,0) + (45:0.23)$) -- node[above] {$\uset{\rightarrow}{l_1}$} ($(0.4,0) + (135:0.23)$);
      \draw ($(-0.4,0) + (0:0.2)$) -- ($(0.4,0) + (180:0.2)$);
      \draw ($(-0.4,0) + (-45:0.23)$) -- node[below] {$\oset{\rightarrow}{l_3}$} ($(0.4,0) + (-135:0.23)$);
      \draw ($(0.4,0) + (45:0.23)$) -- ($(0.4,0) + (45:0.5)$) node[right] {$3$};
      \draw ($(0.4,0) + (-45:0.23)$) -- ($(0.4,0) + (-45:0.5)$) node[right] {$4$};
    \end{tikzpicture}
    +
    %\tikzset{external/force remake}%
    \begin{tikzpicture}
      [line width=1pt,
        baseline={([yshift=-0.5ex]current bounding box.center)},
        font=\scriptsize]
      \fill[black] (-0.4,0) ellipse (0.2 and 0.3);
      \draw[] (0.4,0) ellipse (0.2 and 0.3);
      %\node[fill=white,rotate=90,inner sep=0.1] at (-0.4,0) {\bf\scalebox{.6}{MHV}};
      %\node[fill=white,rotate=90,inner sep=0.1] at (0.4,0) {\bf\scalebox{.6}{$\MHVb$}};
      \draw ($(-0.4,0) + (-135:0.23)$) -- ($(-0.4,0) + (-135:0.5)$) node[left] {$1$};
      \draw ($(-0.4,0) + (135:0.23)$) -- ($(-0.4,0) + (135:0.5)$) node[left] {$2$};
      \draw ($(-0.4,0) + (45:0.23)$) -- node[above] {$\uset{\rightarrow}{l_1}$} ($(0.4,0) + (135:0.23)$);
      \draw ($(-0.4,0) + (0:0.2)$) -- ($(0.4,0) + (180:0.2)$);
      \draw ($(-0.4,0) + (-45:0.23)$) -- node[below] {$\oset{\rightarrow}{l_3}$} ($(0.4,0) + (-135:0.23)$);
      \draw ($(0.4,0) + (45:0.23)$) -- ($(0.4,0) + (45:0.5)$) node[right] {$3$};
      \draw ($(0.4,0) + (-45:0.23)$) -- ($(0.4,0) + (-45:0.5)$) node[right] {$4$};
    \end{tikzpicture}
    =&-\frac{ \ttr(43l_121l_3) }{ s_{2l_1}s_{3(-l_1)}s_{1l_3}s_{4(-l_3)}s_{l_1l_2}s_{l_2l_3} }\kappa\,.
  \end{aligned}
\end{equation}
The trace arises as $\ttr=\trp+\trm$ from the two contributions.

%%%%%%%%%%%%%%%%%%%%%%%%%%%%%%%%%%%%%%%%%%%%%%%%%%
\subsubsection{\texorpdfstring{$\text{MHV}\times\text{MHV}$}{MHV x MHV}}

The computation of two-particle cuts involving two MHV trees is analogous:
\begin{align}
  %\tikzset{external/force remake}
  \begin{tikzpicture}
    [line width=1pt,
    baseline={([yshift=-0.5ex]current bounding box.center)},
    font=\scriptsize]
    \fill[black] (-0.4,0) ellipse (0.2 and 0.3);
    \fill[black] (0.4,0) ellipse (0.2 and 0.3);
    %\node[fill=white,rotate=90,inner sep=0.1] at (-0.4,0) {\bf\scalebox{.6}{MHV}};
    %\node[fill=white,rotate=90,inner sep=0.1] at (0.4,0) {\bf\scalebox{.6}{MHV}};
    \draw ($(-0.4,0) + (-135:0.23)$) -- ($(-0.4,0) + (-135:0.5)$) node[left] {$1$};
    \draw ($(-0.4,0) + (135:0.23)$) -- ($(-0.4,0) + (135:0.5)$) node[left] {$k$};
    \draw[dotted] ($(-0.4,0) + (-155:0.55)$) .. controls ($(-0.4,0) + (-160:0.66)$) and ($(-0.4,0) + (160:0.65)$) .. ($(-0.4,0) + (155:0.55)$);
    \draw ($(-0.4,0) + (45:0.23)$) -- node[above] {$\uset{\rightarrow}{l_1}$} ($(0.4,0) + (135:0.23)$);
    \draw ($(-0.4,0) + (-45:0.23)$) -- node[below] {$\oset{\rightarrow}{l_2}$} ($(0.4,0) + (-135:0.23)$);
    \draw ($(0.4,0) + (45:0.23)$) -- ($(0.4,0) + (45:0.5)$) node[right] {$k+1$};
    \draw[dotted] ($(0.4,0) + (25:0.55)$) .. controls ($(0.4,0) + (20:0.65)$) and ($(0.4,0) + (-20:0.65)$) .. ($(0.4,0) + (-25:0.55)$);
    \draw ($(0.4,0) + (-45:0.23)$) -- ($(0.4,0) + (-45:0.5)$) node[right] {$n$};
  \end{tikzpicture}
  =i\frac{\trp(l_1k(k+1)l_1l_2n1l_2)}{s_{1l_2}s_{kl_1}s_{(k+1)(-l_1)}s_{n(-l_2)}}A_n^{(0),\text{MHV}}\,.
\end{align}
When $k=2$ and $n=4$ we recover the two-particle cut~\eqref{eq:twoCutNeq4} using
$\trp(l_123l_1l_241l_2)=-sts_{3(-l_1)}s_{2l_1}$.
The two MHV Parke-Taylor factors from the trees have been manipulated using
\begin{equation}\label{eqn:PTProductMHVMHV}
  \begin{multlined}
    \frac{1}{\braket{l_21}\cdots\braket{kl_1}\braket{l_1l_2}}\times\frac{1}{\braket{l_1(k+1)}\cdots\braket{nl_2}\braket{l_2l_1}} \\
    = -\frac{[l_1|k|k+1|l_1][l_2|n|1|l_2][l_1l_2]^2}{\underbrace{s_{1l_2}s_{kl_1}s_{(k+1)(-l_1)}s_{n(-l_2)}s_{l_1l_2}^2}_{\text{phys. poles}}}\frac1{\underbrace{\braket{12}\cdots\braket{n1}}_{\text{Parke-Taylor}}}\,,
    \end{multlined}
\end{equation}
and we inserted the supersum computed in~\eqref{eqn:supersumGeneral} (with $\cN=4$ supersymmetries)
\begin{equation}\label{eq:supersumMHVMHV}
  \int\!\d^\cN\eta_{l_1}\d^\cN\eta_{l_2}
  \delta^{2\cN}(Q_{\text{L}})\delta^{2\cN}(Q_{\text{R}})=\braket{l_1l_2}^\cN\delta^{2\cN}(Q)\,.
\end{equation}
The cut with two $\MHVb$ trees is again given by replacing $|i\rangle\leftrightarrow|i]$.

%%%%%%%%%%%%%%%%%%%%%%%%%%%%%%%%%%%%%%%%%%%%%%%%%%
\subsection{\texorpdfstring{$\cN=2$}{N=2} SQCD}

In $\cN=2$ SQCD there is a similar generalization, and we study cuts of the form
\begin{equation}
  %\tikzset{external/force remake}
  \begin{tikzpicture}
    [line width=1pt,
    baseline={([yshift=-0.5ex]current bounding box.center)},
    font=\scriptsize]
    \fill[blob] (-0.4,0) ellipse (0.2 and 0.3);
    \fill[blob] (0.4,0) ellipse (0.2 and 0.3);
    \draw ($(-0.4,0) + (-135:0.23)$) -- ($(-0.4,0) + (-135:0.5)$) node[left] {$1$};
    \draw[dotted] ($(-0.4,0) + (-155:0.55)$) .. controls ($(-0.4,0) + (-160:0.66)$) and ($(-0.4,0) + (160:0.65)$) .. ($(-0.4,0) + (155:0.55)$);
    \draw ($(-0.4,0) + (135:0.23)$) -- ($(-0.4,0) + (135:0.5)$) node[left] {$k$};
    \draw ($(-0.4,0) + (45:0.23)$) -- node[above] {$\uset{\rightarrow}{l_1}$} ($(0.4,0) + (135:0.23)$);
    \draw[dotted] (0,0.13) -- (0,-0.13);
    \draw ($(-0.4,0) + (-45:0.23)$) -- node[below] {$\oset{\rightarrow}{l_m}$} ($(0.4,0) + (-135:0.23)$);
    \draw ($(0.4,0) + (45:0.23)$) -- ($(0.4,0) + (45:0.5)$) node[right] {$k+1$};
    \draw[dotted] ($(0.4,0) + (25:0.55)$) .. controls ($(0.4,0) + (20:0.65)$) and ($(0.4,0) + (-20:0.65)$) .. ($(0.4,0) + (-25:0.55)$);
    \draw ($(0.4,0) + (-45:0.23)$) -- ($(0.4,0) + (-45:0.5)$) node[right] {$n$};
  \end{tikzpicture}
  =\int\!\d^4\eta_{l_1}\cdots \d^4\eta_{l_m}
  \begin{multlined}[t]
    A_{k+m,(ab)(cd)}^{(0)}(1,\ldots,k,l_1,\ldots,l_m)\\
    A_{n-k+m,(ef)(gh)}^{(0)}(k+1,\ldots,n,-l_m,\ldots,-l_1)\,.
  \end{multlined}  
\end{equation}
By analogy to $\kappa_{(ab)(cd)}$, we have introduced a new notation for tree amplitudes to encode the external-state configuration by projecting out external $\cN=2$ states from the $\cN=4$ Parke-Taylor formula~\eqref{eqn:PT}:
\begin{subequations}
\begin{align}
 A_{n,(ab)(cd)}^{(0),\text{MHV}}(1,2,\ldots,n) &=
  i\frac{\delta^4(Q)\eta_a^3 \braket{ab}\eta_b^3 \eta_c^4 \braket{cd} \eta_d^4}{\braket{12}\braket{23}\cdots\braket{n1}}\,,\\
  A_{n,(ab)(cd)}^{(0),\MHVb}(1,2,\ldots,n) &=
  i\frac{\delta^4(\bar{Q})\bar{\eta}_{a,3}[ab]\bar{\eta}_{b,3}\bar{\eta}_{c,4}[cd]\bar{\eta}_{d,4}}{[12][23]\cdots[n1]}\,,
\end{align}
\end{subequations}
where in the former $a$, $b$, $c$, and $d$ mark the legs carrying negative-helicity partons;
in the latter these indices mark the positive-helicity partons.
We study the same two possibilities as in $\cN=4$ SYM:
$\MHVb\times\text{MHV}$ and $\text{MHV}\times\text{MHV}$.

%%%%%%%%%%%%%%%%%%%%%%%%%%%%%%%%%%%%%%%%%%%%%%%%%%
\subsubsection{\texorpdfstring{$\MHVb\times\text{MHV}$}{MHV-bar x MHV}}

Once again specializing to $k=2$ without loss of generality,
we find the iterative structure of the cut as
\begin{equation}
  %\tikzset{external/force remake}
  \begin{tikzpicture}
    [line width=1pt,
    baseline={([yshift=-0.5ex]current bounding box.center)},
    font=\scriptsize]
    \draw[] (-0.4,0) ellipse (0.2 and 0.3);
    \fill[black] (0.4,0) ellipse (0.2 and 0.3);
    %\node[fill=white,rotate=90,inner sep=0.1] at (-0.4,0) {\bf\scalebox{.6}{$\MHVb$}};
    %\node[fill=white,rotate=90,inner sep=0.1] at (0.4,0) {\bf\scalebox{.6}{MHV}};
    \draw ($(-0.4,0) + (-135:0.23)$) -- ($(-0.4,0) + (-135:0.5)$) node[left] {$1$};
    \draw ($(-0.4,0) + (135:0.23)$) -- ($(-0.4,0) + (135:0.5)$) node[left] {$2$};
    \draw ($(-0.4,0) + (45:0.23)$) -- node[above] {$\uset{\rightarrow}{l_1}$} ($(0.4,0) + (135:0.23)$);
    \draw[dotted] (0,0.13) -- (0,-0.13);
    \draw ($(-0.4,0) + (-45:0.23)$) -- node[below] {$\oset{\rightarrow}{l_m}$} ($(0.4,0) + (-135:0.23)$);
    \draw ($(0.4,0) + (45:0.23)$) -- ($(0.4,0) + (45:0.5)$) node[right] {$3$};
    \draw[dotted] ($(0.4,0) + (25:0.55)$) .. controls ($(0.4,0) + (20:0.65)$) and ($(0.4,0) + (-20:0.65)$) .. ($(0.4,0) + (-25:0.55)$);
    \draw ($(0.4,0) + (-45:0.23)$) -- ($(0.4,0) + (-45:0.5)$) node[right] {$n$};
  \end{tikzpicture}
    = \frac{ i\,s_{12}\braket{2|3|l_1|2}\braket{1|l_m|n|1}[ab][cd]
      \braket{ef}\braket{gh}[qr][st] }
    { s_{2l_1}s_{3(-l_1)}s_{1l_m}s_{n(-l_m)}s_{l_1l_2} \cdots s_{l_{r-1}l_r} }
    \frac{A^{(0),\text{MHV}}_{n,(qr)(st)}}{s_{qr}s_{st}}\,.
\end{equation}
To obtain this we have used the same superspace integral given earlier~(\ref{eq:supersumMHVbMHV}),
this time with $\cN=2$ supersymmetries.
By Lorentz invariance it is clear that the spinor-helicity objects always close to form Dirac traces;
again, the opposite configuration is related by a CPT conjugation.

Further specializing to $r=2$, the two-particle formula~\eqref{eq:supersumN2A} is recovered.
For a final expression with $n=4$ it is possible to cancel the unwanted $s_{23}$ pole sitting in the tree-level factor and reintroduce $\kappa_{(qr)(st)}$:
\begin{equation}
  \begin{aligned}
    %\tikzset{external/force remake}
    \begin{tikzpicture}
      [line width=1pt,
        baseline={([yshift=-0.5ex]current bounding box.center)},
        font=\scriptsize]
      \draw[] (-0.4,0) ellipse (0.2 and 0.3);
      \fill[black] (0.4,0) ellipse (0.2 and 0.3);
      %\node[fill=white,rotate=90,inner sep=0.1] at (-0.4,0) {\bf\scalebox{.6}{$\MHVb$}};
      %\node[fill=white,rotate=90,inner sep=0.1] at (0.4,0) {\bf\scalebox{.6}{MHV}};
      \draw ($(-0.4,0) + (-135:0.23)$) -- ($(-0.4,0) + (-135:0.5)$) node[left] {$1$};
      \draw ($(-0.4,0) + (135:0.23)$) -- ($(-0.4,0) + (135:0.5)$) node[left] {$2$};
      \draw ($(-0.4,0) + (45:0.23)$) -- node[above] {$\uset{\rightarrow}{l_1}$} ($(0.4,0) + (135:0.23)$);
      \draw[dotted] (0,0.13) -- (0,-0.13);
      \draw ($(-0.4,0) + (-45:0.23)$) -- node[below] {$\oset{\rightarrow}{l_m}$} ($(0.4,0) + (-135:0.23)$);
      \draw ($(0.4,0) + (45:0.23)$) -- ($(0.4,0) + (45:0.5)$) node[right] {$3$};
      \draw ($(0.4,0) + (-45:0.23)$) -- ($(0.4,0) + (-45:0.5)$) node[right] {$4$};
    \end{tikzpicture}
    %&= \frac{ i\,s_{12}\braket{2|3|r|2}\braket{1|s|4|1}[ij][kl]
    %          \braket{mn}\braket{op}[ab][cd] }
    %        { s_{2r}s_{3(-r)}s_{1s}s_{4(-s)} s_{r,r+1} \cdots s_{s-1,s} }
    %   \frac{A^{(0),\text{MHV}}_{n,(ab)(cd)}}{s_{ab}s_{cd}}\\
    % &= \frac{ \braket{12}\braket{23}[3|l_1|2\rangle\langle1|l_r|4]\braket{41}
    %           [12][ij][kl]\braket{mn}\braket{op}[ab][cd] }
    %         { s_{2l_1}s_{3(-l_1)}s_{1l_r}s_{4(-l_r)} s_{l_1l_2}\cdots s_{l_{r-1}l_r}
    %           s_{12}\braket{23}[32] }
    %     \frac{\kappa_{(ab)(cd)}}{s_{ab}s_{cd}}\\
    %&=-\frac{ \braket{12}[3|r|2\rangle\langle1|s|4]\braket{43}[ij][kl]
    %          \braket{mn}\braket{op}[ab][cd] }
    %        { s_{2r}s_{3(-r)}s_{1s}s_{4(-s)} s_{r,r+1}\cdots s_{s-1,s}s_{12} }
    %   \frac{\kappa_{(ab)(cd)}}{s_{ab}s_{cd}}\\
    &=-\frac{ \braket{12}\braket{1|l_m|4|3|l_1|2}[ab][cd]\braket{ef}\braket{gh}[qr][st] }
            {s_{2l_1}s_{3(-l_1)}s_{1l_m}s_{4(-l_m)} s_{l_1l_2}\cdots s_{l_{m-1}l_m}s_{12} }
       \frac{\kappa_{(qr)(st)}}{s_{qr}s_{st}}\,.
  \end{aligned}
\end{equation}
These cuts do not introduce any new (unphysical) poles except the ones already found in the two-particle cut, which we identified as residues of the spinor-helicity notation (see \sec{sec:locality}).
%% r=3 formula
%% \begin{equation}
%%   \begin{aligned}
%%     %\tikzset{external/force remake}%
%%     \begin{tikzpicture}
%%       [line width=1pt,
%%         baseline={([yshift=-0.5ex]current bounding box.center)},
%%         font=\scriptsize]
%%       \fill[blob] (-0.4,0) ellipse (0.2 and 0.3);
%%       \fill[blob] (0.4,0) ellipse (0.2 and 0.3);
%%       \node[fill=white,rotate=90,inner sep=0.1] at (-0.4,0) {\bf\scalebox{.6}{$\MHVb$}};
%%       \node[fill=white,rotate=90,inner sep=0.1] at (0.4,0) {\bf\scalebox{.6}{MHV}};
%%       \draw ($(-0.4,0) + (-135:0.23)$) -- ($(-0.4,0) + (-135:0.5)$) node[left] {$1$};
%%       \draw ($(-0.4,0) + (135:0.23)$) -- ($(-0.4,0) + (135:0.5)$) node[left] {$2$};
%%       \draw ($(-0.4,0) + (45:0.23)$) -- node[above] {$\underset{\rightarrow}{l_1}$} ($(0.4,0) + (135:0.23)$);
%%       \draw ($(-0.4,0) + (0:0.2)$) -- ($(0.4,0) + (180:0.2)$);
%%       \draw ($(-0.4,0) + (-45:0.23)$) -- node[below] {$\overset{\rightarrow}{l_3}$} ($(0.4,0) + (-135:0.23)$);
%%       \draw ($(0.4,0) + (45:0.23)$) -- ($(0.4,0) + (45:0.5)$) node[right] {$3$};
%%       \draw ($(0.4,0) + (-45:0.23)$) -- ($(0.4,0) + (-45:0.5)$) node[right] {$4$};
%%     \end{tikzpicture}
%%     &= \int \d^2\eta_r\cdots\d^2\eta_s A^{(0),\MHVb}_{5,(ij)(kl)}A^{(0),\text{MHV}}_{5,(mn)(op)}\\
%%     &= -\frac{ \braket{12}\braket{1|l_3|4|3|l_1|2}[ij][kl]
%%                \braket{mn}\braket{op}[ab][cd] }
%%              { s_{2l_1}s_{3(-l_1)}s_{1l_3}s_{4(-l_3)}s_{l_1l_2}s_{l_2l_3}s_{12} }
%%         \kT_{(ab)(cd)}\,.
%%   \end{aligned}
%% \end{equation}

%%%%%%%%%%%%%%%%%%%%%%%%%%%%%%%%%%%%%%%%%%%%%%%%%%
\subsubsection{\texorpdfstring{$\text{MHV}\times\text{MHV}$}{MHV x MHV}}

In this last example we obtain
\begin{equation}\!
  %\tikzset{external/force remake}
  \begin{tikzpicture}
    [line width=1pt,
    baseline={([yshift=-0.5ex]current bounding box.center)},
    font=\scriptsize]
    \fill[black] (-0.4,0) ellipse (0.2 and 0.3);
    \fill[black] (0.4,0) ellipse (0.2 and 0.3);
    %\node[fill=white,rotate=90,inner sep=0.1] at (-0.4,0) {\bf\scalebox{.6}{MHV}};
    %\node[fill=white,rotate=90,inner sep=0.1] at (0.4,0) {\bf\scalebox{.6}{MHV}};
    \draw ($(-0.4,0) + (-135:0.23)$) -- ($(-0.4,0) + (-135:0.5)$) node[left] {$1$};
    \draw ($(-0.4,0) + (135:0.23)$) -- ($(-0.4,0) + (135:0.5)$) node[left] {$k$};
    \draw[dotted] ($(-0.4,0) + (-155:0.55)$) .. controls ($(-0.4,0) + (-160:0.66)$) and ($(-0.4,0) + (160:0.65)$) .. ($(-0.4,0) + (155:0.55)$);
    \draw ($(-0.4,0) + (45:0.23)$) -- node[above] {$\uset{\rightarrow}{l_1}$} ($(0.4,0) + (135:0.23)$);
    \draw ($(-0.4,0) + (-45:0.23)$) -- node[below] {$\oset{\rightarrow}{l_2}$} ($(0.4,0) + (-135:0.23)$);
    \draw ($(0.4,0) + (45:0.23)$) -- ($(0.4,0) + (45:0.5)$) node[right] {$k+1$};
    \draw[dotted] ($(0.4,0) + (25:0.55)$) .. controls ($(0.4,0) + (20:0.65)$) and ($(0.4,0) + (-20:0.65)$) .. ($(0.4,0) + (-25:0.55)$);
    \draw ($(0.4,0) + (-45:0.23)$) -- ($(0.4,0) + (-45:0.5)$) node[right] {$n$};
  \end{tikzpicture}
  = i\frac{[l_1|k|(k+1)|l_1][l_2|n|1|l_2]\braket{ab}\braket{cd}\braket{ef}\braket{gh}[qr][st]}{s_{1l_2}s_{kl_1}s_{(k+1)(-l_1)}s_{n(-l_2)}}\frac{A_{n,(qr)(st)}^{(0),\text{MHV}}}{s_{qr}s_{st}}\,,
\end{equation}
where the superspace integral~\eqref{eq:supersumMHVMHV} with $\cN=2$ supersymmetries is used.
Again, $\MHVb\times\MHVb$ is related by CPT conjugation.

%%%%%%%%%%%%%%%%%%%%%%%%%%%%%%%%%%%%%%%%%%%%%%%%%%
%%%%%%%%%%%%%%%%%%%%%%%%%%%%%%%%%%%%%%%%%%%%%%%%%%
\section{Conclusions and outlook}
\label{sec:outro}

In this paper we have developed an iterative method for calculating two-particle cuts in $\cN=2$ SQCD --- in essence, we have generalized the $\cN=4$ SYM rung rule to $\cN=2$ supersymmetries.
The new technology allows us to write down expressions for all iterated two-particle cuts in four dimensions using simple diagrammatic rules.
This eliminates the need for explicit state summation (Grassmann integration).
Moreover, by factorizing physical propagators it expresses the cuts in a form that assigns contributions to individual diagrams and thus suggests their natural off-shell uplift.
Armed with this new technology, we have found color-dual representations of all four-point massless $\cN=2$ SQCD amplitudes up to two loops.
We have also described extensions of the technology to multi-particle cuts.

The basic principle of the iteration is simple: when two tree amplitudes are glued together by summing over intermediate states, the result is always proportional to another tree amplitude.
This means that the Grassmann integration can be performed once, and then re-used with each iteration.
Propagators are exposed, so the expressions for contributing numerators can be lifted off shell, often without the need for ans\"atze.
We expect this to work to all loop orders, inviting us to progress to three-loop $\cN=2$ SQCD amplitudes.

We also expect the construction to work for lower-degree ($\cN<2$) supersymmetry.
The generalization of the supersum formula~\eqref{eq:supersumN2} to arbitrary $\cN$ is discussed in \rcite{Kalin1305181}.
The remaining challenge is to eliminate the square root that appears, similarly to \eqn{eq:supersumSR}, in the general formula.
We anticipate that, while it may not be possible to find diagrammatic rules, such a construction will nevertheless make the propagator structure manifest.
This should make it easier to lift expressions off shell, and we are encouraged to attempt two-loop $\cN=0$ QCD examples.

With the ability to lift numerators directly from their cuts, we have seen hints of a close connection to the IR structure of the gauge theory.
For instance, the one-loop box numerator with internal matter and external vectors, given in \eqn{eq:1Lchiralbox}, is completely IR regulated.
Local integrands of this kind have already been studied by Badger, Peraro and one of the present authors~\cite{Badger:2016ozq,Badger:2016egz}.
Similarly the one-loop mixed numerators vanish when loop momenta carried by internal hypermultiplets become soft.
In both cases, the appearance of Dirac traces naturally induce these properties.

The two-loop color-dual solution with four external vector multiplets, previously determined by Johansson and two of the present authors in \rcite{Johansson:2017bfl}, also exhibits good IR structure.
It contains chiral double-box numerators introduced by Caron-Huot and Larsen~\cite{CaronHuot:2012ab}, which are IR finite when integrated to all orders in $\eps=(4-D)/2$.
Their integrands are closely related to the chiral, BCFW-derived ``local integrands'' developed in planar $\cN=4$ SYM, which also have good IR properties
\cite{ArkaniHamed:2010kv,ArkaniHamed:2010gh,Bourjaily:2015jna}.
We also wonder whether such simplicity persists at three loops.

However, in other cases we encountered obstacles to finding simple color-dual integrands.
We first noticed this in the one-loop solution with external matter multiplets --- while a simple color-dual representation exists, it does not exhibit the IR properties we have come to expect.
This persisted at two loops: in both cases with external matter, while we always found compact expressions for the cuts, extending them to off-shell color-dual numerators required more work.
Nevertheless, knowing the terms appearing in different cuts allowed us to restrict our ans\"{a}tze to certain terms, thus simplifying the computation.

Such obstacles are not uncommon in the pursuit of color-dual loop integrands.
At three loops in $\cN=4$ SYM, the rung rule does not give four-point color-dual numerators~\cite{Bern:1997nh,ArkaniHamed:2010kv,ArkaniHamed:2010gh,Bern:2010ue,Bern:2012cd};
instead, it gives a representation matching the one in \rcite{ArkaniHamed:2010kv}.
Efforts to find a color-dual representation of the four-point, five-loop $\cN=4$ integrand have faced similar impediments~\cite{Bern:2012uc} --- however, the need for such a representation to perform the double copy has now been circumvented to compute UV divergences in $\cN=8$ supergravity~\cite{Bern:2017yxu,Bern:2017ucb,Bern:2018jmv}.
Another conspicuous example is the five-gluon two-loop all-plus integrand: while non-local color-dual numerators were found requiring twelve powers of loop momentum~\cite{Mogull:2015adi}, a planar local-integrand-based presentation is far more compact, with only four non-zero numerators~\cite{Badger:2016ozq,Badger:2016egz}.

This suggests that our requirement of color-kinematics duality is, in some cases, creating tension between the off-shell numerators,
so if our objective is not to use the double copy, we should consider relaxing it.
Doing so would allow us to directly lift expressions for the numerators from their cuts; however, each numerator would then need to be computed separately.
It would also be necessary to ensure that expressions for numerators coming from different cuts overlap with each other.
The reward may be more chiral integrand structures for the two-loop solutions with external hypermultiplets.

A study of non-planar structures would also be required --- we believe these can be isolated from iterated two-particle cuts.
Local-integrand-like non-planar integrands have already been developed for $\cN=4$ SYM~\cite{Arkani-Hamed:2014bca}; the relevant integrals have logarithmic singularities manifested by expressing them in $\d log$ forms~\cite{Bern:2014kca,Bern:2015ple}.
In this paper, we have found examples of non-planar chiral integrands in the two-loop four-vector amplitude, which warrant further study.

We intend to explore these concepts further in an upcoming work,
in which our main objective will be a better understanding of the interplay between local integrand representations and the IR structure of two-loop $\cN=2$ SQCD integrands.

%%%%%%%%%%%%%%%%%%%%%%%%%%%%%%%%%%%%%%%%%%%%%%%%%%%%
\begin{acknowledgments}

We thank Marco Chiodaroli, Lance Dixon, and Ben Page for helpful discussions. We are especially grateful to Marco Chiodaroli for access to unpublished notes about his one-loop results, and to Henrik Johansson for comments on the draft of this paper. The research of GK and GM is supported by the Swedish Research Council under grant 621-2014-5722, the Knut and Alice Wallenberg Foundation under grant KAW 2013.0235, and the Ragnar S\"{o}derberg Foundation under grant S1/16. AO has received funding from the European Union's Horizon 2020 research and innovation programme under the Marie Sk{\l}odowska-Curie grant agreement 746138.

\end{acknowledgments}
%%%%%%%%%%%%%%%%%%%%%%%%%%%%%%%%%%%%%%%%%%%%%%%%%%%%

\appendix

% ----------------------------------------------------------------------
\section{Superspace calculus}
\label{app:superspace}

In this section
we detail the derivation of \eqnss{eq:supersumMHVMHV}{eq:supersumN4}{eq:supersumN2}.
Grassmann variables are widely used in the literature
to trivialize supersymmetric state sums,
see notably \rcites{Drummond:2008bq,Bern:2009xq}.
Here we also restore some relative signs that we chose to omit in the main text.

We start by pointing out that already in \eqn{eq:deltasusy}
there is a sign ambiguity due to an unspecified order of Grassmann multiplication
inside $\delta^{2\cN}(Q)$. We fix that sign by taking the right-hand side of \eqn{eq:deltasusy} as the definition of the Grassmann delta function.
Then we can use the Schouten identity to derive the following identity
valid for any $p$ and $q$ such that $\braket{pq} \neq 0$:
\be\label{eq:deltasusyproperty}
\delta^{2\cN}\!\bigg( \sum_{i=1}^n \ket{i} \eta_i \bigg)
\equiv \prod_{I=1}^\cN\sum_{i<j}^n\braket{i\,j}\eta_i^I\eta_j^I
= \frac{1}{\braket{pq}^{\cN}}\,
\delta^{\cN}\!\bigg( \sum_{i=1}^n \braket{pi} \eta_i \bigg)
\delta^{\cN}\!\bigg( \sum_{i=1}^n \braket{qi} \eta_i \bigg) \,.\!
\ee
Using this identity twice, we compute
the supersum~\eqref{eq:supersumMHVMHV}
relevant for a general two-particle $\text{MHV}\times\text{MHV}$:
\begin{align}\label{eqn:supersumGeneral}
   \int &\d^\cN\!\eta_{l_1}\d^\cN\!\eta_{l_2}
      \delta^{2\cN}(Q_{\text{L}}) \delta^{2\cN}(Q_{\text{R}}) \\ &
    = \int\!\d^\cN\!\eta_{l_1}\d^\cN\!\eta_{l_2}
      \delta^{2\cN}\!\bigg(\! \ket{l_1}\eta_{l_1} + \ket{l_2}\eta_{l_2}
                           + \sum_{i=1}^{k}\ket{i}\eta_i \!\bigg)
      \delta^{2\cN}\!\bigg(\!\!-\ket{l_1}\eta_{l_1} - \ket{l_2}\eta_{l_2}
                           +\!\sum_{i=k+1}^n\!\ket{i}\eta_i \!\bigg) \nn \\ &
    = \frac{1}{\braket{l_1l_2}^{2\cN}}\!
      \int\!\d^\cN\!\eta_{l_1}\d^\cN\!\eta_{l_2}
      \delta^{\cN}\!\bigg( \braket{l_1l_2}\eta_{l_2}
                         + \sum_{i=1}^{k} \braket{l_1i}\eta_i \bigg)
      \delta^{\cN}\!\bigg( \braket{l_2l_1}\eta_{l_1}
                         + \sum_{i=1}^{k} \braket{l_2i}\eta_i \bigg) \nn \\* &
      \qquad \qquad \qquad \qquad~~\:\,\times
      \delta^{\cN}\!\bigg( \braket{l_1l_2}\eta_{l_2}
                         -\!\sum_{i=k+1}^{n}\!\braket{l_1i}\eta_i \bigg)
      \delta^{\cN}\!\bigg( \braket{l_2l_1}\eta_{l_1}
                         -\!\sum_{i=k+1}^{n}\!\braket{l_2i}\eta_i \bigg) \nn \\ &
    = \delta^{\cN}\!\bigg( \sum_{i=1}^n\braket{l_1i}\eta_i \bigg)
      \delta^{\cN}\!\bigg( \sum_{i=1}^n\braket{l_2i}\eta_i \bigg)
    = \braket{l_1l_2}^\cN\delta^{2\cN}(Q)\,. \nn
\end{align}
It is now effortless to verify the $\cN=4$ supersum in \eqn{eq:supersumN4}:
\begin{align}
   \int\!\d^4\eta_{l_1} d^4\eta_{l_2} \kappa(1,2,l_1,l_2) \kappa(3,4,-l_2,-l_1) &
    = \frac{[12][34][l_1 l_2]^2}{\braket{12}\braket{34}\braket{l_1 l_2}^2}\!
      \int\!\d^4\eta_{l_1} d^4\eta_{l_2}\:\!
      \delta^8(Q_{\text{L}})\delta^8(Q_{\text{R}}) \nn \\
    = \frac{[12][34]\braket{l_1 l_2}^2[l_1 l_2]^2}{\braket{12}\braket{34}}
      \delta^{8}\;\!\!\bigg( \sum_{i=1}^4 \ket{i} \eta_i \bigg) &
    = s_{l_1 l_2}^2 \kappa(1,2,3,4)\,.
\end{align}
The $\cN=2$ supersum in \eqn{eq:supersumN2} is handled similarly
\begin{align}
 & \int\!\d^4\eta_{l_1} \d^4\eta_{l_2}
      \kappa_{(ab)(cd)}(1,2,l_1,l_2) \kappa_{(ef)(gh)}(3,4,-l_2,-l_1) \nn \\ &
    = \frac{[12][34][l_1 l_2]^2}{\braket{12}\braket{34}\braket{l_1 l_2}^2}\!
         \int\!\d^4\eta_{l_1}\d^4\eta_{l_2}
         \delta^4(Q_{\text{L}})
         \braket{ab} \braket{cd} \eta_a^3 \eta_b^3 \eta_c^4 \eta_d^4
         \times\delta^4(Q_{\text{R}})
         \braket{ef} \braket{gh} \eta_e^3 \eta_f^3 \eta_g^4 \eta_h^4 \nn \\ &
       = \frac{[12][34][l_1 l_2]^2}{\braket{12}\braket{34}\braket{l_1 l_2}^2}
         \int\!\d^2\eta_{l_1}\d^2\eta_{l_2}
         \delta^4(Q_{\text{L}})\delta^4(Q_{\text{R}}) \\ &
         \qquad \qquad \qquad \qquad\:\,\times
         \int\!\d\eta_{l_1}^4\d\eta_{l_1}^3\d\eta_{l_2}^4\d\eta_{l_2}^3\,
         (\eta_a^3 \eta_b^3 \eta_c^4 \eta_d^4) (\eta_e^3 \eta_f^3 \eta_g^4 \eta_h^4)
         \braket{ab} \braket{cd} \braket{ef} \braket{gh} \nn \\ &
       = \frac{[12][34][l_1 l_2]^2}{\braket{12}\braket{34}}\delta^4(Q)
         \int\!\d\eta_{l_1}^4\d\eta_{l_1}^3\d\eta_{l_2}^4\d\eta_{l_2}^3\,
         (\eta_a^3 \eta_b^3 \eta_c^4 \eta_d^4) (\eta_e^3 \eta_f^3  \eta_g^4 \eta_h^4)
         \braket{ab} \braket{cd} \braket{ef} \braket{gh} \nn\\ &
       = \sgn(abcd)\sgn(efgh)[l_1 l_2]^2
         \braket{ab}\braket{cd}\braket{ef}\braket{gh}[qr][st]
         \frac{\kappa_{(qr)(st)}}{s_{qr}s_{st}} , \nn
\end{align}
up to the last step in the derivation, where we have used the fact that
the broken-superspace variables must factorize
onto the external ones comprising $\kappa_{(qr)(st)}$
and the internal ones annihilated by the remaining Grassmann integration,
\be
   (\eta_a^3 \eta_b^3 \eta_c^4 \eta_d^4) (\eta_e^3 \eta_f^3  \eta_g^4 \eta_h^4)
    = (\eta_{l_1}^3 \eta_{l_1}^4 \eta_{l_2}^3 \eta_{l_2}^4)
      (\eta_q^3 \eta_r^3 \eta_s^4 \eta_t^4) .
\ee
The signs $\sgn(abcd)$ and $\sgn(efgh)$ are determined by the permutation signatures
with respect to $\{q,r,l_1,l_2\}$ and $\{s,t,l_1,l_2\}$, respectively.

% ----------------------------------------------------------------------
\section{All integrands summarized}\label{sec:allRes}
In this section we summarize the full color-dual representation of all one- and two-loop integrands for $\cN=2$ SQCD.
The one-loop results are short enough to be explicitly written out here.
The two-loop integrands are quite lengthy, but they, as well as their one-loop counterpart, can be downloaded as ancillary files.
Table~\ref{tbl:properties} summarizes which representations fulfill the properties discussed in \sec{sec:bcj}.

\begin{table}
  \centering
  \begin{tabular}{l | c c c c}
    & Two-term id. & Manifest CPT & Matter reversal & $\cN=4$\\
    \midrule
    1-loop vectors & \checkmark & \checkmark & \checkmark & \checkmark\\
    1-loop mixed & \checkmark & \checkmark & \checkmark & \checkmark\\
    1-loop matter & \checkmark & \checkmark & \checkmark & \checkmark\\
    1-loop matter alt. & \checkmark & \checkmark & $\times^\ast$ & $\times$\\
    2-loop vectors & \checkmark & \checkmark & \checkmark & \checkmark\\
    2-loop mixed & \checkmark & \checkmark & \checkmark & \checkmark\\
    2-loop matter & \checkmark & \checkmark & $\times^\ast$ & $\times$\\
    \bottomrule
  \end{tabular}  
  \caption{\small Properties of the various solutions summarized: two-term identities (see \sec{sec:2term}), manifest CPT invariance (see \sec{sec:CPT}), matter-reversal symmetry (see \sec{sec:matterReverse}), and adding up to $\cN=4$ (see \sec{sec:Neq4ident}). ${}^\ast$Matter-reversal symmetry works for all numerators except for some of those with matter tadpoles. The symmetry can still be used to reduce the set of masters for all other topologies.}
  \label{tbl:properties}
\end{table}

All representations are attached in a machine-readable format to the arXiv submission of this paper.
The ancillary files for each solution are named:
\begin{itemize}
\item One-loop external vectors: \texttt{ancillaryLeq1Vectors.m}
\item One-loop mixed: \texttt{ancillaryLeq1Mixed.m}
\item One-loop external matter: \texttt{ancillaryLeq1Matter.m}
\item Two-loop external vectors: \texttt{ancillaryLeq2Vectors.m}
\item Two-loop mixed: \texttt{ancillaryLeq2Mixed.m}
\item Two-loop external matter: \texttt{ancillaryLeq2Matter.m}
\end{itemize}
The files are optimized for usage with Mathematica, but the format is general enough to allow for an import into any other computer algebra system. Each file contains a short overview of its contents in the start.

\subsection{One-loop external vectors}
\begin{subequations}
\begin{align}
%\tikzset{external/force remake}%
   n\!\left(\gBox[yshift=-1,eLA=$1$,eLB=$2$,eLC=$3$,eLD=$4$,
                  iA=aquark,iB=aquark,iD=aquark,iC=aquark,
                  iLD=$\uset{\rightarrow}{\ell}$]{}\right) & =
   \begin{aligned}[t]
      \kT_{13}\trm(1(\ell-p_1)(\ell+p_4)3)
    + \kT_{24}\trp(1(\ell-p_1)(\ell+p_4)3) & \\
    + \mu^2 \big(s(\kT_{12}\!+\!\kT_{34}) + t(\kT_{23}\!+\!\kT_{14})
                                          + u(\kT_{13}\!+\!\kT_{24})\big) & \,,
   \end{aligned} \\\!\!\!
   %%%
   n\!\left(\gTriC[eLA=$2$,eLB=$3$,eLC=$4$,eLD=$1$,iB=aquark,iD=aquark,iC=aquark,
                   iLC=$\ell\uparrow$]{}\right) & =
   \begin{aligned}[t]
       & (\kT_{13}+\kT_{34})\trm(1(\ell-p_1)(\ell+p_4)3) \\ &
       + (\kT_{12}+\kT_{24})\trp(1(\ell-p_1)(\ell+p_4)3)
       + (\kT_{12}+\kT_{34})t\ell^2\,,
   \end{aligned} \\\!\!\!
   %%%
   n\bigg(\gBubB[yshift=-3,eLA=$1$,eLB=$2$,eLC=$3$,eLD=$4$,iB=aquark,iC=aquark,
                   iLC=$\uset{\rightarrow}{\ell}$]{}\bigg) &
    = 2\ell\!\cdot\!(p_{12}-\ell)
      \big[t(\kT_{23}+\kT_{14}) - u(\kT_{13}+\kT_{24})\big] \,, \\\!\!\!
   %%%
   n\bigg(\!\gBubC[yshift=-4,eLA=$1$,eLB=$2$,eLC=$3$,eLD=$4$,iC=aquark,iD=aquark,
                   iLD=$\oset{\ell}{\rightarrow}$]{}\bigg) &
    = 2\ell\!\cdot\!(p_4+\ell)
      \big[u(\kT_{13}+\kT_{24}) - t(\kT_{14}+\kT_{23})\big] \,, \\\!\!\!
   %%%
   n\!\left(\!\gTadA[eLA=1,eLB=2,eLC=3,eLD=4,iD=aquark,
                   iLD=$\downarrow\!\ell$]{}\right) &
    = 4(\ell\cdot p_4) \big[u(\kT_{13}+\kT_{24}) - t(\kT_{14}+\kT_{23})\big] \,, \\\!\!\!
   %%%
   n\!\left(\gTadB[eLA=1,eLB=2,eLC=3,eLD=4,iC=aquark,
                   iLC=$\downarrow\!\ell$]{}\right) &
    = 4(\ell\cdot p_{34})
      \big[u(\kT_{13}+\kT_{24}) - t(\kT_{14}+\kT_{23})\big] \,, \\\!\!\!
   %%%
   n\!\left(\gBox[eLA=$1$,eLB=$2$,eLC=$3$,eLD=$4$]{}\right) &
   = n^{[\cN=4]}\left(\gBox[eLA=$1$,eLB=$2$,eLC=$3$,eLD=$4$]{}\right)
   - 2n\!\left(\gBox[eLA=$1$,eLB=$2$,eLC=$3$,eLD=$4$,iA=aquark,iB=aquark,
                     iD=aquark,iC=aquark]{}\right) , \\\!\!\!
   %%%
   n\!\left(\gTriC[eLD=$1$,eLA=$2$,eLB=$3$,eLC=$4$,iLC=$\ell\uparrow$]{}\right) &
    = -2n\!\left(\gTriC[eLA=$2$,eLB=$3$,eLC=$4$,eLD=$1$,iB=aquark,iD=aquark,
                        iC=aquark,iLC=$\ell\uparrow$]{}\right) , \\\!\!\!
   %%%
   n\bigg(\gBubB[yshift=-3,eLA=$1$,eLB=$2$,eLC=$3$,eLD=$4$,
     iLC=$\uset{\rightarrow}{\ell}$]{}\bigg) &
    = -2n\bigg(\gBubB[yshift=-3,eLA=$1$,eLB=$2$,eLC=$3$,eLD=$4$,iB=aquark,iC=aquark,
                        iLC=$\uset{\rightarrow}{\ell}$]{}\bigg) , \quad
   %%%
   n\bigg(\!\gBubC[yshift=-4,eLA=$1$,eLB=$2$,eLC=$3$,eLD=$4$,
     iLD=$\oset{\ell}{\rightarrow}$]{}\bigg)
    = -2n\bigg(\!\gBubC[yshift=-4,eLA=$1$,eLB=$2$,eLC=$3$,eLD=$4$,iC=aquark,iD=aquark,
                        iLD=$\overset{\ell}{\rightarrow}$]{}\bigg) ,\!\\\!\!\!
   %%%
   n\!\left(\!\gTadA[eLA=1,eLB=2,eLC=3,eLD=4,iLD=$\downarrow\!\ell$]{}\right) &
    = -2n\!\left(\!\gTadA[eLA=1,eLB=2,eLC=3,eLD=4,iD=aquark,
                        iLD=$\downarrow\!\ell$]{}\right) , \quad~~\,
   %%%
    n\bigg(\gTadB[yshift=-3,eLA=1,eLB=2,eLC=3,eLD=4,iLC=$\downarrow\!\ell$]{}\bigg)
    = -2n\bigg(\gTadB[yshift=-3,eLA=1,eLB=2,eLC=3,eLD=4,iC=aquark,
                        iLC=$\downarrow\!\ell$]{}\bigg) .\!
\end{align} \label{eqn:solComplVectors}%
\end{subequations}

\subsection{One-loop external vectors + matter}
\begin{subequations}
\begin{align}
%\tikzset{external/force remake}%
   n\!\left(\gBox[yshift=-1,eLA=$1$,eLB=$2$,eLC=$3$,eLD=$4$,
                  iA=quark,iD=quark,iC=quark,eB=quark,eC=aquark,
                  iLD=$\uset{\rightarrow}{\ell}$]{}\right) &
    = \kT_{(12)(13)}\trp(4\ell12) + \kT_{(24)(34)}\trm(4\ell12)\,,\\\!\!\!
   %%%
   n\!\left(\gBox[yshift=-1,eLA=$1$,eLB=$2$,eLC=$3$,eLD=$4$,iA=quark,iD=quark,
                  eB=quark,eD=aquark,iLD=$\uset{\rightarrow}{\ell}$]{}\right) &
    = \kT_{(12)(14)}\trp(3\ell12) + \kT_{(23)(34)}\trm(3\ell12)\,,\\\!\!\!
    %%%
   n\!\left(\gBox[yshift=-1,eLA=$1$,eLB=$2$,eLC=$3$,eLD=$4$,iD=quark,
                  eA=quark,eD=aquark,iLD=$\uset{\rightarrow}{\ell}$]{}\right) &
    = \kT_{(12)(24)}\trp(3\ell21) + \kT_{(13)(34)}\trm(3\ell21)\,,\\\!\!\!
   %%%
   n\!\left(\gTriC[eLA=$2$,eLB=$3$,eLC=$4$,eLD=$1$,iB=quark,iD=quark,iC=quark,
                   eA=quark,eB=aquark,iLC=$\ell\uparrow$]{}\right) & = -\frac{1}{2}
   n\!\left(\gTriC[eLA=$2$,eLB=$3$,eLC=$4$,eLD=$1$,
                   eA=quark,eB=aquark,iLC=$\ell\uparrow$]{}\right)
    = \kT_{(12)(13)}\trp(4\ell12) + \kT_{(24)(34)}\trm(4\ell12)\,.\!\!
\end{align} \label{eqn:solComplMixed}%
\end{subequations}

\subsection{One-loop external matter}
\begin{subequations}
\begin{align}
%\tikzset{external/force remake}%
   n\!\left(\gBox[eLA=$1$,eLB=$2$,eLC=$3$,eLD=$4$,eA=quark,eB=aquark,eC=quark,
                  eD=aquark,iB=quark,iD=quark]{}\right) &
     = -su\,\kT_{(13)(24)}\,, \qquad \qquad \qquad \quad
   n\!\left(\gBox[eLA=$1$,eLB=$2$,eLC=$3$,eLD=$4$,eA=quark,eB=quark,eC=aquark,
                  eD=aquark,iB=aquark,iD=quark]{}\right)
     = s^2\,\kT_{(12)(34)}\,, \\
   %%%
   n\!\left(\gTriC[eLA=$1$,eLB=$2$,eLC=$3$,eLD=$4$,eA=quark,eB=aquark,eC=quark,
                   eD=aquark,iB=quark,iD=quark]{}\right) &
    = -n\!\left(\gTriC[eLA=$1$,eLB=$2$,eLC=$3$,eLD=$4$,eA=quark,eB=aquark,eC=quark,
                   eD=aquark,iC=aquark]{}\right)
    = -su\,\kT_{(13)(24)}\,, \\
   %%%
   n\!\left(\gBubB[eLA=$1$,eLB=$2$,eLC=$3$,eLD=$4$,eA=quark,eB=aquark,eC=quark,
                   eD=aquark,iB=quark,iC=quark]{}\right) & = -\frac{1}{2} 
   n\!\left(\gBubB[eLA=$1$,eLB=$2$,eLC=$3$,eLD=$4$,iB=agluon,iC=agluon,eA=quark,
                   eB=aquark,eC=quark,eD=aquark]{}\right)
     =-su\,\kT_{(13)(24)}\,.
\end{align} \label{eqn:solComplMatter}%
\end{subequations}

\bibliographystyle{JHEP}
\bibliography{references}

\providecommand{\href}[2]{#2}\begingroup\raggedright\begin{thebibliography}{100}

\bibitem{Caron-Huot:2016owq}
S.~Caron-Huot, L.~J. Dixon, A.~McLeod and M.~von Hippel, \emph{{Bootstrapping a
  Five-Loop Amplitude Using Steinmann Relations}},
  \href{https://doi.org/10.1103/PhysRevLett.117.241601}{\emph{Phys. Rev. Lett.}
  {\bfseries 117} (2016) 241601}
  [\href{https://arxiv.org/abs/1609.00669}{{\ttfamily 1609.00669}}].

\bibitem{Anastasiou:2003kj}
C.~Anastasiou, Z.~Bern, L.~J. Dixon and D.~A. Kosower, \emph{{Planar amplitudes
  in maximally supersymmetric Yang-Mills theory}},
  \href{https://doi.org/10.1103/PhysRevLett.91.251602}{\emph{Phys. Rev. Lett.}
  {\bfseries 91} (2003) 251602}
  [\href{https://arxiv.org/abs/hep-th/0309040}{{\ttfamily hep-th/0309040}}].

\bibitem{Bern:2005iz}
Z.~Bern, L.~J. Dixon and V.~A. Smirnov, \emph{{Iteration of planar amplitudes
  in maximally supersymmetric Yang-Mills theory at three loops and beyond}},
  \href{https://doi.org/10.1103/PhysRevD.72.085001}{\emph{Phys. Rev.}
  {\bfseries D72} (2005) 085001}
  [\href{https://arxiv.org/abs/hep-th/0505205}{{\ttfamily hep-th/0505205}}].

\bibitem{ArkaniHamed:2010kv}
N.~Arkani-Hamed, J.~L. Bourjaily, F.~Cachazo, S.~Caron-Huot and J.~Trnka,
  \emph{{The All-Loop Integrand For Scattering Amplitudes in Planar N=4 SYM}},
  \href{https://doi.org/10.1007/JHEP01(2011)041}{\emph{JHEP} {\bfseries 1101}
  (2011) 041} [\href{https://arxiv.org/abs/1008.2958}{{\ttfamily 1008.2958}}].

\bibitem{ArkaniHamed:2010gh}
N.~Arkani-Hamed, J.~L. Bourjaily, F.~Cachazo and J.~Trnka, \emph{{Local
  Integrals for Planar Scattering Amplitudes}},
  \href{https://doi.org/10.1007/JHEP06(2012)125}{\emph{JHEP} {\bfseries 06}
  (2012) 125} [\href{https://arxiv.org/abs/1012.6032}{{\ttfamily 1012.6032}}].

\bibitem{ArkaniHamed:2012nw}
N.~Arkani-Hamed, J.~L. Bourjaily, F.~Cachazo, A.~B. Goncharov, A.~Postnikov and
  J.~Trnka, \emph{{Grassmannian Geometry of Scattering Amplitudes}}. Cambridge
  University Press, 2016,
  \href{https://doi.org/10.1017/CBO9781316091548}{10.1017/CBO9781316091548},
  [\href{https://arxiv.org/abs/1212.5605}{{\ttfamily 1212.5605}}].

\bibitem{Bourjaily:2015jna}
J.~L. Bourjaily and J.~Trnka, \emph{{Local Integrand Representations of All
  Two-Loop Amplitudes in Planar SYM}},
  \href{https://doi.org/10.1007/JHEP08(2015)119}{\emph{JHEP} {\bfseries 08}
  (2015) 119} [\href{https://arxiv.org/abs/1505.05886}{{\ttfamily
  1505.05886}}].

\bibitem{Bern:2007hh}
Z.~Bern, J.~Carrasco, L.~J. Dixon, H.~Johansson, D.~Kosower et~al.,
  \emph{{Three-Loop Superfiniteness of N=8 Supergravity}},
  \href{https://doi.org/10.1103/PhysRevLett.98.161303}{\emph{Phys.Rev.Lett.}
  {\bfseries 98} (2007) 161303}
  [\href{https://arxiv.org/abs/hep-th/0702112}{{\ttfamily hep-th/0702112}}].

\bibitem{Bern:2010tq}
Z.~Bern, J.~Carrasco, L.~J. Dixon, H.~Johansson and R.~Roiban, \emph{{The
  Complete Four-Loop Four-Point Amplitude in N=4 Super-Yang-Mills Theory}},
  \href{https://doi.org/10.1103/PhysRevD.82.125040}{\emph{Phys.Rev.} {\bfseries
  D82} (2010) 125040} [\href{https://arxiv.org/abs/1008.3327}{{\ttfamily
  1008.3327}}].

\bibitem{Carrasco:2011mn}
J.~J.~M. Carrasco and H.~Johansson, \emph{{Five-Point Amplitudes in N=4
  Super-Yang-Mills Theory and N=8 Supergravity}},
  \href{https://doi.org/10.1103/PhysRevD.85.025006}{\emph{Phys.Rev.} {\bfseries
  D85} (2012) 025006} [\href{https://arxiv.org/abs/1106.4711}{{\ttfamily
  1106.4711}}].

\bibitem{Bern:2012uc}
Z.~Bern, J.~Carrasco, H.~Johansson and R.~Roiban, \emph{{The Five-Loop
  Four-Point Amplitude of N=4 super-Yang-Mills Theory}},
  \href{https://doi.org/10.1103/PhysRevLett.109.241602}{\emph{Phys.Rev.Lett.}
  {\bfseries 109} (2012) 241602}
  [\href{https://arxiv.org/abs/1207.6666}{{\ttfamily 1207.6666}}].

\bibitem{Bern:2015ple}
Z.~Bern, E.~Herrmann, S.~Litsey, J.~Stankowicz and J.~Trnka, \emph{{Evidence
  for a Nonplanar Amplituhedron}},
  \href{https://doi.org/10.1007/JHEP06(2016)098}{\emph{JHEP} {\bfseries 06}
  (2016) 098} [\href{https://arxiv.org/abs/1512.08591}{{\ttfamily
  1512.08591}}].

\bibitem{Henn:2016jdu}
J.~M. Henn and B.~Mistlberger, \emph{{Four-Gluon Scattering at Three Loops,
  Infrared Structure, and the Regge Limit}},
  \href{https://doi.org/10.1103/PhysRevLett.117.171601}{\emph{Phys. Rev. Lett.}
  {\bfseries 117} (2016) 171601}
  [\href{https://arxiv.org/abs/1608.00850}{{\ttfamily 1608.00850}}].

\bibitem{Anastasiou:2000kg}
C.~Anastasiou, E.~W.~N. Glover, C.~Oleari and M.~E. Tejeda-Yeomans,
  \emph{{Two-loop QCD corrections to the scattering of massless distinct
  quarks}}, \href{https://doi.org/10.1016/S0550-3213(01)00079-7}{\emph{Nucl.
  Phys.} {\bfseries B601} (2001) 318}
  [\href{https://arxiv.org/abs/hep-ph/0010212}{{\ttfamily hep-ph/0010212}}].

\bibitem{Anastasiou:2000ue}
C.~Anastasiou, E.~W.~N. Glover, C.~Oleari and M.~E. Tejeda-Yeomans, \emph{{Two
  loop QCD corrections to massless identical quark scattering}},
  \href{https://doi.org/10.1016/S0550-3213(01)00080-3}{\emph{Nucl. Phys.}
  {\bfseries B601} (2001) 341}
  [\href{https://arxiv.org/abs/hep-ph/0011094}{{\ttfamily hep-ph/0011094}}].

\bibitem{Anastasiou:2001sv}
C.~Anastasiou, E.~W.~N. Glover, C.~Oleari and M.~E. Tejeda-Yeomans, \emph{{Two
  loop QCD corrections to massless quark gluon scattering}},
  \href{https://doi.org/10.1016/S0550-3213(01)00195-X}{\emph{Nucl. Phys.}
  {\bfseries B605} (2001) 486}
  [\href{https://arxiv.org/abs/hep-ph/0101304}{{\ttfamily hep-ph/0101304}}].

\bibitem{Glover:2001af}
E.~W.~N. Glover, C.~Oleari and M.~E. Tejeda-Yeomans, \emph{{Two loop QCD
  corrections to gluon-gluon scattering}},
  \href{https://doi.org/10.1016/S0550-3213(01)00210-3}{\emph{Nucl. Phys.}
  {\bfseries B605} (2001) 467}
  [\href{https://arxiv.org/abs/hep-ph/0102201}{{\ttfamily hep-ph/0102201}}].

\bibitem{Garland:2001tf}
L.~W. Garland, T.~Gehrmann, E.~W.~N. Glover, A.~Koukoutsakis and E.~Remiddi,
  \emph{{The Two loop QCD matrix element for e+ e- $\to$ 3 jets}},
  \href{https://doi.org/10.1016/S0550-3213(02)00057-3}{\emph{Nucl. Phys.}
  {\bfseries B627} (2002) 107}
  [\href{https://arxiv.org/abs/hep-ph/0112081}{{\ttfamily hep-ph/0112081}}].

\bibitem{Garland:2002ak}
L.~W. Garland, T.~Gehrmann, E.~W.~N. Glover, A.~Koukoutsakis and E.~Remiddi,
  \emph{{Two loop QCD helicity amplitudes for e+ e- $\to$ three jets}},
  \href{https://doi.org/10.1016/S0550-3213(02)00627-2}{\emph{Nucl. Phys.}
  {\bfseries B642} (2002) 227}
  [\href{https://arxiv.org/abs/hep-ph/0206067}{{\ttfamily hep-ph/0206067}}].

\bibitem{Catani:2011qz}
S.~Catani, L.~Cieri, D.~de~Florian, G.~Ferrera and M.~Grazzini, \emph{{Diphoton
  production at hadron colliders: a fully-differential QCD calculation at
  NNLO}}, \href{https://doi.org/10.1103/PhysRevLett.108.072001}{\emph{Phys.
  Rev. Lett.} {\bfseries 108} (2012) 072001}
  [\href{https://arxiv.org/abs/1110.2375}{{\ttfamily 1110.2375}}].

\bibitem{Gehrmann:2011aa}
T.~Gehrmann, M.~Jaquier, E.~Glover and A.~Koukoutsakis, \emph{{Two-Loop QCD
  Corrections to the Helicity Amplitudes for $H \to$ 3 partons}},
  \href{https://doi.org/10.1007/JHEP02(2012)056}{\emph{JHEP} {\bfseries 1202}
  (2012) 056} [\href{https://arxiv.org/abs/1112.3554}{{\ttfamily 1112.3554}}].

\bibitem{Czakon:2013goa}
M.~Czakon, P.~Fiedler and A.~Mitov, \emph{{Total Top-Quark Pair-Production
  Cross Section at Hadron Colliders Through $O(\alpha\frac{4}{S})$}},
  \href{https://doi.org/10.1103/PhysRevLett.110.252004}{\emph{Phys.Rev.Lett.}
  {\bfseries 110} (2013) 252004}
  [\href{https://arxiv.org/abs/1303.6254}{{\ttfamily 1303.6254}}].

\bibitem{Grazzini:2013bna}
M.~Grazzini, S.~Kallweit, D.~Rathlev and A.~Torre, \emph{{$Z\gamma$ production
  at hadron colliders in NNLO QCD}},
  \href{https://doi.org/10.1016/j.physletb.2014.02.037}{\emph{Phys.Lett.}
  {\bfseries B731} (2014) 204}
  [\href{https://arxiv.org/abs/1309.7000}{{\ttfamily 1309.7000}}].

\bibitem{Cascioli:2014yka}
F.~Cascioli, T.~Gehrmann, M.~Grazzini, S.~Kallweit, P.~Maierhöfer et~al.,
  \emph{{ZZ production at hadron colliders in NNLO QCD}},
  \href{https://doi.org/10.1016/j.physletb.2014.06.056}{\emph{Phys.Lett.}
  {\bfseries B735} (2014) 311}
  [\href{https://arxiv.org/abs/1405.2219}{{\ttfamily 1405.2219}}].

\bibitem{Gehrmann:2014fva}
T.~Gehrmann, M.~Grazzini, S.~Kallweit, P.~Maierhöfer, A.~von Manteuffel
  et~al., \emph{{$W^+W^-$ Production at Hadron Colliders in Next to Next to
  Leading Order QCD}},
  \href{https://doi.org/10.1103/PhysRevLett.113.212001}{\emph{Phys.Rev.Lett.}
  {\bfseries 113} (2014) 212001}
  [\href{https://arxiv.org/abs/1408.5243}{{\ttfamily 1408.5243}}].

\bibitem{Chen:2014gva}
X.~Chen, T.~Gehrmann, E.~Glover and M.~Jaquier, \emph{{Precise QCD predictions
  for the production of Higgs + jet final states}},
  \href{https://doi.org/10.1016/j.physletb.2014.11.021}{\emph{Phys.Lett.}
  {\bfseries B740} (2015) 147}
  [\href{https://arxiv.org/abs/1408.5325}{{\ttfamily 1408.5325}}].

\bibitem{Caola:2014iua}
F.~Caola, J.~M. Henn, K.~Melnikov, A.~V. Smirnov and V.~A. Smirnov,
  \emph{{Two-loop helicity amplitudes for the production of two off-shell
  electroweak bosons in quark-antiquark collisions}},
  \href{https://doi.org/10.1007/JHEP11(2014)041}{\emph{JHEP} {\bfseries 1411}
  (2014) 041} [\href{https://arxiv.org/abs/1408.6409}{{\ttfamily 1408.6409}}].

\bibitem{Czakon:2014xsa}
M.~Czakon, P.~Fiedler and A.~Mitov, \emph{{Resolving the Tevatron Top Quark
  Forward-Backward Asymmetry Puzzle: Fully Differential
  Next-to-Next-to-Leading-Order Calculation}},
  \href{https://doi.org/10.1103/PhysRevLett.115.052001}{\emph{Phys. Rev. Lett.}
  {\bfseries 115} (2015) 052001}
  [\href{https://arxiv.org/abs/1411.3007}{{\ttfamily 1411.3007}}].

\bibitem{Gehrmann:2015ora}
T.~Gehrmann, A.~von Manteuffel and L.~Tancredi, \emph{{The two-loop helicity
  amplitudes for $ q\overline{q}^{\prime}\to {V}_1{V}_2\to 4 $ leptons}},
  \href{https://doi.org/10.1007/JHEP09(2015)128}{\emph{JHEP} {\bfseries 09}
  (2015) 128} [\href{https://arxiv.org/abs/1503.04812}{{\ttfamily
  1503.04812}}].

\bibitem{Caola:2015ila}
F.~Caola, J.~M. Henn, K.~Melnikov, A.~V. Smirnov and V.~A. Smirnov,
  \emph{{Two-loop helicity amplitudes for the production of two off-shell
  electroweak bosons in gluon fusion}},
  \href{https://doi.org/10.1007/JHEP06(2015)129}{\emph{JHEP} {\bfseries 1506}
  (2015) 129} [\href{https://arxiv.org/abs/1503.08759}{{\ttfamily
  1503.08759}}].

\bibitem{vonManteuffel:2015msa}
A.~von Manteuffel and L.~Tancredi, \emph{{The two-loop helicity amplitudes for
  $gg \to V_1 V_2 \to 4~\mathrm{leptons}$}},
  \href{https://doi.org/10.1007/JHEP06(2015)197}{\emph{JHEP} {\bfseries 1506}
  (2015) 197} [\href{https://arxiv.org/abs/1503.08835}{{\ttfamily
  1503.08835}}].

\bibitem{Grazzini:2015nwa}
M.~Grazzini, S.~Kallweit and D.~Rathlev, \emph{{$W\gamma$ and $Z\gamma$
  production at the LHC in NNLO QCD}},
  \href{https://doi.org/10.1007/JHEP07(2015)085}{\emph{JHEP} {\bfseries 07}
  (2015) 085} [\href{https://arxiv.org/abs/1504.01330}{{\ttfamily
  1504.01330}}].

\bibitem{Boughezal:2015dva}
R.~Boughezal, C.~Focke, X.~Liu and F.~Petriello, \emph{{$W$-boson production in
  association with a jet at next-to-next-to-leading order in perturbative
  QCD}}, \href{https://doi.org/10.1103/PhysRevLett.115.062002}{\emph{Phys. Rev.
  Lett.} {\bfseries 115} (2015) 062002}
  [\href{https://arxiv.org/abs/1504.02131}{{\ttfamily 1504.02131}}].

\bibitem{Boughezal:2015dra}
R.~Boughezal, F.~Caola, K.~Melnikov, F.~Petriello and M.~Schulze, \emph{{Higgs
  boson production in association with a jet at next-to-next-to-leading
  order}}, \href{https://doi.org/10.1103/PhysRevLett.115.082003}{\emph{Phys.
  Rev. Lett.} {\bfseries 115} (2015) 082003}
  [\href{https://arxiv.org/abs/1504.07922}{{\ttfamily 1504.07922}}].

\bibitem{Boughezal:2015aha}
R.~Boughezal, C.~Focke, W.~Giele, X.~Liu and F.~Petriello, \emph{{Higgs boson
  production in association with a jet using jettiness subtraction}},
  \href{https://doi.org/10.1016/j.physletb.2015.06.055}{\emph{Phys.Lett.}
  {\bfseries B748} (2015) 5}
  [\href{https://arxiv.org/abs/1505.03893}{{\ttfamily 1505.03893}}].

\bibitem{Ridder:2015dxa}
A.~Gehrmann-De~Ridder, T.~Gehrmann, E.~W.~N. Glover, A.~Huss and T.~A. Morgan,
  \emph{{Precise QCD predictions for the production of a Z boson in association
  with a hadronic jet}},
  \href{https://doi.org/10.1103/PhysRevLett.117.022001}{\emph{Phys. Rev. Lett.}
  {\bfseries 117} (2016) 022001}
  [\href{https://arxiv.org/abs/1507.02850}{{\ttfamily 1507.02850}}].

\bibitem{Anastasiou:2015vya}
C.~Anastasiou, C.~Duhr, F.~Dulat, F.~Herzog and B.~Mistlberger, \emph{{Higgs
  Boson Gluon-Fusion Production in QCD at Three Loops}},
  \href{https://doi.org/10.1103/PhysRevLett.114.212001}{\emph{Phys.Rev.Lett.}
  {\bfseries 114} (2015) 212001}
  [\href{https://arxiv.org/abs/1503.06056}{{\ttfamily 1503.06056}}].

\bibitem{Bern:1994zx}
Z.~Bern, L.~J. Dixon, D.~C. Dunbar and D.~A. Kosower, \emph{{One-Loop n-Point
  Gauge Theory Amplitudes, Unitarity and Collinear Limits}},
  \href{https://doi.org/10.1016/0550-3213(94)90179-1}{\emph{Nucl. Phys.}
  {\bfseries B425} (1994) 217}
  [\href{https://arxiv.org/abs/hep-ph/9403226}{{\ttfamily hep-ph/9403226}}].

\bibitem{Bern:1994cg}
Z.~Bern, L.~J. Dixon, D.~C. Dunbar and D.~A. Kosower, \emph{{Fusing gauge
  theory tree amplitudes into loop amplitudes}},
  \href{https://doi.org/10.1016/0550-3213(94)00488-Z}{\emph{Nucl.Phys.}
  {\bfseries B435} (1995) 59}
  [\href{https://arxiv.org/abs/hep-ph/9409265}{{\ttfamily hep-ph/9409265}}].

\bibitem{Britto:2004nc}
R.~Britto, F.~Cachazo and B.~Feng, \emph{{Generalized unitarity and one-loop
  amplitudes in N=4 super-Yang-Mills}},
  \href{https://doi.org/10.1016/j.nuclphysb.2005.07.014}{\emph{Nucl.Phys.}
  {\bfseries B725} (2005) 275}
  [\href{https://arxiv.org/abs/hep-th/0412103}{{\ttfamily hep-th/0412103}}].

\bibitem{Britto:2004ap}
R.~Britto, F.~Cachazo and B.~Feng, \emph{{New recursion relations for tree
  amplitudes of gluons}},
  \href{https://doi.org/10.1016/j.nuclphysb.2005.02.030}{\emph{Nucl.Phys.}
  {\bfseries B715} (2005) 499}
  [\href{https://arxiv.org/abs/hep-th/0412308}{{\ttfamily hep-th/0412308}}].

\bibitem{Britto:2005fq}
R.~Britto, F.~Cachazo, B.~Feng and E.~Witten, \emph{{Direct proof of tree-level
  recursion relation in Yang-Mills theory}},
  \href{https://doi.org/10.1103/PhysRevLett.94.181602}{\emph{Phys.Rev.Lett.}
  {\bfseries 94} (2005) 181602}
  [\href{https://arxiv.org/abs/hep-th/0501052}{{\ttfamily hep-th/0501052}}].

\bibitem{Forde:2007mi}
D.~Forde, \emph{{Direct extraction of one-loop integral coefficients}},
  \href{https://doi.org/10.1103/PhysRevD.75.125019}{\emph{Phys.Rev.} {\bfseries
  D75} (2007) 125019} [\href{https://arxiv.org/abs/0704.1835}{{\ttfamily
  0704.1835}}].

\bibitem{Anastasiou:2006jv}
C.~Anastasiou, R.~Britto, B.~Feng, Z.~Kunszt and P.~Mastrolia,
  \emph{{D-dimensional unitarity cut method}},
  \href{https://doi.org/10.1016/j.physletb.2006.12.022}{\emph{Phys. Lett.}
  {\bfseries B645} (2007) 213}
  [\href{https://arxiv.org/abs/hep-ph/0609191}{{\ttfamily hep-ph/0609191}}].

\bibitem{Giele:2008ve}
W.~T. Giele, Z.~Kunszt and K.~Melnikov, \emph{{Full one-loop amplitudes from
  tree amplitudes}},
  \href{https://doi.org/10.1088/1126-6708/2008/04/049}{\emph{JHEP} {\bfseries
  0804} (2008) 049} [\href{https://arxiv.org/abs/0801.2237}{{\ttfamily
  0801.2237}}].

\bibitem{Badger:2013gxa}
S.~Badger, H.~Frellesvig and Y.~Zhang, \emph{{A Two-Loop Five-Gluon Helicity
  Amplitude in QCD}},
  \href{https://doi.org/10.1007/JHEP12(2013)045}{\emph{JHEP} {\bfseries 1312}
  (2013) 045} [\href{https://arxiv.org/abs/1310.1051}{{\ttfamily 1310.1051}}].

\bibitem{Badger:2015lda}
S.~Badger, G.~Mogull, A.~Ochirov and D.~O'Connell, \emph{{A Complete Two-Loop,
  Five-Gluon Helicity Amplitude in Yang-Mills Theory}},
  \href{https://doi.org/10.1007/JHEP10(2015)064}{\emph{JHEP} {\bfseries 10}
  (2015) 064} [\href{https://arxiv.org/abs/1507.08797}{{\ttfamily
  1507.08797}}].

\bibitem{Gehrmann:2015bfy}
T.~Gehrmann, J.~M. Henn and N.~A. Lo~Presti, \emph{{Analytic form of the
  two-loop planar five-gluon all-plus-helicity amplitude in QCD}},
  \href{https://doi.org/10.1103/PhysRevLett.116.189903,
  10.1103/PhysRevLett.116.062001}{\emph{Phys. Rev. Lett.} {\bfseries 116}
  (2016) 062001} [\href{https://arxiv.org/abs/1511.05409}{{\ttfamily
  1511.05409}}].

\bibitem{Badger:2017jhb}
S.~Badger, C.~Bronnum-Hansen, H.~B. Hartanto and T.~Peraro, \emph{{First look
  at two-loop five-gluon scattering in QCD}},
  \href{https://doi.org/10.1103/PhysRevLett.120.092001}{\emph{Phys. Rev. Lett.}
  {\bfseries 120} (2018) 092001}
  [\href{https://arxiv.org/abs/1712.02229}{{\ttfamily 1712.02229}}].

\bibitem{Abreu:2017hqn}
S.~Abreu, F.~Febres~Cordero, H.~Ita, B.~Page and M.~Zeng, \emph{{Planar
  Two-Loop Five-Gluon Amplitudes from Numerical Unitarity}},
  \href{https://doi.org/10.1103/PhysRevD.97.116014}{\emph{Phys. Rev.}
  {\bfseries D97} (2018) 116014}
  [\href{https://arxiv.org/abs/1712.03946}{{\ttfamily 1712.03946}}].

\bibitem{Chawdhry:2018awn}
H.~A. Chawdhry, M.~A. Lim and A.~Mitov, \emph{{Two-loop five-point massless QCD
  amplitudes within the IBP approach}},
  \href{https://arxiv.org/abs/1805.09182}{{\ttfamily 1805.09182}}.

\bibitem{Badger:2018gip}
S.~Badger, C.~Bronnum-Hansen, T.~Gehrmann, H.~B. Hartanto, J.~Henn, N.~A.
  Lo~Presti et~al., \emph{{Applications of integrand reduction to two-loop
  five-point scattering amplitudes in QCD}},
  \href{https://doi.org/10.22323/1.303.0006}{\emph{PoS} {\bfseries LL2018}
  (2018) 006} [\href{https://arxiv.org/abs/1807.09709}{{\ttfamily
  1807.09709}}].

\bibitem{Abreu:2018jgq}
S.~Abreu, F.~Febres~Cordero, H.~Ita, B.~Page and V.~Sotnikov, \emph{{Planar
  Two-Loop Five-Parton Amplitudes from Numerical Unitarity}},
  \href{https://arxiv.org/abs/1809.09067}{{\ttfamily 1809.09067}}.

\bibitem{Dixon:1996wi}
L.~J. Dixon, \emph{{Calculating scattering amplitudes efficiently}},  in
  \emph{{QCD and beyond. Proceedings, Theoretical Advanced Study Institute in
  Elementary Particle Physics, TASI-95, Boulder, USA, June 4-30, 1995}},
  pp.~539--584, 1996, \href{https://arxiv.org/abs/hep-ph/9601359}{{\ttfamily
  hep-ph/9601359}},
  \href{http://www-public.slac.stanford.edu/sciDoc/docMeta.aspx?slacPubNumber=SLAC-PUB-7106}{http://www-public.slac.stanford.edu/sciDoc/docMeta.aspx?slacPubNumber=SLAC-PUB-7106}.

\bibitem{Dixon:2010ik}
L.~J. Dixon, J.~M. Henn, J.~Plefka and T.~Schuster, \emph{{All tree-level
  amplitudes in massless QCD}},
  \href{https://doi.org/10.1007/JHEP01(2011)035}{\emph{JHEP} {\bfseries 1101}
  (2011) 035} [\href{https://arxiv.org/abs/1010.3991}{{\ttfamily 1010.3991}}].

\bibitem{Melia:2013epa}
T.~Melia, \emph{{Getting more flavor out of one-flavor QCD}},
  \href{https://doi.org/10.1103/PhysRevD.89.074012}{\emph{Phys.Rev.} {\bfseries
  D89} (2014) 074012} [\href{https://arxiv.org/abs/1312.0599}{{\ttfamily
  1312.0599}}].

\bibitem{Bern:1993mq}
Z.~Bern, L.~J. Dixon and D.~A. Kosower, \emph{{One loop corrections to five
  gluon amplitudes}},
  \href{https://doi.org/10.1103/PhysRevLett.70.2677}{\emph{Phys.Rev.Lett.}
  {\bfseries 70} (1993) 2677}
  [\href{https://arxiv.org/abs/hep-ph/9302280}{{\ttfamily hep-ph/9302280}}].

\bibitem{Bern:2002zk}
Z.~Bern, A.~De~Freitas, L.~J. Dixon and H.~L. Wong, \emph{{Supersymmetric
  regularization, two loop QCD amplitudes and coupling shifts}},
  \href{https://doi.org/10.1103/PhysRevD.66.085002}{\emph{Phys. Rev.}
  {\bfseries D66} (2002) 085002}
  [\href{https://arxiv.org/abs/hep-ph/0202271}{{\ttfamily hep-ph/0202271}}].

\bibitem{Bern:1997nh}
Z.~Bern, J.~S. Rozowsky and B.~Yan, \emph{{Two loop four gluon amplitudes in
  N=4 superYang-Mills}},
  \href{https://doi.org/10.1016/S0370-2693(97)00413-9}{\emph{Phys. Lett.}
  {\bfseries B401} (1997) 273}
  [\href{https://arxiv.org/abs/hep-ph/9702424}{{\ttfamily hep-ph/9702424}}].

\bibitem{Bern:1998ug}
Z.~Bern, L.~J. Dixon, D.~C. Dunbar, M.~Perelstein and J.~S. Rozowsky, \emph{{On
  the relationship between Yang-Mills theory and gravity and its implication
  for ultraviolet divergences}},
  \href{https://doi.org/10.1016/S0550-3213(98)00420-9}{\emph{Nucl. Phys.}
  {\bfseries B530} (1998) 401}
  [\href{https://arxiv.org/abs/hep-th/9802162}{{\ttfamily hep-th/9802162}}].

\bibitem{Bern:2008qj}
Z.~Bern, J.~Carrasco and H.~Johansson, \emph{{New Relations for Gauge-Theory
  Amplitudes}},
  \href{https://doi.org/10.1103/PhysRevD.78.085011}{\emph{Phys.Rev.} {\bfseries
  D78} (2008) 085011} [\href{https://arxiv.org/abs/0805.3993}{{\ttfamily
  0805.3993}}].

\bibitem{Bern:2010ue}
Z.~Bern, J.~J.~M. Carrasco and H.~Johansson, \emph{{Perturbative Quantum
  Gravity as a Double Copy of Gauge Theory}},
  \href{https://doi.org/10.1103/PhysRevLett.105.061602}{\emph{Phys.Rev.Lett.}
  {\bfseries 105} (2010) 061602}
  [\href{https://arxiv.org/abs/1004.0476}{{\ttfamily 1004.0476}}].

\bibitem{Johansson:2015oia}
H.~Johansson and A.~Ochirov, \emph{{Color-Kinematics Duality for QCD
  Amplitudes}}, \href{https://doi.org/10.1007/JHEP01(2016)170}{\emph{JHEP}
  {\bfseries 01} (2016) 170}
  [\href{https://arxiv.org/abs/1507.00332}{{\ttfamily 1507.00332}}].

\bibitem{Chiodaroli:2013upa}
M.~Chiodaroli, Q.~Jin and R.~Roiban, \emph{{Color/kinematics duality for
  general abelian orbifolds of N=4 super Yang-Mills theory}},
  \href{https://doi.org/10.1007/JHEP01(2014)152}{\emph{JHEP} {\bfseries 1401}
  (2014) 152} [\href{https://arxiv.org/abs/1311.3600}{{\ttfamily 1311.3600}}].

\bibitem{Johansson:2014zca}
H.~Johansson and A.~Ochirov, \emph{{Pure Gravities via Color-Kinematics Duality
  for Fundamental Matter}},
  \href{https://doi.org/10.1007/JHEP11(2015)046}{\emph{JHEP} {\bfseries 11}
  (2015) 046} [\href{https://arxiv.org/abs/1407.4772}{{\ttfamily 1407.4772}}].

\bibitem{Johansson:2017bfl}
H.~Johansson, G.~Kälin and G.~Mogull, \emph{{Two-loop supersymmetric QCD and
  half-maximal supergravity amplitudes}},
  \href{https://doi.org/10.1007/JHEP09(2017)019}{\emph{JHEP} {\bfseries 09}
  (2017) 019} [\href{https://arxiv.org/abs/1706.09381}{{\ttfamily
  1706.09381}}].

\bibitem{Bern:2011rj}
Z.~Bern, C.~Boucher-Veronneau and H.~Johansson, \emph{{N $\ge$ 4 Supergravity
  Amplitudes from Gauge Theory at One Loop}},
  \href{https://doi.org/10.1103/PhysRevD.84.105035}{\emph{Phys.Rev.} {\bfseries
  D84} (2011) 105035} [\href{https://arxiv.org/abs/1107.1935}{{\ttfamily
  1107.1935}}].

\bibitem{Carrasco:2012ca}
J.~J.~M. Carrasco, M.~Chiodaroli, M.~Gunaydin and R.~Roiban, \emph{{One-loop
  four-point amplitudes in pure and matter-coupled ${\cal N} \le 4$
  supergravity}}, \href{https://doi.org/10.1007/JHEP03(2013)056}{\emph{JHEP}
  {\bfseries 1303} (2013) 056}
  [\href{https://arxiv.org/abs/1212.1146}{{\ttfamily 1212.1146}}].

\bibitem{Chiodaroli:2014xia}
M.~Chiodaroli, M.~Günaydin, H.~Johansson and R.~Roiban, \emph{{Scattering
  amplitudes in $ \mathcal{N}=2 $ Maxwell-Einstein and Yang-Mills/Einstein
  supergravity}}, \href{https://doi.org/10.1007/JHEP01(2015)081}{\emph{JHEP}
  {\bfseries 1501} (2015) 081}
  [\href{https://arxiv.org/abs/1408.0764}{{\ttfamily 1408.0764}}].

\bibitem{Chiodaroli:2015rdg}
M.~Chiodaroli, M.~Gunaydin, H.~Johansson and R.~Roiban, \emph{{Spontaneously
  Broken Yang-Mills-Einstein Supergravities as Double Copies}},
  \href{https://doi.org/10.1007/JHEP06(2017)064}{\emph{JHEP} {\bfseries 06}
  (2017) 064} [\href{https://arxiv.org/abs/1511.01740}{{\ttfamily
  1511.01740}}].

\bibitem{Chiodaroli:2015wal}
M.~Chiodaroli, M.~Gunaydin, H.~Johansson and R.~Roiban, \emph{{Complete
  construction of magical, symmetric and homogeneous N=2 supergravities as
  double copies of gauge theories}},
  \href{https://doi.org/10.1103/PhysRevLett.117.011603}{\emph{Phys. Rev. Lett.}
  {\bfseries 117} (2016) 011603}
  [\href{https://arxiv.org/abs/1512.09130}{{\ttfamily 1512.09130}}].

\bibitem{Anastasiou:2016csv}
A.~Anastasiou, L.~Borsten, M.~J. Duff, M.~J. Hughes, A.~Marrani, S.~Nagy
  et~al., \emph{{Twin supergravities from Yang-Mills theory squared}},
  \href{https://doi.org/10.1103/PhysRevD.96.026013}{\emph{Phys. Rev.}
  {\bfseries D96} (2017) 026013}
  [\href{https://arxiv.org/abs/1610.07192}{{\ttfamily 1610.07192}}].

\bibitem{Johansson:2017srf}
H.~Johansson and J.~Nohle, \emph{{Conformal Gravity from Gauge Theory}},
  \href{https://arxiv.org/abs/1707.02965}{{\ttfamily 1707.02965}}.

\bibitem{Chiodaroli:2017ehv}
M.~Chiodaroli, M.~Gunaydin, H.~Johansson and R.~Roiban, \emph{{Gauged
  Supergravities and Spontaneous Supersymmetry Breaking from the Double Copy
  Construction}},
  \href{https://doi.org/10.1103/PhysRevLett.120.171601}{\emph{Phys. Rev. Lett.}
  {\bfseries 120} (2018) 171601}
  [\href{https://arxiv.org/abs/1710.08796}{{\ttfamily 1710.08796}}].

\bibitem{Johansson:2018ues}
H.~Johansson, G.~Mogull and F.~Teng, \emph{{Unraveling conformal gravity
  amplitudes}}, \href{https://doi.org/10.1007/JHEP09(2018)080}{\emph{JHEP}
  {\bfseries 09} (2018) 080}
  [\href{https://arxiv.org/abs/1806.05124}{{\ttfamily 1806.05124}}].

\bibitem{Glover:2008tu}
E.~W.~N. Glover, V.~V. Khoze and C.~Williams, \emph{{Component MHV amplitudes
  in N=2 SQCD and in N=4 SYM at one loop}},
  \href{https://doi.org/10.1088/1126-6708/2008/08/033}{\emph{JHEP} {\bfseries
  08} (2008) 033} [\href{https://arxiv.org/abs/0805.4190}{{\ttfamily
  0805.4190}}].

\bibitem{Andree:2010na}
R.~Andree and D.~Young, \emph{{Wilson Loops in N=2 Superconformal Yang-Mills
  Theory}}, \href{https://doi.org/10.1007/JHEP09(2010)095}{\emph{JHEP}
  {\bfseries 09} (2010) 095} [\href{https://arxiv.org/abs/1007.4923}{{\ttfamily
  1007.4923}}].

\bibitem{Leoni:2014fja}
M.~Leoni, A.~Mauri and A.~Santambrogio, \emph{{Four-point amplitudes in
  $\mathcal{N}=2$ SCQCD}}, \href{https://doi.org/10.1007/JHEP09(2014)017,
  10.1007/JHEP02(2015)022}{\emph{JHEP} {\bfseries 09} (2014) 017}
  [\href{https://arxiv.org/abs/1406.7283}{{\ttfamily 1406.7283}}].

\bibitem{Leoni:2015zxa}
M.~Leoni, A.~Mauri and A.~Santambrogio, \emph{{On the amplitude/Wilson loop
  duality in N=2 SCQCD}},
  \href{https://doi.org/10.1016/j.physletb.2015.06.013}{\emph{Phys. Lett.}
  {\bfseries B747} (2015) 325}
  [\href{https://arxiv.org/abs/1502.07614}{{\ttfamily 1502.07614}}].

\bibitem{Badger:2016ozq}
S.~Badger, G.~Mogull and T.~Peraro, \emph{{Local integrands for two-loop
  all-plus Yang-Mills amplitudes}},
  \href{https://doi.org/10.1007/JHEP08(2016)063}{\emph{JHEP} {\bfseries 08}
  (2016) 063} [\href{https://arxiv.org/abs/1606.02244}{{\ttfamily
  1606.02244}}].

\bibitem{Bern:2000dn}
Z.~Bern, L.~J. Dixon and D.~Kosower, \emph{{A Two loop four gluon helicity
  amplitude in QCD}},
  \href{https://doi.org/10.1088/1126-6708/2000/01/027}{\emph{JHEP} {\bfseries
  0001} (2000) 027} [\href{https://arxiv.org/abs/hep-ph/0001001}{{\ttfamily
  hep-ph/0001001}}].

\bibitem{Dunbar:2016aux}
D.~C. Dunbar and W.~B. Perkins, \emph{{Two-loop five-point all plus helicity
  Yang-Mills amplitude}},
  \href{https://doi.org/10.1103/PhysRevD.93.085029}{\emph{Phys. Rev.}
  {\bfseries D93} (2016) 085029}
  [\href{https://arxiv.org/abs/1603.07514}{{\ttfamily 1603.07514}}].

\bibitem{Dunbar:2016cxp}
D.~C. Dunbar, G.~R. Jehu and W.~B. Perkins, \emph{{The two-loop n-point
  all-plus helicity amplitude}},
  \href{https://doi.org/10.1103/PhysRevD.93.125006}{\emph{Phys. Rev.}
  {\bfseries D93} (2016) 125006}
  [\href{https://arxiv.org/abs/1604.06631}{{\ttfamily 1604.06631}}].

\bibitem{Dunbar:2016gjb}
D.~C. Dunbar, G.~R. Jehu and W.~B. Perkins, \emph{{Two-loop six gluon all plus
  helicity amplitude}},
  \href{https://doi.org/10.1103/PhysRevLett.117.061602}{\emph{Phys. Rev. Lett.}
  {\bfseries 117} (2016) 061602}
  [\href{https://arxiv.org/abs/1605.06351}{{\ttfamily 1605.06351}}].

\bibitem{Bern:1996ja}
Z.~Bern, L.~J. Dixon, D.~C. Dunbar and D.~A. Kosower, \emph{{One loop selfdual
  and N=4 super Yang-Mills}},
  \href{https://doi.org/10.1016/S0370-2693(96)01676-0}{\emph{Phys.Lett.}
  {\bfseries B394} (1997) 105}
  [\href{https://arxiv.org/abs/hep-th/9611127}{{\ttfamily hep-th/9611127}}].

\bibitem{Nair:1988bq}
V.~Nair, \emph{{A Current Algebra for Some Gauge Theory Amplitudes}},
  \href{https://doi.org/10.1016/0370-2693(88)91471-2}{\emph{Phys.Lett.}
  {\bfseries B214} (1988) 215}.

\bibitem{BjerrumBohr:2007vu}
N.~Bjerrum-Bohr, D.~C. Dunbar and W.~B. Perkins, \emph{{Analytic structure of
  three-mass triangle coefficients}},
  \href{https://doi.org/10.1088/1126-6708/2008/04/038}{\emph{JHEP} {\bfseries
  0804} (2008) 038} [\href{https://arxiv.org/abs/0709.2086}{{\ttfamily
  0709.2086}}].

\bibitem{Ochirov:2013oca}
A.~Ochirov, \emph{{All one-loop NMHV gluon amplitudes in N=1 SYM}},
  \href{https://doi.org/10.1007/JHEP12(2013)080}{\emph{JHEP} {\bfseries 12}
  (2013) 080} [\href{https://arxiv.org/abs/1311.1491}{{\ttfamily 1311.1491}}].

\bibitem{Parke:1986gb}
S.~J. Parke and T.~Taylor, \emph{{An Amplitude for $n$ Gluon Scattering}},
  \href{https://doi.org/10.1103/PhysRevLett.56.2459}{\emph{Phys.Rev.Lett.}
  {\bfseries 56} (1986) 2459}.

\bibitem{Elvang:2015rqa}
H.~Elvang and Y.-t. Huang, \emph{{Scattering Amplitudes in Gauge Theory and
  Gravity}}. Cambridge University Press, 2015.

\bibitem{Ochirov:2016ewn}
A.~Ochirov and B.~Page, \emph{{Full Colour for Loop Amplitudes in Yang-Mills
  Theory}}, \href{https://doi.org/10.1007/JHEP02(2017)100}{\emph{JHEP}
  {\bfseries 02} (2017) 100}
  [\href{https://arxiv.org/abs/1612.04366}{{\ttfamily 1612.04366}}].

\bibitem{Kalin:2017oqr}
G.~Kälin, \emph{{Cyclic Mario worlds — color-decomposition for one-loop
  QCD}}, \href{https://doi.org/10.1007/JHEP04(2018)141}{\emph{JHEP} {\bfseries
  04} (2018) 141} [\href{https://arxiv.org/abs/1712.03539}{{\ttfamily
  1712.03539}}].

\bibitem{Kiermaier}
M.~Kiermaier, \emph{{Gravity as the square of gauge theory}},  talk at
  conference Amplitudes, 2010,
  \href{http://strings.ph.qmul.ac.uk/~theory/Amplitudes2010/Talks/MK2010.pdf}{http://strings.ph.qmul.ac.uk/~theory/Amplitudes2010/Talks/MK2010.pdf}.

\bibitem{BjerrumBohr:2010hn}
N.~Bjerrum-Bohr, P.~H. Damgaard, T.~Sondergaard and P.~Vanhove, \emph{{The
  Momentum Kernel of Gauge and Gravity Theories}},
  \href{https://doi.org/10.1007/JHEP01(2011)001}{\emph{JHEP} {\bfseries 1101}
  (2011) 001} [\href{https://arxiv.org/abs/1010.3933}{{\ttfamily 1010.3933}}].

\bibitem{BjerrumBohr:2009rd}
N.~Bjerrum-Bohr, P.~H. Damgaard and P.~Vanhove, \emph{{Minimal Basis for Gauge
  Theory Amplitudes}},
  \href{https://doi.org/10.1103/PhysRevLett.103.161602}{\emph{Phys.Rev.Lett.}
  {\bfseries 103} (2009) 161602}
  [\href{https://arxiv.org/abs/0907.1425}{{\ttfamily 0907.1425}}].

\bibitem{Stieberger:2009hq}
S.~Stieberger, \emph{{Open \& Closed vs. Pure Open String Disk Amplitudes}},
  \href{https://arxiv.org/abs/0907.2211}{{\ttfamily 0907.2211}}.

\bibitem{Feng:2010my}
B.~Feng, R.~Huang and Y.~Jia, \emph{{Gauge Amplitude Identities by On-shell
  Recursion Relation in S-matrix Program}},
  \href{https://doi.org/10.1016/j.physletb.2010.11.011}{\emph{Phys.Lett.}
  {\bfseries B695} (2011) 350}
  [\href{https://arxiv.org/abs/1004.3417}{{\ttfamily 1004.3417}}].

\bibitem{delaCruz:2015dpa}
L.~de~la Cruz, A.~Kniss and S.~Weinzierl, \emph{{Proof of the fundamental BCJ
  relations for QCD amplitudes}},
  \href{https://doi.org/10.1007/JHEP09(2015)197}{\emph{JHEP} {\bfseries 09}
  (2015) 197} [\href{https://arxiv.org/abs/1508.01432}{{\ttfamily
  1508.01432}}].

\bibitem{Chiodaroli:2016jqw}
M.~Chiodaroli, \emph{{Simplifying amplitudes in Maxwell-Einstein and
  Yang-Mills-Einstein supergravities}},  in \emph{{Jochen Brüning, Matthias
  Staudacher (Eds.), Space – Time – Matter: Analytic and Geometric
  Structures. Berlin, Boston: De Gruyter.}}, pp.~266--287, 2018,
  \href{https://arxiv.org/abs/1607.04129}{{\ttfamily 1607.04129}},
  \href{https://doi.org/10.1515/9783110452150-011}{DOI}.

\bibitem{Bern:2004kq}
Z.~Bern, L.~J. Dixon and D.~A. Kosower, \emph{{N=4 super-Yang-Mills theory, QCD
  and collider physics}},
  \href{https://doi.org/10.1016/j.crhy.2004.09.007}{\emph{Comptes Rendus
  Physique} {\bfseries 5} (2004) 955}
  [\href{https://arxiv.org/abs/hep-th/0410021}{{\ttfamily hep-th/0410021}}].

\bibitem{Green:1982sw}
M.~B. Green, J.~H. Schwarz and L.~Brink, \emph{{N=4 Yang-Mills and N=8
  Supergravity as Limits of String Theories}},
  \href{https://doi.org/10.1016/0550-3213(82)90336-4}{\emph{Nucl. Phys.}
  {\bfseries B198} (1982) 474}.

\bibitem{Bern:2012cd}
Z.~Bern, S.~Davies, T.~Dennen and Y.-t. Huang, \emph{{Absence of Three-Loop
  Four-Point Divergences in N=4 Supergravity}},
  \href{https://doi.org/10.1103/PhysRevLett.108.201301}{\emph{Phys.Rev.Lett.}
  {\bfseries 108} (2012) 201301}
  [\href{https://arxiv.org/abs/1202.3423}{{\ttfamily 1202.3423}}].

\bibitem{Catani:1996jh}
S.~Catani and M.~H. Seymour, \emph{{The Dipole formalism for the calculation of
  QCD jet cross-sections at next-to-leading order}},
  \href{https://doi.org/10.1016/0370-2693(96)00425-X}{\emph{Phys. Lett.}
  {\bfseries B378} (1996) 287}
  [\href{https://arxiv.org/abs/hep-ph/9602277}{{\ttfamily hep-ph/9602277}}].

\bibitem{Catani:1996vz}
S.~Catani and M.~H. Seymour, \emph{{A General algorithm for calculating jet
  cross-sections in NLO QCD}},
  \href{https://doi.org/10.1016/S0550-3213(96)00589-5,
  10.1016/S0550-3213(98)81022-5}{\emph{Nucl. Phys.} {\bfseries B485} (1997)
  291} [\href{https://arxiv.org/abs/hep-ph/9605323}{{\ttfamily
  hep-ph/9605323}}].

\bibitem{Catani:1998bh}
S.~Catani, \emph{{The Singular behavior of QCD amplitudes at two loop order}},
  \href{https://doi.org/10.1016/S0370-2693(98)00332-3}{\emph{Phys. Lett.}
  {\bfseries B427} (1998) 161}
  [\href{https://arxiv.org/abs/hep-ph/9802439}{{\ttfamily hep-ph/9802439}}].

\bibitem{deWit:2002vz}
B.~de~Wit, \emph{{Supergravity}},  in \emph{{Unity from duality: Gravity, gauge
  theory and strings. Proceedings, NATO Advanced Study Institute, Euro Summer
  School, 76th session, Les Houches, France, July 30-August 31, 2001}},
  pp.~1--135, 2002, \href{https://arxiv.org/abs/hep-th/0212245}{{\ttfamily
  hep-th/0212245}}.

\bibitem{Cheung:2009dc}
C.~Cheung and D.~O'Connell, \emph{{Amplitudes and Spinor-Helicity in Six
  Dimensions}},
  \href{https://doi.org/10.1088/1126-6708/2009/07/075}{\emph{JHEP} {\bfseries
  0907} (2009) 075} [\href{https://arxiv.org/abs/0902.0981}{{\ttfamily
  0902.0981}}].

\bibitem{Boels:2009bv}
R.~Boels, \emph{{Covariant representation theory of the Poincare algebra and
  some of its extensions}},
  \href{https://doi.org/10.1007/JHEP01(2010)010}{\emph{JHEP} {\bfseries 01}
  (2010) 010} [\href{https://arxiv.org/abs/0908.0738}{{\ttfamily 0908.0738}}].

\bibitem{Dennen:2009vk}
T.~Dennen, Y.-t. Huang and W.~Siegel, \emph{{Supertwistor space for 6D maximal
  super Yang-Mills}},
  \href{https://doi.org/10.1007/JHEP04(2010)127}{\emph{JHEP} {\bfseries 04}
  (2010) 127} [\href{https://arxiv.org/abs/0910.2688}{{\ttfamily 0910.2688}}].

\bibitem{Bern:2010qa}
Z.~Bern, J.~J. Carrasco, T.~Dennen, Y.-t. Huang and H.~Ita, \emph{{Generalized
  Unitarity and Six-Dimensional Helicity}},
  \href{https://doi.org/10.1103/PhysRevD.83.085022}{\emph{Phys.Rev.} {\bfseries
  D83} (2011) 085022} [\href{https://arxiv.org/abs/1010.0494}{{\ttfamily
  1010.0494}}].

\bibitem{Huang:2011um}
Y.-t. Huang, \emph{{Non-Chiral S-Matrix of N=4 Super Yang-Mills}},
  \href{https://arxiv.org/abs/1104.2021}{{\ttfamily 1104.2021}}.

\bibitem{Elvang:2011fx}
H.~Elvang, Y.-t. Huang and C.~Peng, \emph{{On-shell superamplitudes in N $<$ 4
  SYM}}, \href{https://doi.org/10.1007/JHEP09(2011)031}{\emph{JHEP} {\bfseries
  1109} (2011) 031} [\href{https://arxiv.org/abs/1102.4843}{{\ttfamily
  1102.4843}}].

\bibitem{Collins:1984xc}
J.~C. Collins, \emph{{Renormalization}}, vol.~26 of \emph{Cambridge Monographs
  on Mathematical Physics}. Cambridge University Press, Cambridge, 1986,
  \href{https://doi.org/10.1017/CBO9780511622656}{10.1017/CBO9780511622656}.

\bibitem{tHooft:1972tcz}
G.~'t~Hooft and M.~J.~G. Veltman, \emph{{Regularization and Renormalization of
  Gauge Fields}},
  \href{https://doi.org/10.1016/0550-3213(72)90279-9}{\emph{Nucl. Phys.}
  {\bfseries B44} (1972) 189}.

\bibitem{Gnendiger:2017pys}
C.~Gnendiger et~al., \emph{{To ${d}$, or not to ${d}$: recent developments and
  comparisons of regularization schemes}},
  \href{https://doi.org/10.1140/epjc/s10052-017-5023-2}{\emph{Eur. Phys. J.}
  {\bfseries C77} (2017) 471}
  [\href{https://arxiv.org/abs/1705.01827}{{\ttfamily 1705.01827}}].

\bibitem{Bern:2013yya}
Z.~Bern, S.~Davies, T.~Dennen, Y.-t. Huang and J.~Nohle,
  \emph{{Color-Kinematics Duality for Pure Yang-Mills and Gravity at One and
  Two Loops}}, \href{https://doi.org/10.1103/PhysRevD.92.045041}{\emph{Phys.
  Rev.} {\bfseries D92} (2015) 045041}
  [\href{https://arxiv.org/abs/1303.6605}{{\ttfamily 1303.6605}}].

\bibitem{Nohle:2013bfa}
J.~Nohle, \emph{{Color-Kinematics Duality in One-Loop Four-Gluon Amplitudes
  with Matter}},
  \href{https://doi.org/10.1103/PhysRevD.90.025020}{\emph{Phys.Rev.} {\bfseries
  D90} (2014) 025020} [\href{https://arxiv.org/abs/1309.7416}{{\ttfamily
  1309.7416}}].

\bibitem{Ochirov:2013xba}
A.~Ochirov and P.~Tourkine, \emph{{BCJ duality and double copy in the closed
  string sector}}, \href{https://doi.org/10.1007/JHEP05(2014)136}{\emph{JHEP}
  {\bfseries 05} (2014) 136} [\href{https://arxiv.org/abs/1312.1326}{{\ttfamily
  1312.1326}}].

\bibitem{Badger:2016egz}
S.~Badger, G.~Mogull and T.~Peraro, \emph{{Local integrands for two-loop QCD
  amplitudes}},  in \emph{{13th DESY Workshop on Elementary Particle Physics:
  Loops and Legs in Quantum Field Theory (LL2016) Leipzig, Germany, April
  24-29, 2016}}, 2016, \href{https://arxiv.org/abs/1607.00311}{{\ttfamily
  1607.00311}},
  \href{http://inspirehep.net/record/1473353/files/arXiv:1607.00311.pdf}{http://inspirehep.net/record/1473353/files/arXiv:1607.00311.pdf}.

\bibitem{CaronHuot:2012ab}
S.~Caron-Huot and K.~J. Larsen, \emph{{Uniqueness of two-loop master
  contours}}, \href{https://doi.org/10.1007/JHEP10(2012)026}{\emph{JHEP}
  {\bfseries 10} (2012) 026} [\href{https://arxiv.org/abs/1205.0801}{{\ttfamily
  1205.0801}}].

\bibitem{Bern:2009xq}
Z.~Bern, J.~Carrasco, H.~Ita, H.~Johansson and R.~Roiban, \emph{{On the
  Structure of Supersymmetric Sums in Multi-Loop Unitarity Cuts}},
  \href{https://doi.org/10.1103/PhysRevD.80.065029}{\emph{Phys.Rev.} {\bfseries
  D80} (2009) 065029} [\href{https://arxiv.org/abs/0903.5348}{{\ttfamily
  0903.5348}}].

\bibitem{Kalin1305181}
G.~Kälin, \emph{Scattering Amplitudes in Supersymmetric Quantum Chromodynamics
  and Gravity}, Ph.D. thesis, Acta Universitatis Upsaliensis, 2019.
\newblock
  \href{http://urn.kb.se/resolve?urn=urn:nbn:se:uu:diva-381772}{http://urn.kb.se/resolve?urn=urn:nbn:se:uu:diva-381772}.

\bibitem{Bern:2017yxu}
Z.~Bern, J.~J. Carrasco, W.-M. Chen, H.~Johansson and R.~Roiban, \emph{{Gravity
  Amplitudes as Generalized Double Copies of Gauge-Theory Amplitudes}},
  \href{https://doi.org/10.1103/PhysRevLett.118.181602}{\emph{Phys. Rev. Lett.}
  {\bfseries 118} (2017) 181602}
  [\href{https://arxiv.org/abs/1701.02519}{{\ttfamily 1701.02519}}].

\bibitem{Bern:2017ucb}
Z.~Bern, J.~J.~M. Carrasco, W.-M. Chen, H.~Johansson, R.~Roiban and M.~Zeng,
  \emph{{Five-loop four-point integrand of $N=8$ supergravity as a generalized
  double copy}}, \href{https://doi.org/10.1103/PhysRevD.96.126012}{\emph{Phys.
  Rev.} {\bfseries D96} (2017) 126012}
  [\href{https://arxiv.org/abs/1708.06807}{{\ttfamily 1708.06807}}].

\bibitem{Bern:2018jmv}
Z.~Bern, J.~J. Carrasco, W.-M. Chen, A.~Edison, H.~Johansson, J.~Parra-Martinez
  et~al., \emph{{Ultraviolet Properties of $\mathcal N = 8$ Supergravity at
  Five Loops}}, \href{https://doi.org/10.1103/PhysRevD.98.086021}{\emph{Phys.
  Rev.} {\bfseries D98} (2018) 086021}
  [\href{https://arxiv.org/abs/1804.09311}{{\ttfamily 1804.09311}}].

\bibitem{Mogull:2015adi}
G.~Mogull and D.~O'Connell, \emph{{Overcoming Obstacles to Colour-Kinematics
  Duality at Two Loops}},
  \href{https://doi.org/10.1007/JHEP12(2015)135}{\emph{JHEP} {\bfseries 12}
  (2015) 135} [\href{https://arxiv.org/abs/1511.06652}{{\ttfamily
  1511.06652}}].

\bibitem{Arkani-Hamed:2014bca}
N.~Arkani-Hamed, J.~L. Bourjaily, F.~Cachazo, A.~Postnikov and J.~Trnka,
  \emph{{On-Shell Structures of MHV Amplitudes Beyond the Planar Limit}},
  \href{https://doi.org/10.1007/JHEP06(2015)179}{\emph{JHEP} {\bfseries 1506}
  (2015) 179} [\href{https://arxiv.org/abs/1412.8475}{{\ttfamily 1412.8475}}].

\bibitem{Bern:2014kca}
Z.~Bern, E.~Herrmann, S.~Litsey, J.~Stankowicz and J.~Trnka, \emph{{Logarithmic
  Singularities and Maximally Supersymmetric Amplitudes}},
  \href{https://doi.org/10.1007/JHEP06(2015)202}{\emph{JHEP} {\bfseries 06}
  (2015) 202} [\href{https://arxiv.org/abs/1412.8584}{{\ttfamily 1412.8584}}].

\bibitem{Drummond:2008bq}
J.~Drummond, J.~Henn, G.~Korchemsky and E.~Sokatchev, \emph{{Generalized
  unitarity for N=4 super-amplitudes}},
  \href{https://doi.org/10.1016/j.nuclphysb.2012.12.009}{\emph{Nucl.Phys.}
  {\bfseries B869} (2013) 452}
  [\href{https://arxiv.org/abs/0808.0491}{{\ttfamily 0808.0491}}].

\end{thebibliography}\endgroup
\end{document}